\documentclass[reprint,pre,aps,showpacs,tightenlines,superscriptaddress,amsmath,amssymb,showkeys,10pt]{revtex4-2}
\usepackage{slashed}
\usepackage{bm}
\usepackage[colorlinks,linkcolor=blue,citecolor=blue,urlcolor=blue]{hyperref}
\usepackage{xcolor}
\usepackage{graphicx}
\usepackage{dcolumn}
\usepackage{graphicx}
\usepackage{subcaption}
\usepackage{ragged2e}
\usepackage{bm}
\usepackage{svg}
\usepackage[normalem]{ulem}
\usepackage{soul}

\usepackage{xcolor} 
\usepackage{mathrsfs}

\definecolor{winered}{RGB}{128, 0, 32}

\newcommand{\be}{\begin{equation}}\newcommand{\ee}{\end{equation}}
\newcommand{\bea}{\begin{eqnarray}}\newcommand{\eea}{\end{eqnarray}}
\newcommand{\brr}{\begin{array}}\newcommand{\err}{\end{array}}
\newcommand{\bit}{\begin{itemize}}\newcommand{\eit}{\end{itemize}}
\newcommand{\ben}{\begin{enumerate}}\newcommand{\een}{\end{enumerate}}

\newcommand{\ba}{\begin{array}}
\newcommand{\ea}{\end{array}}

\definecolor{darkred}{rgb}{.8,0,0}

\definecolor{darkblue}{rgb}{0,0,0}

\def\1{{_{1}}}\def\2{{_{2}}}

\def\noHe0{:\;\!\!\;\!\!:H_e(0):\;\!\!\;\!\!:}
\def\noHm0{:\;\!\!\;\!\!:H_\mu(0):\;\!\!\;\!\!:}

\begin{document}


\title{
On the Practical Estimation and Interpretation of Rényi Transfer Entropy
}

\author{Zlata Tabachov\'{a}}
\email{zlatatabachova@gmail.com}
\thanks{Corresponding author.}
\affiliation{FNSPE,
Czech Technical University in Prague, B\v{r}ehov\'{a} 7, 115 19, Prague, Czech Republic}
\affiliation{Department of Complex Systems, Institute of Computer Science of the Czech
Academy of Sciences, Pod Vodárenskou věží 2, 182 00 Prague 8, Czech Republic}

\author{Petr Jizba}
\email{p.jizba@fjfi.cvut.cz}
\affiliation{FNSPE,
Czech Technical University in Prague, B\v{r}ehov\'{a} 7, 115 19, Prague, Czech Republic}

\author{Hynek Lavi\v{c}ka}
\email{hynek.lavicka@gmail.com}
\affiliation{FNSPE,
Czech Technical University in Prague, B\v{r}ehov\'{a} 7, 115 19, Prague, Czech Republic}
\affiliation{Cerabyte GmbH, Rundfunkplatz 2, 80335 Munich, Germany }

\author{Milan Palu\v{s}}
\email{mp@cs.cas.cz}
\affiliation{Department of Complex Systems, Institute of Computer Science of the Czech
Academy of Sciences, Pod Vodárenskou věží 2, 182 00 Prague 8, Czech Republic}

%
%


\date{\today}

\begin{abstract}
Rényi transfer entropy (RTE) is a generalization of classical transfer entropy that replaces Shannon's entropy with Rényi's information measure. This, in turn, introduces a new tunable parameter $\alpha$, which accounts for sensitivity to low- or high-probability events.
Although RTE shows strong potential for analyzing causal relations in complex, non-Gaussian systems, its practical use is limited, primarily due to challenges related to its accurate estimation and interpretation. These difficulties are especially pronounced when working with finite,  high-dimensional, or heterogeneous datasets.
In this paper, we systematically study the performance of a $k$-nearest neighbor estimator for both Rényi entropy (RE) and RTE using various synthetic data sets with clear cause-and-effect relationships inherent to their construction.  We test the estimator across a broad range of parameters, including sample size, dimensionality, memory length, and Rényi order $\alpha$. 
In particular, we apply the estimator to a set of simulated processes with increasing structural complexity, ranging from linear dynamics to nonlinear systems with multi-source couplings.
To address interpretational challenges arising from potentially negative RE and RTE values, we introduce three reliability conditions and formulate practical guidelines for tuning the estimator parameters. 
%
%
%
%
%
We show that when the reliability conditions are met and the parameters are calibrated accordingly, the resulting effective RTE estimates accurately capture directional information flow across a broad range of scenarios. 
Results obtained show that the explanatory power of RTE depends sensitively on the choice of the R\'{e}nyi parameter $\alpha$. This highlights the usefulness of the RTE framework for identifying the drivers of extreme behavior in complex systems.


\end{abstract}

\maketitle


\section{Introduction}

Understanding the causal mechanisms in complex systems represents a fundamental and ongoing challenge across a broad range of scientific fields, including neuroscience, energetics, climate science, finance, engineering, and nonlinear dynamical systems~\cite{kantz2003nonlinear,lizier2010differentiating}. These systems are often characterized by high-dimensional, non-stationary, and nonlinear interactions, which complicate the identification of causal or effective relationships among their constituent components. 
%
When the processes under study evolve over time and are represented by sequences of measurements recorded at successive time points --- commonly arranged as time series --- it becomes possible to apply quantitative methods to infer causal relationships between variables across different datasets.
One of the foundational approaches to causality in time series analysis is {\em the Granger causality} (GC), which was originally introduced by Granger~\cite{Granger} and  based on earlier ideas about predictability proposed by Wiener~\cite{Wiener}. Granger's formulation is grounded in linear autoregressive models, and it has been widely used due to its mathematical tractability and  conceptual simplicity.

In practice, however, many real-world systems exhibit complex, nonlinear dynamics that fall outside the scope of linear models. To address this limitation, a variety of nonlinear extensions of GC have been proposed in the literature, drawing on ideas from the theory of nonlinear dynamical systems ~\cite{Quiroga:2000,Ye:2015,Palus:2018c}, data compression efficiency~\cite{Kathpalia:2022},  machine learning~\cite{Huang:2020}, and information theory~\cite{ay2008information,jizba2012renyi,JLT:22,hlavavckova2007causality}, among others.  
In particular, information theory provides powerful tools and conceptually transparent frameworks for analyzing causal relationships through directional dependencies in information flow.
In this context, one of the most widely adopted tools for detecting and quantifying the directional information flow between time-dependent processes is {\em transfer entropy} (TE) or equally conditional mutual information (CMI) --- an information-theoretic measure introduced by  Schreiber~\cite{schreiber2000measuring}, Kantz et al.~\cite{marschinski2002analysing}, and Paluš et al.~\cite{paluvs2001synchronization}.

In essence, TE quantifies the amount of directed (time-asymmetric) information flow from a source process to a target process. It does so by measuring the reduction in uncertainty about the target's future state when the past states of the source are considered, beyond what is already explained by the target's own history.
Unlike methods based on explicit modeling assumptions, TE is model-independent and does not require prior knowledge of the underlying dynamics. This makes TE  particularly suitable for analyzing complex systems, where interactions may be highly nonlinear or otherwise difficult to parametrize.  Consequently, TE  has been successfully applied in diverse scientific disciplines~\cite{Gencaga:18,Bossomaier}, including the Earth sciences~\cite{paluvs2024causes,Jajcay:2018}, neurosciences~\cite{PEREDA20051,Vicente:2011} and financial markets~\cite{marschinski2002analysing,Permuter:2011,jizba2012renyi}.

Although most TE applications rely on the entropy framework introduced by Shannon, recent studies have begun to explore the utility of Rényi's information theory~\cite{Renyi:1976a,renyi1961measures,JA,paluvs2024causes}. The latter provides a natural generalization of conventional TE and, in doing so, it  introduces a new tunable parameter, $\alpha$, which modulates sensitivity to rare versus frequent events. The resulting extension, known as Rényi transfer entropy (RTE), offers several advantages,
particularly for systems exhibiting heavy-tailed distributions, skewness, or the presence of outliers that play a critical role in the dynamics of the system~\cite{jizba2012renyi,JLT:22,paluvs2024causes,ZHANG2023,BEHRENDT2019}.  
The RTE framework was originally proposed in Ref.~\cite{jizba2012renyi} to investigate causal information transfer between bivariate financial time series. The underlying idea was that specific values of the Rényi parameter $\alpha$ could selectively amplify only certain regions of the underlying probability distribution function (PDF) while strongly attenuating others. Since extreme events are typically associated with the tails of PDFs, RTE seems to be particularly well-suited for identifying the causal drivers of such rare events. This expectation has been confirmed in a number of works~\cite{jizba2012renyi,JLT:22,paluvs2024causes,ZHANG2023,Korbel:2019,BEHRENDT2019}. 
Yet, despite its theoretical appeal, RTE remains underutilized, largely due to a lack of systematic studies addressing its estimation and interpretability, especially in the context of empirical or high-dimensional data.

In this paper, we conduct a comprehensive evaluation of the $k$-nearest neighbors ($k$-NN) estimator for RE and RTE, originally proposed by Leonenko et al.~\cite{leonenko2008class}. Our analysis proceeds in two stages. First, we systematically investigate how key parameters --- such as sample size, number of neighbors $k$, data dimensionality and memory length --- affect estimation accuracy across a broad range of Rényi $\alpha$-parameters, with particular attention to the regime $\alpha < 1$, which emphasizes tail events. To explore diverse PDF behaviors, we benchmark the estimator against the RE of the multidimensional $t$-distribution, which encompasses both the normal and Cauchy distributions as limiting cases with exponentially and polynomially decaying tails, respectively.

In the second stage, we apply the estimator to a set of simulated coupled stochastic processes with explicitly defined causal structures. This setup enables a controlled evaluation of the estimator’s performance in capturing directional information flow. Specifically, we estimate and analyze RTE for four synthetic process pairs with increasing structural complexity, including linear, nonlinear, and multi-source couplings.


The paper is organized as follows. In the following section, we present the fundamental concepts underlying RTE and the $k$-NN–based estimator. 
Section~\ref{sec_estimation_RE} examines the estimator's performance across key parameters for both normal and heavy-tailed distributions, specifically multivariate Student's $t$-distributions. In Section~\ref{sec_est_RTE}, we evaluate RTE for four synthetic coupled stochastic processes. A detailed discussion of the results concerning both the RE estimator and the synthetic stochastic processes employed is presented in Section~\ref{sec_discussion}. This section also provides practical guidelines for parameter tuning, summarizes the reliability conditions for RTE estimation, and addresses key challenges ---particularly those associated with small values of $\alpha$. Section~\ref{sec_conclusion} concludes the paper by summarizing the main findings and outlining potential directions for future research. Additional technical details and notational clarifications are provided in {five} accompanying appendices.

\section{Rényi transfer entropy}\label{sec_methods}
\subsection{Some Fundamenetals of RTE}

Let us consider three stochastic processes 
\begin{eqnarray}
X = \{x_i,\}_{i=1}^N\, , \;\; Y = \{y_i,\}_{i=1}^N \;\;\ \mbox{and}\  \;\; Z = \{z_i,\}_{i=1}^N\, , ~~~~~~
\end{eqnarray}
with respective Markov orders $r$, $l$ and $s$. We assume that these processes are directionally dependent, in particular that (apart of its own past) $X$ is influenced by $Y$ and $Z$, i.e. $X=f(X,Y,Z)$. To quantify directional information flow between stochastic processes, Rényi transfer entropy offers a natural extension of the standard Shannon-based information transfer~\cite{schreiber2000measuring,marschinski2002analysing,palus2003}. 
In particular, it quantifies the extent to which the future state of a target process $X$ is determined by the past of a source process $Y$ (with the influence of a third process $Z$ incorporated below), conditioned on the past of $X$ itself. Formally, RTE is defined as follows~\cite{jizba2012renyi} 
\begin{eqnarray}
  \label{RTE}
    T^R_{\alpha,Y\rightarrow X}(r,l)
    &=&I_{\alpha}(x_{t+1}:y_t^{(l)}|x_t^{(r)})\nonumber \\[2mm]
    &=&
    H_{\alpha}(x_{t+1}|x_t^{(r)})   -H_{\alpha}(x_{t+1}|x_t^{(r)},y_t^{(l)})\,,~~~~
    \label{apparent_rte}
\end{eqnarray}
where $I_\alpha$ and $H_\alpha$ are Rényi conditional mutual information and Rényi entropy, respectively. The parameter $\alpha \in \mathbb{R}_{0}^{+}$ specifies the order of the entropy, and $y_t^{(l)}=y_t,y_{t-1},\ldots,y_{t-l+1}$. For formal definitions of $I_{\alpha}$ and $H_{\alpha}$ we refer the reader to  Appendix \ref{Appendix_RE}. 

When $\alpha = 1$, the RTE reduces to the Shannon-based transfer entropy originally introduced by Schreiber~\cite{schreiber2000measuring}. Consequently, RTE represents a generalization of the conventional Shannon transfer entropy and possesses the following essential properties:

\begin{itemize}

    \item In general $T^R_{\alpha,Y\rightarrow X}\neq T^R_{\alpha,X\rightarrow Y}$, reflecting its directional nature.

    \item $T^R_{\alpha=1,Y\rightarrow X} \geq 0$, as the Shannon TE is a special case defined via the Kullback--Leibler divergence.

    \item $T^R_{\alpha=1,Y\rightarrow X} = 0$ implies statistical independence between $X$ and $Y$.
    \item For $\alpha\neq 1$, RTE can take negative values, as additional conditioning, e.g. via process $y_t^{(l)}$, can increase RE, i.e. $ H_{\alpha}(x_{t+1}|x_t^{(r)},y_t^{(l)}) > H_{\alpha}(x_{t+1}|x_t^{(r)})$.
\end{itemize}

Due to its inherent directionality, transfer entropy is widely used as a proxy for \textit{information flow} \cite{ay2008information}, serving as an information-theoretic indicator of causality. However, it is important to distinguish between predictive power and causal influence: TE primarily measures how much knowing the source reduces uncertainty about the target, whereas information flow is concerned with how changes in the source directly affect the target. As shown in \cite{lizier2010differentiating}, under proper conditioning and with appropriate memory parameters, \textit{complete transfer entropy} can serve as a robust approximation to actual information flow. Although a full theoretical treatment of information flow in the context of Rényi entropy is beyond the scope of this paper, we emphasize its relevance, as our empirical findings suggest strong parallels. 

Following the framework proposed by Prokopenko and Lizier \cite{lizier2010differentiating}, who distinguish between \textit{apparent} and \textit{complete} Shannon transfer entropy, we extend this distinction to the Rényi setting.

When the target process is described as $X = f(X, Y, Z)$,  computing RTE using Equation~\ref{apparent_rte}—without accounting for all influencing sources—yields only the apparent transfer entropy from $Y$ to $X$. To obtain the complete transfer entropy, one must condition on all sources influencing the target, including $Z$. Thus, the \textit{complete Rényi transfer entropy} from $Y$ to $X$ is defined as:
\begin{eqnarray}
  \label{complete_rte}
    \bar{T}^{R}_{\alpha,Y\rightarrow X|Z}(r,l,s)
    &=&I_{\alpha}(x_{t+1}:y_t^{(l)}|x_t^{(r)},z_t^{(s)})\nonumber \\[2mm]
    &=&
    H_{\alpha}(x_{t+1}|x_t^{(r)},z_t^{(s)}) \nonumber \\[2mm]
    &-&H_{\alpha}(x_{t+1}|x_t^{(r)},z_t^{(s)},y_t^{(l)})\, .  
\end{eqnarray}
In general, when the target process is directly influenced by multiple sources $Z_1, \ldots, Z_n$, it is essential to condition on all of them to accurately capture the underlying causal structure. Equally important is the careful selection of memory parameters — $r$ and $l$ for the target and source processes, and $s_1, \ldots, s_n$ for the additional sources. These parameters determine the temporal depth of dependencies and are critical for obtaining the complete Rényi transfer entropy rather than a merely apparent one. Therefore, accurate identification of all relevant source processes, along with appropriate memory lengths, is crucial for approximating the true information flow.

Finally, in cases where the target process $X$ is influenced by only a single source $Y$, the complete and apparent RTE are equivalent, that is, $\bar{T}^{R}_{\alpha,Y\rightarrow X} \equiv T^{R}_{\alpha,Y\rightarrow X}$. Additionally, we note that the definition $\bar{T}^{R}_{\alpha}$ corresponds to the \textit{conditional transfer entropy} or \textit{causation entropy} introduced in~\cite{sun2014causation}. For clarity, we will use the notation $T^{CR}_{\alpha}$ to denote conditional RTE explicitly. If conditioning is incomplete or memory parameters are chosen incorrectly, the resulting conditional RTE should be interpreted as an apparent rather than complete measure of information transfer. Therefore, while the mathematical definitions of RTE, complete RTE, and conditional RTE may coincide in form, their interpretation as either complete or apparent depends on the specific structure of the process under consideration.

\subsection{Estimation of RTE}

Using the expression given in Eq.~\eqref{complete_rte} RTE can be equivalently rewritten as:
\begin{widetext}
\begin{eqnarray}
      T^R_{\alpha,Y\rightarrow X|Z}(r,l,s)\ &=& \ \underbrace{H_\alpha(x_{t+1},x_t^{(r)},z_t^{(s)})}_{\mathscr{I}_{1}}\ - \  \underbrace{\ H_\alpha(x_t^{(r)},z_t^{(s)})}_{\mathscr{I}_{2}}
        \nonumber \\[3mm] &-& \ \underbrace{H_{\alpha}(x_{t+1},x_t^{(r)},z_t^{(s)},y_t^{(l)})}_{\mathscr{I}_{3}}\ +  \ \ \underbrace{H_{\alpha}(x_{t}^{(r)},z_t^{(s)},y_t^{(l)})}_{\mathscr{I}_{4}}\, .~~~~~
        \label{complete_H}
\end{eqnarray}
\end{widetext}
Thus, computing RTE reduces to estimating four REs, which we denote for simplicity as $\mathscr{I}_1$, $\mathscr{I}_2$, $\mathscr{I}_3$, and $\mathscr{I}_4$. The dimensionality of each entropy term depends on the Markov orders $r$, $l$, and $s$ associated with the target, source, and additional source processes, respectively.

To estimate these entropy terms, we adopt the $k$-NN estimator of the differential RE, i.e. 
\begin{eqnarray}
H_\alpha\left(f_X\right)=\frac{1}{1-\alpha} \log_2 \int_{\mathbb{R}^d} f^{\alpha}(x)d^dx\, ,
\end{eqnarray}
developed by Leonenko et al.~\cite{leonenko2008class} for a random vector $X\in \mathbb{R}^d$ with the probability density function  $f_X$ on bounded support, which is well-defined with respect to the $d$-dimensional Lebesgue measure. This estimator is given by:
\begin{widetext}
    \begin{equation}
        \hat{H}_\alpha(N,k,\alpha,d) \ = \  \frac{1}{1-\alpha}\log_2 \left(  \frac{N-1}{N}B_d^{1-\alpha}C_k^{1-\alpha} \sum_{i=1}^N \frac{\rho_{k,i}^{d(1-\alpha)}}{(N-1)^\alpha}\right),
        \label{eq_leonenko}
    \end{equation}
    
\end{widetext}
where $C_k = \left[\frac{\Gamma(k)}{\Gamma(k+1-\alpha)}\right]^{1/(1-\alpha)}$, and $B_d = \frac{\pi^{d/2}}{\Gamma(d/2+1)}$ denotes the volume of the $d$-dimensional unit ball. The term $\rho_{k,i}$ represents the {$d$-dimensional} Euclidean distance between the $i$th point and its $k$th nearest neighbor~\footnote{While the estimator uses Euclidean distance by default, alternative distance measures such as the Mahalanobis distance can be applied when appropriate, particularly for spherically distributed data.}.

The definition of the estimator for Shannon entropy (i.e., when $\alpha = 1$) is also provided in~\cite{leonenko2008class}. Nonetheless, since Rényi entropy is a continuous function of $\alpha$, its value at $\alpha = 1$ can also be approximated via interpolation.

It is important to note that Leonenko's estimator is derived for differential Rényi entropy, i.e., the Rényi entropy of a continuous probability density function. In contrast to the discrete case, non-negativity of differential Rényi entropy is not guaranteed --- not even for Shannon's entropy~\cite{cover1999elements,van2013detection}. Negativity can occur for degenerate distributions or highly leptokurtic full-dimensional distributions.

Because Rényi transfer entropy is originally defined for discrete processes, it is important to assess how the use of a continuous estimator may affect the interpretation of the measure. The relationship between the discrete and continuous forms of Rényi entropy is given by~\cite{Renyi:1976a,jizba2004world}:
\begin{equation}
H_\alpha\!\left(f_X\right) \ = \  \lim_{M \rightarrow\infty} \left[ H_\alpha(X) \ - \ d\log M\right]\, .
\label{eq_RE_cont_to_discr}
\end{equation}
Here, $M$ denotes the number of $d$-dimensional boxes used to tessellate the continuous space. For example, in one dimension, $M$ corresponds to the number of intervals or bins. 

Relation~(\ref{eq_RE_cont_to_discr}) indicates that the transition from the discrete to the continuous case introduces a constant offset, which depends on both the dimensionality of the space and the discretization scale.

\section{Parameters in Estimation of Rényi entropy using $k$-NN method }\label{sec_estimation_RE}

\subsection{Sample Size Parameter $N$.}

\begin{figure}[] 
    {\includegraphics[width=0.48\textwidth]{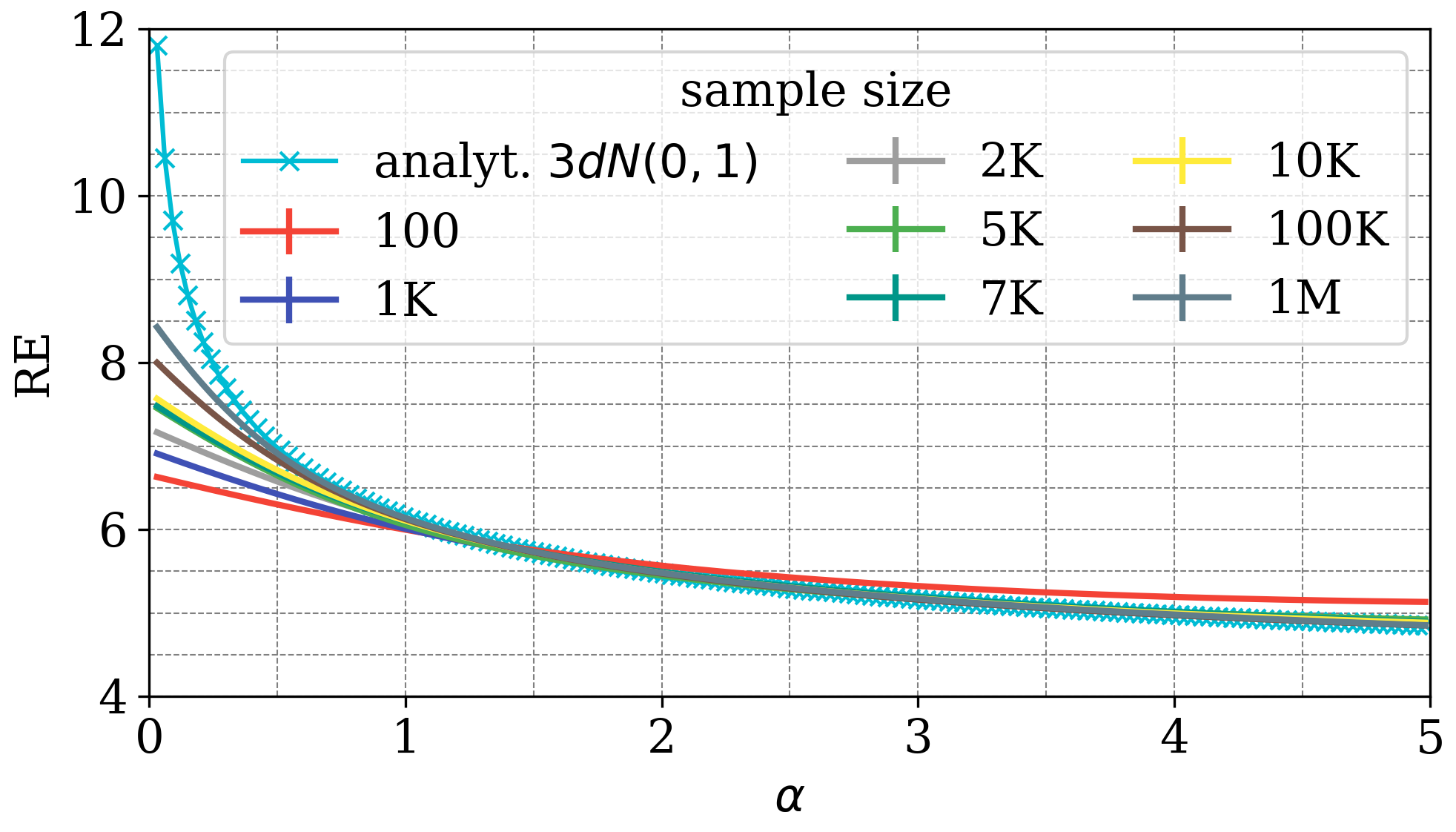}}
    \caption{\justifying \footnotesize {\em Effect of Sample Size $N$ on the Estimation of Rényi Entropy for a 3D Normal Distribution.} The $x$-axis represents the Rényi parameter $\alpha$, while the $y$-axis shows the estimated Rényi entropy for a 3-dimensional normal distribution with identity covariance. Analytical values are indicated by cyan crosses. Solid lines show estimates obtained using $k = 10$ for sample sizes $N = 100$, $10^3$, $2 \times 10^3$, $5 \times 10^3$, $7 \times 10^3$, $10^4$, $10^5$, and $10^6$, corresponding to red, blue, light gray, green, turquoise, yellow, brown, and dark gray, respectively. For $\alpha > 1$, accurate estimates are achieved even for moderate sample sizes (e.g., $N = 10^3$). In contrast, estimates for small $\alpha$—which emphasize low-probability events—are highly sensitive to $N$, and become reliable only for $\alpha > 0.5$ even at $N = 10^6$. This reduced accuracy at low $\alpha$ is partly attributable to challenges in estimating densities on unbounded support. }
    \label{fig_est_norm_size}
\end{figure}

The accuracy of the $k$-NN estimator is highly sensitive to the sample size. Larger sample sizes yield more reliable estimates of Rényi entropy (RE), as they more effectively capture the underlying distributional structure. This effect is illustrated in Fig.~\ref{fig_est_norm_size}, which shows how estimates of RE for a three-dimensional normal distribution converge to the analytical values as $N$ increases (and similarly for Cauchy distribution as shown in Fig.\ref{fig_cauchy_size}). 

The precision of RE estimates is also closely linked to the Rényi parameter $\alpha$. In particular, estimates for small $\alpha$, in the example in Fig.~\ref{fig_est_norm_size} for $\alpha<0.7$,  exhibit higher variability and systematic bias, even for moderately large samples. This is due in part to the fact that smaller $\alpha$ values emphasize low-probability (tail) events, which are often underrepresented in finite samples. A detailed summary of challenges associated with estimating RE for $\alpha<1$ is provided at the end of this section.

In the context of transfer entropy, the adverse effects of limited sample size can be mitigated using the \textit{effective Rényi transfer entropy} (ERTE) \cite{marschinski2002analysing}, defined as:
\begin{equation}
    T^{\text{eff}}_{\alpha,Y\rightarrow X}(r,l)\ \equiv \ T_{\alpha,Y\rightarrow X}(r,l) \ - \   T_{\alpha,Y_\text{sh} \rightarrow X}(r,l)\, .~~
    \label{eff}
\end{equation}
Here, the shuffled version of the source process $Y_{\text{sh}}$ preserves marginal statistics (e.g., mean, variance, or power spectrum), but destroys temporal dependencies using, e.g., Fisher--Yates algorithm \textcolor{red}~\cite{fischer-yates:1938}. Consequently, the transfer entropy computed from $Y_{\text{sh}}$ to $X$ should, in theory, vanish. Any non-zero estimate of this quantity reflects bias introduced by finite sample effects. By subtracting this baseline, ERTE provides a bias-corrected measure of directional entropy transfer. Note that the target process can also be shuffled instead of the source process. The decision on which process to shuffle should be based on identifying the main contributors to estimation bias, which typically arise from high dimensionality or heavy-tailed distributions.

Additionally, the concept of effective transfer entropy allows for the application of Leonenko's estimator of differential RE in the assessment of transfer entropy originally defined for discrete systems. When Eq.~(\ref{eq_RE_cont_to_discr}) is substituted into the expression for ERTE~(\ref{eff}), the dimension- and partition-dependent constants cancel out. As a result, only the estimated differential RE terms contribute to the final value. This cancellation justifies the use of differential entropy estimators within the framework of effective transfer entropy, even when the original formulation is discrete.

\subsection{Nearest-Neighbor Parameter $k$}

Among the key parameters in Rényi entropy estimation, the choice of the nearest-neighbor parameter $k$ is particularly nuanced and challenging. Although its optimal value naturally depends on the sample size $N$, a widely cited heuristic suggests setting $k=\sqrt{N}$, see~\cite{kantz2003nonlinear}. However, this rule does not take into account the local structure of the data distribution across the phase space.

Intuitively, densely populated regions can support larger $k$ values, as more neighbors are available within a small radius. In contrast, in sparsely populated regions, a smaller $k$ is preferable to avoid overestimating the local probability density. Choosing an excessively large $k$ in such areas may artificially inflate the estimated entropy, while too small $k$ in dense regions may lead to extremely short distances 
$\rho_{k,i}$, driving the argument of the logarithm in Eq.~\ref{eq_leonenko} close to zero and thus producing negative entropy estimates.

The sensitivity of the estimator to $k$ also varies with the Rényi parameter $\alpha$. For small $\alpha$, which gives greater weight to low-probability (tail) events, smaller $k$ values are more appropriate. Conversely, larger $\alpha$ values emphasize high-probability events and benefit from using higher $k$.

\begin{figure}[] 
    {\includegraphics[width=0.48\textwidth]{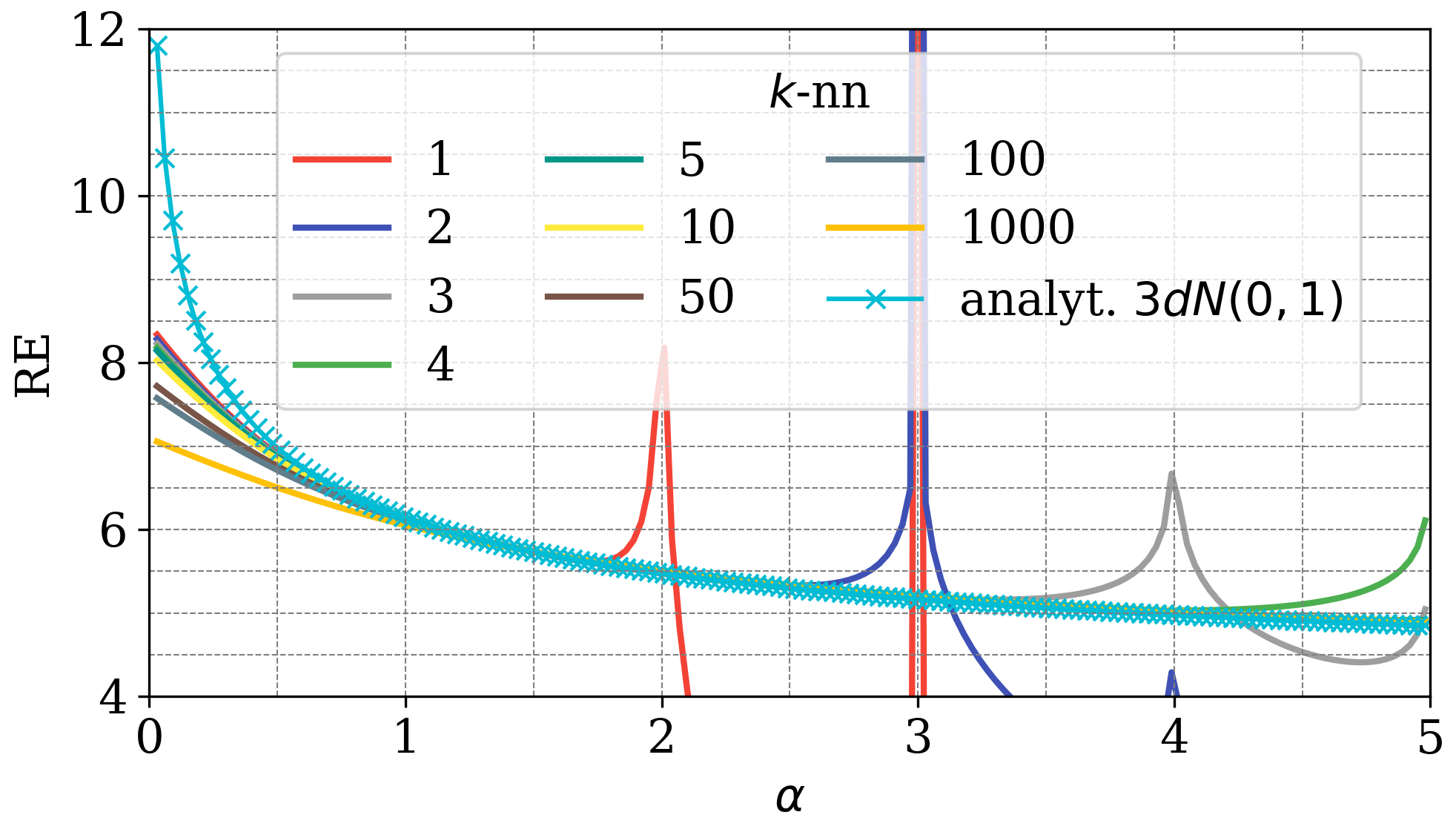}}
    \caption{\justifying \footnotesize {\em Influence of the $k$-Nearest Neighbor Parameter on Estimation Accuracy for 3D Normal Distribution.} The $x$-axis shows RE parameter $\alpha$ and the $y$-axis shows RE of 3-dim normal distribution with identity covariance matrix. Analytical values are indicated by cyan crosses and estimates for $k=1,2,3,4,5,10,50,100,1000$ are indicated by solid lines. Estimates are obtained on sample with $N=10^5$ data points. As the estimator is dependent on $\Gamma(k+1-\alpha)$ results diverge for particular combinations of $\alpha$ and $k$ for which the argument of the gamma function approaches zero and turns negative. For $\alpha < 1$, the estimates diverge at larger values of $k$, while higher values of $\alpha$ yield more stable results. This suggests that the optimal choice of $k$ may vary across different $\alpha$-intervals. Specifically, for $\alpha < 1$, the best performance is achieved with $k = 1$.
}
    \label{fig_est_norm_knn}
\end{figure}

These dependencies are demonstrated in Fig.~\ref{fig_est_norm_knn}, which shows RE estimates on a 3-dimensional normally distributed dataset of $N = 10^5$ samples, using a range of $k$ values: $k=1,2,3,4,5,10,50,100$ and $1000$ (for Cauchy distribution, see Fig.~\ref{fig_cauchy_neigh}). The most accurate estimates for small $\alpha$, i.e. $\alpha<1$,  are obtained with 
$k=1$, for which estimate converges to the theoretical values starting from $\alpha \approx 0.5$. Further improvements in precision can be achieved by increasing the sample size. Nevertheless, as highlighted in \cite{leonenko2008class}, for distributions with unbounded support—such as the Normal distribution—errors will arise in the limit  $\alpha \to 0$, as the unboundedness violates preconditions for convergence due to the increasing influence of extreme tail behavior. For $\alpha \geq 1$ high $k$ yields accurate estimates. Note, that there is a restriction for particular combinations of $k$ and $\alpha$ arising from the presence of function $\Gamma(k+1-\alpha)$ in the estimator. This is why results for small $k$ in Fig.~\ref{fig_est_norm_knn} diverge for $\alpha\geq k+1$.

\subsection{Sample Dimension Parameter $d$.}

\begin{figure}[] 
    {\includegraphics[width=0.48\textwidth]{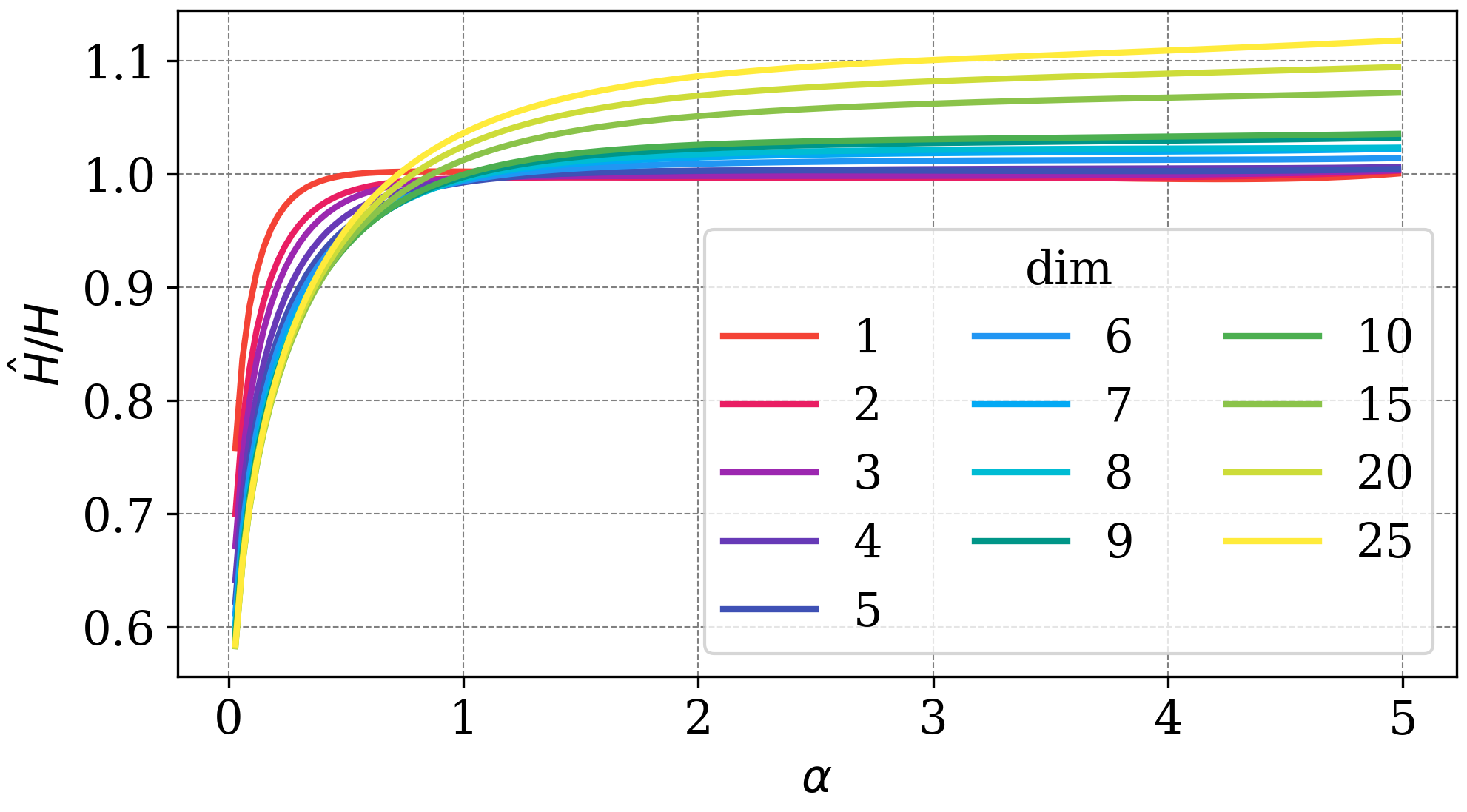}}

    \caption{\footnotesize\justifying {\em Effect of Sample Dimensionality $d$ on Estimation Accuracy.}
The $x$-axis represents the Rényi parameter $\alpha$, and the $y$-axis shows the ratio of estimated to analytical Rényi entropy for $d$-dimensional normal distributions with identity covariance matrices. Estimates are computed using samples of size $N = 10^5$. A ratio close to 1 indicates high estimation accuracy. The results demonstrate that increasing dimensionality significantly degrades estimator performance, particularly for smaller values of $\alpha$, where sensitivity to low-probability regions is greater. }
    \label{fig_est_dims}
\end{figure}

As illustrated in Fig.~\ref{fig_est_dims} and Fig.~\ref{fig_cauchy_dims2}, the accuracy of Rényi entropy estimation deteriorates as the dimensionality of the data increases. This effect is a manifestation of the well-known \textit{curse of dimensionality} \cite{bellman1961adaptive, bishop2006pattern, hastie2009elements}
, which limits the reliability of non-parametric estimators like the $k$-nearest neighbor method in high-dimensional spaces.

In the context of RTE estimation, the dimensionality of the entropy terms $\mathscr{I}_1$ through $\mathscr{I}_4$ is determined by the memory (or Markov) parameters $r$, $l$ and $s$ that define the lagged state vectors of the target, source, and additional processes, respectively, in Eq.~\ref{complete_H}. For instance, when $r=l=s=1$, the entropies take the following forms: $\mathscr{I}_1=H_\alpha(x_{t+1},x_t,z_t)$, $\mathscr{I}_2=H_\alpha(x_t,z_t)$, $\mathscr{I}_3=H_{\alpha}(x_{t+1},x_t,z_t,y_t)$ and $\mathscr{I}_4=H_{\alpha}(x_{t+1},x_t,z_t,y_t)$ of dimensions 3, 2, 4 and 3, respectively. 

Among these, $\mathscr{I}_3$ consistently has the highest dimensionality, and thus often contributes the most to estimation error. As the dimensionality increases, the density of points in the phase space decreases, reducing the reliability of local neighbor-based estimates. Therefore, careful selection of memory parameters is critical—not only for capturing the system's dynamics but also for ensuring estimator stability and interpretability.

\begin{figure*}[t] 

    \begin{subcaptionbox}{ \footnotesize \justifying {\em Estimated REs of 3-dim Multivariate $t$-distribution for various degrees of freedom, $\nu$.} The $x$-axis shows the Rényi entropy parameter $\alpha$, and the $y$-axis represents Rényi entropy (RE) in bits. Analytical results are marked with \texttt{x}-symbols, while solid lines indicate estimates obtained using $N = 10^6$ samples and $k = 1$, as this setting provides the most accurate estimates for $\alpha < 1$. For $\nu = 1$, the underlying distribution is Cauchy, which has the heaviest tails. In contrast, as $\nu \to +\infty$, the distribution converges to the multivariate normal distribution with exponentially decaying tails. For $\nu = 1000$, the RE is already quantitatively close to that of a 3-dimensional normal distribution (see Fig.~\ref{fig_est_norm_size} for comparison). Deviations between analytical and estimated values are discussed in panel (b).
    \label{fig_student_nu}}
        {\includegraphics[width=0.48\textwidth]{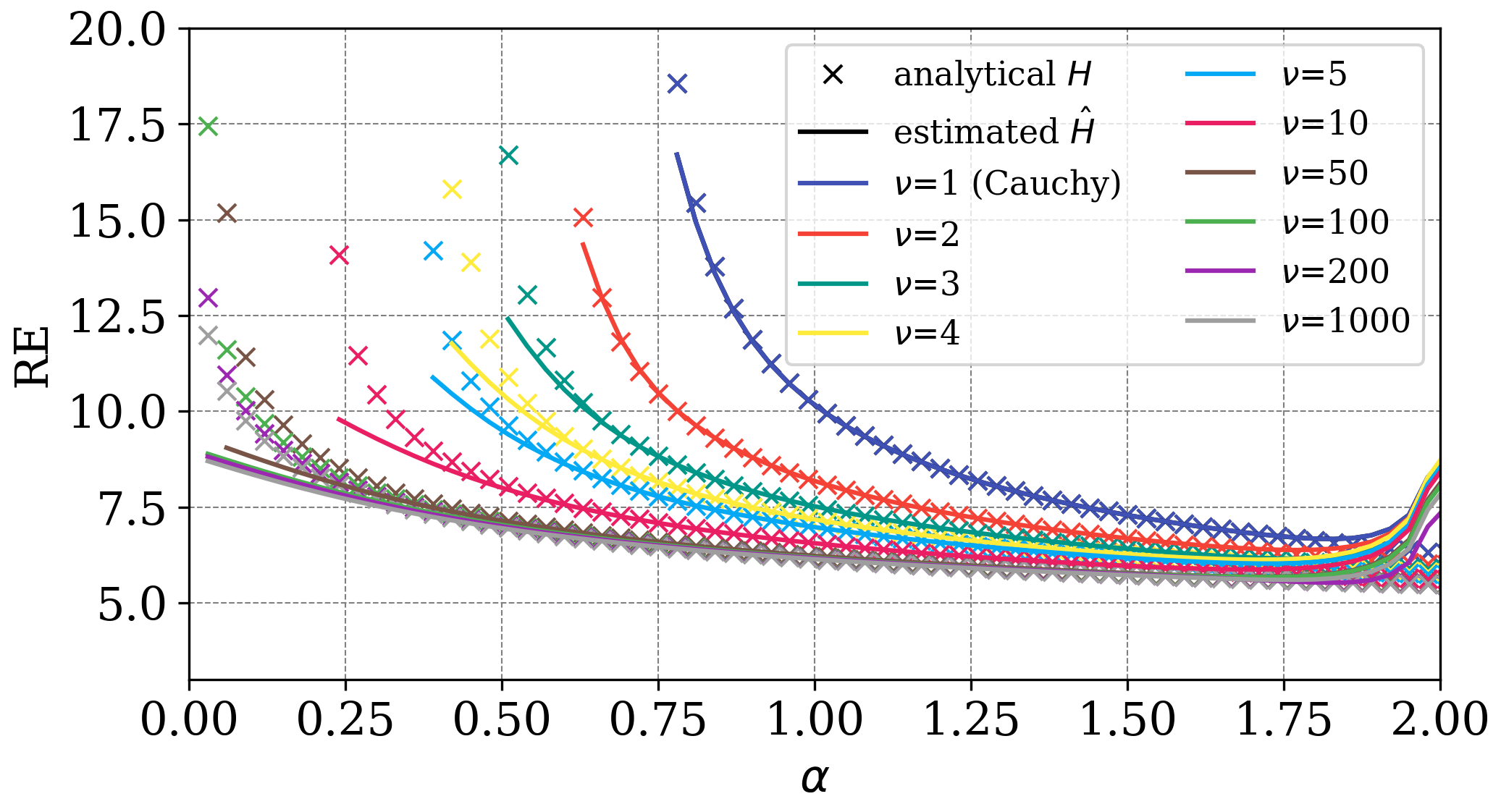}}
    \end{subcaptionbox}
    \hfill
    \begin{subcaptionbox}{\footnotesize\justifying {\em Estimate-to-truth ratios for Multivariate $t$-distribution.} RE parameter $\alpha \in (0,2)$ is on the log-scaled $x$-axis, and ratio of estimated and analytical values from Fig.\ref{fig_student_nu} is on the $y$-axis. To reach the best precision estimates are obtained on sample size $N=10^6$ and for $k=1$, which as is shown in Fig.\ref{fig_est_norm_size} and Fig.\ref{fig_est_norm_knn} improve accuracy for small $\alpha$. The heavier the tails, the earlier the deviation occurs for $\alpha <1$. For example, when $\nu=1$ (Cauchy distribution), the estimate deviates by $3\%$ at $\alpha=0.8$, while for $\nu=1000$, the deviation is only $0.001\%$. The ratio equals $1$ for $\alpha \geq 0.9$, indicating high precision of the estimator on this interval, but start to diverge before reaching $2$. This happens because for $k=1$ the estimator is not well defined for high $\alpha$ and higher choices of $k$ should be used here.
    \label{fig_student_nu_error}}
        {\includegraphics[width=0.48\textwidth]{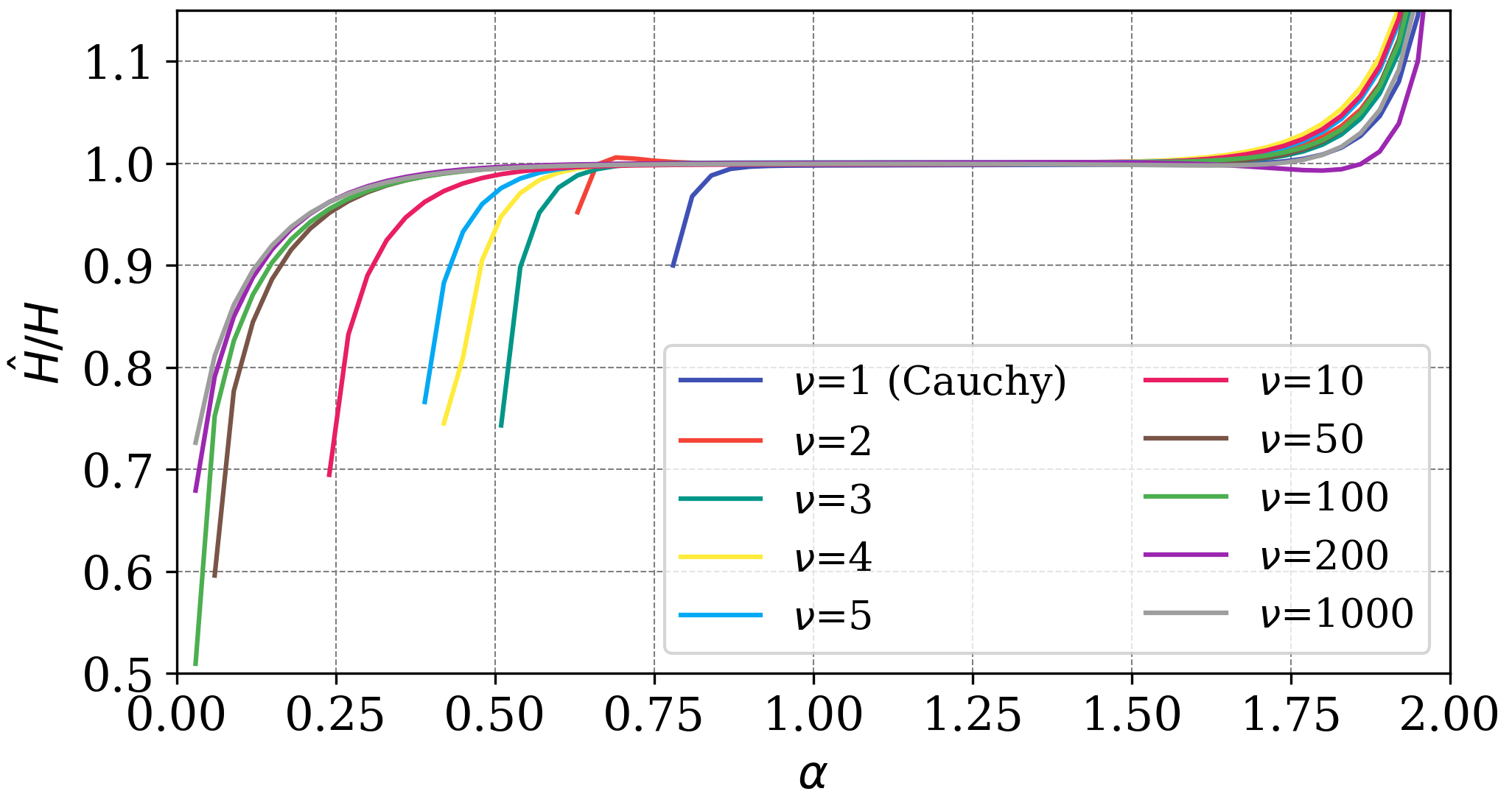}}
    \end{subcaptionbox}
    \caption{\small {Estimator behavior under varying tail heaviness.}}
    \label{fig_student_t_estimates}
\end{figure*}

\subsection{Estimation for $\alpha < 1$}

Figs.~\ref{fig_est_norm_size}–\ref{fig_est_dims} demonstrate that Rényi entropy (RE) estimates tend to deviate more significantly from analytical values as $\alpha$ decreases below 1. The most accurate results in this regime are obtained with large sample sizes ($N$), small nearest-neighbor values ($k$), and low-dimensional data ($d$). Conversely, high dimensionality and limited data substantially degrade estimation accuracy.

For $\alpha < 1$, RE gives greater weight to low-probability events while de-emphasizing high-probability regions. In the limiting cases, as $\alpha \to +\infty$, the entropy is dominated by the highest-probability event, whereas as $\alpha \to 0$, all events are treated as equally probable, corresponding to a uniform distribution. However, uniform distribution is not defined on unbounded domains --- characteristic of heavy-tailed distributions --- posing a challenge for RE estimation in such contexts.

We illustrate this behavior using the multivariate Student’s $t$-distribution, $\mathcal{P_S}$, where the degrees of freedom parameter $\nu$ controls the heaviness of the distribution’s tails. A formal derivation of RE for $\mathcal{P_S}$ is provided in Appendix~\ref{Appendix_RE}. The integral expression for RE is analytically solvable only when the condition $\left(\alpha\frac{\nu + d}{2} - \frac{d}{2}\right) > 0$ is satisfied. This constraint introduces a lower bound on admissible values of $\alpha$ that depends on the dimensionality $d$ and the degrees of freedom $\nu$ of the distribution.

Fig.~\ref{fig_student_nu} shows both analytical and empirical estimates of RE for a 3-dimensional $t$-distribution with identity scale matrix and varying degrees of freedom ($\nu = 1, 2, 3, 4, 5, 10, 50, 100, 200, 1000$). Analytical values are shown with \texttt{x}-markers; empirical estimates are computed using $N = 10^6$ samples and $k = 1$, which is optimal for $\alpha < 1$. Notably, for $k = 1$, the estimator is valid only for $\alpha < 2$ (see Eq.~\ref{eq_leonenko}), setting the right-hand boundary of the figure. When $\nu = 1$ (i.e., the Cauchy distribution), the RE is undefined for most of the $\alpha < 1$ range due to extremely heavy tails. As $\nu$ increases and the distribution approaches normality, the domain of valid $\alpha$ values expands.

Estimation errors, expressed as ratios of estimated to analytical RE values, are illustrated in Fig.~\ref{fig_student_nu_error}, where the $x$-axis is log-scaled to emphasize the region $\alpha \in (0, 1)$. The results reveal that as the tail heaviness increases (lower $\nu$), divergence from the analytical values occurs at progressively higher $\alpha$ on interval $(0,1)$. In other words, for any fixed $\alpha < 1$, the accuracy of RE estimates improves as the distribution becomes less heavy-tailed (i.e., as $\nu$ increases). Divergence for $\alpha > 1.5$ is again due to the small choice of $k=1$.

\subsection{Reliability Conditions for RTE Estimation}\label{rel_conds}

When applying the $k$-NN estimator to differential Rényi entropy, negative values may arise due to two primary reasons. The first stems from the estimator itself --- such values can occur as artifacts of finite-sample effects, estimator bias, rounding errors, or numerical instabilities under certain parameter settings. The second possibility is that the underlying process genuinely exhibits very low entropy. Both scenarios can lead to incorrect assessments of effective RTE, resulting in false positives or false negatives. To support robust interpretation, we recommend incorporating the following diagnostic checks or the \textit{reliability conditions} on the component entropies defined in Eq.~\eqref{complete_H} as part of the RTE estimation procedure.
 
\begin{itemize}
    \item  Non-negativity of all individual entropies estimates: $\hat{\mathscr{I}}_1, \hat{\mathscr{I}}_2, \hat{\mathscr{I}}_3,\hat{\mathscr{I}}_4 \geq 0$.
    \item Non-negativity of all conditional entropies estimates:
    $\hat{\mathscr{I}}_{12}\equiv\hat{\mathscr{I}}_1-\hat{\mathscr{I}}_2 \geq 0$  and $\hat{\mathscr{I}}_{34}\equiv\hat{\mathscr{I}}_3-\hat{\mathscr{I}}_4 \geq 0$.
    \item The estimate of Shannon transfer entropy (corresponding to $\alpha=1$) must be non-negative, i.e. $T^R_{\alpha=1}\geq 0$ . 
\end{itemize}

These checks are based on known theoretical bounds for discrete Rényi entropy, specifically: $0 \leq H_\alpha(X) \leq \log_2 N$ and $0 \leq H_\alpha(X|Y) \leq \log_2 N$, where $N$ denotes the sample size. If any of these conditions are violated, the resulting estimates may require closer inspection or appropriate correction to ensure their interpretability.

\section{Estimation of Rényi Transfer Entropy on stochastic systems with linear and nonlinear coupling}\label{sec_est_RTE}

The primary challenge in estimating the RTE arises from the need to manage  high-dimensional, correlated joint probability distributions. Moreover,  the underlying processes are typically non-stationary and exhibit 
heavy-tailed behavior. These features introduce substantial uncertainty  in the appropriate selection of the $k$-nearest neighbor and memory parameters. To develop intuition about how these parameter choices affect the estimation, 
we apply the method to four illustrative toy-model processes. We start with a 
simple Gaussian system featuring linear coupling and a single driving process, 
then extend the analysis to a system with nonlinear dependencies. The complexity 
is further increased by introducing additional sources and combining both linear 
and nonlinear couplings. Furthermore, for Gaussian systems, the RTE can be derived 
analytically, which allows us to validate the numerical estimates against the 
exact theoretical values.

\begin{figure*}[t] 
    \begin{subcaptionbox}{\justifying \footnotesize {\em Effective Rényi transfer entropy  in both directions between $A$ and $B^I$.} The $x$-axis shows the Rényi parameter $\alpha$, while the $y$-axis presents both estimated and analytical Rényi transfer entropy values between two processes in bits. A value of $k=3$ is used for $\alpha<1.1$ and $k=50$ on the rest of the interval; the transition is marked by a vertical gray band. The memory parameters $r=l=1$ are determined by the process definition. The blue, red, and green quantile bands -- 1st–99th, 5th–95th, and 25th–75th with a line indicating median over 30 runs --    (from bottom to top) correspond to estimated effective transfer entropies in the causal direction, i.e. from $A$ to $B^I$ for coupling strengths $\eta = 0.2$, $0.5$, and $0.8$, respectively. Greater coupling strength leads to higher information transfer, with estimates closely matching analytical results. Gray lines near zero depict transfer entropy in the reverse direction ($B^I$ to $A$), where no causality is expected. Since both $A$ and $B^I$ are normally distributed, the transfer entropy is independent of $\alpha$.
    \label{fig_rte_process1}}
        {\includegraphics[width=0.48\textwidth]{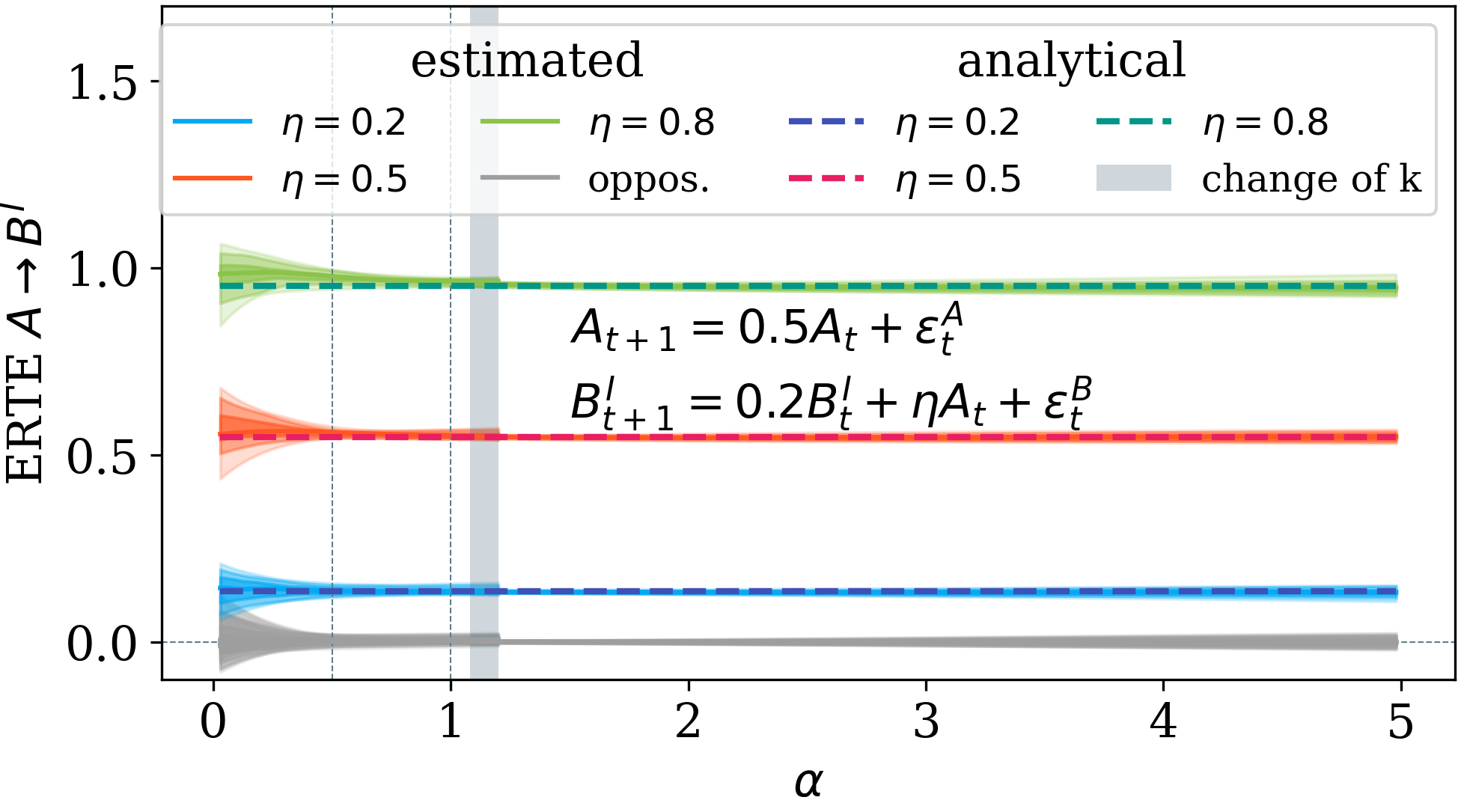}}
    \end{subcaptionbox}
    \hfill
    \begin{subcaptionbox}{\justifying \footnotesize {\em Rényi entropies $\mathscr{I}_1, \mathscr{I}_2, \mathscr{I}_3, \mathscr{I}_4, \mathscr{I}_{12}$ and $\mathscr{I}_{34}$ underlying the RTE $A \rightarrow B^I$ for coupling parameter $\eta = 0.5$.} The $x$-axis represents the Rényi parameter $\alpha$, and the $y$-axis shows both estimated and analytical values of the corresponding Rényi entropies measured in bits. Dashed lines indicate analytical results for $\mathscr{I}_1-\mathscr{I}_4$, while quantile bands show estimated results over 30 runs. The entropies $\mathscr{I}_1$, $\mathscr{I}_2$, $\mathscr{I}_3$, and $\mathscr{I}_4$ (with dimensions 2,1,3,2) are represented by orange, magenta, blue, and green colors, respectively. The estimates closely match the analytical values, but diverge for smaller values, especially when $\alpha < 0.5$. Conditional entropies $\mathscr{I}_{12}=\mathscr{I}_1 - \mathscr{I}_2$ and $\mathscr{I}_{34}=\mathscr{I}_3 - \mathscr{I}_4$ are shown in cyan and yellow colors. All reliability conditions for the estimates are satisfied, i.e $\hat{\mathscr{I}}_1, \hat{\mathscr{I}}_2, \hat{\mathscr{I}}_3, \hat{\mathscr{I}}_4, \hat{\mathscr{I}}_{12}$ and $\hat{\mathscr{I}}_{34}$, are positive.  Additionally, as $\hat{\mathscr{I}}_{12} >\hat{\mathscr{I}}_{34}$ for $\alpha=1$ Shannon transfer entropy is positive as well. This means that the resulting RTE and ERTE are interpretable. 
    \label{fig_re_process1}}
        {\includegraphics[width=0.48\textwidth]{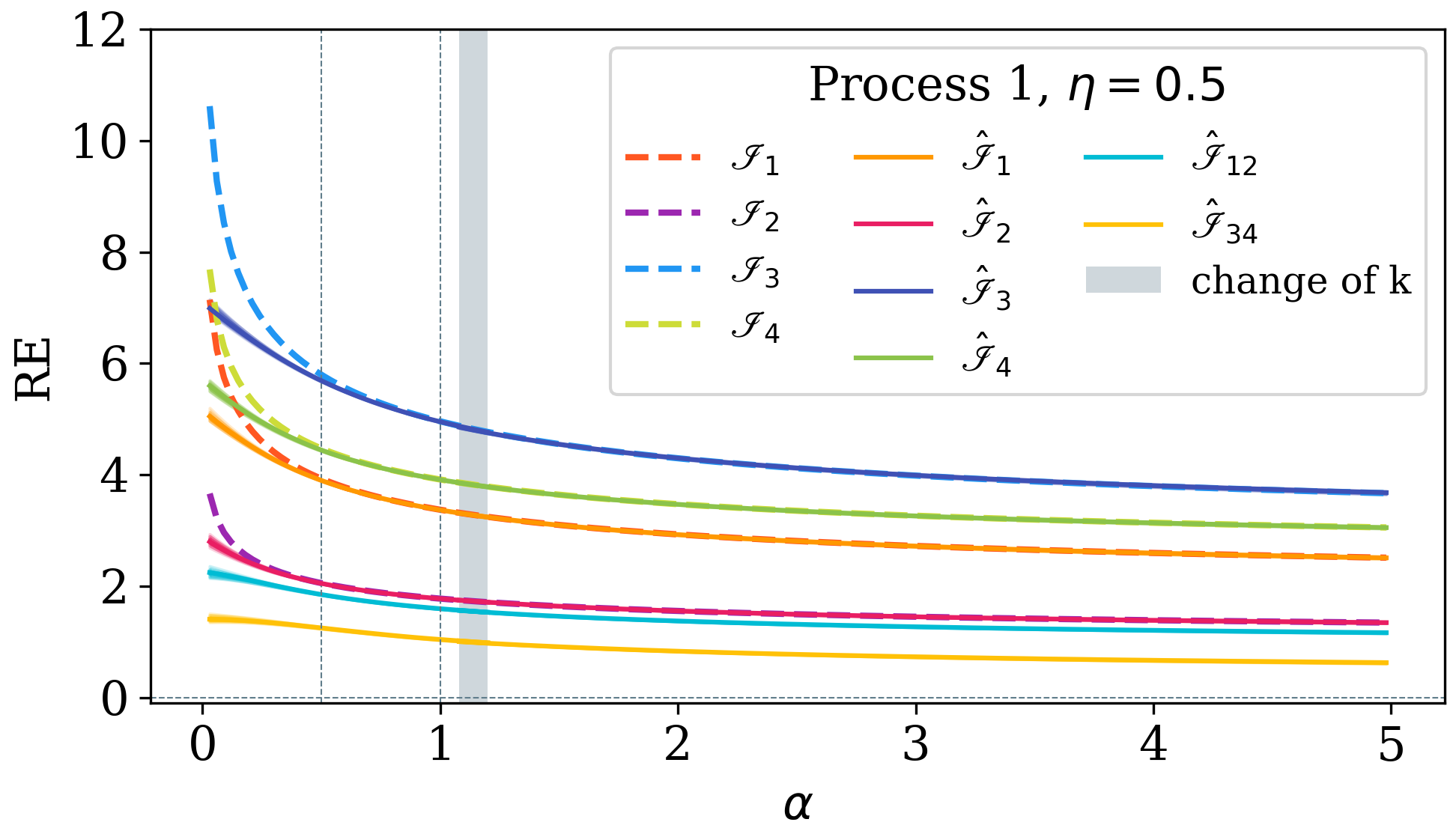}}
    \end{subcaptionbox}
    \caption{\small Process 1: Estimated Rényi Entropies and Effective Transfer Entropies for Linear Single-Source Coupling}
    \label{fig:both}
\end{figure*}

\subsection{Process 1: linear coupling with one source}

The first process considered is a first-order linear autoregressive model (AR(1)) 
with Gaussian noise. In this configuration, $A$ acts as the source (driving or master) 
process, while $B^I$ represents the target (or driven) process. The system evolves 
according to the following equations:
\begin{eqnarray}
&&A_{t+1} \ = \ a A_{t} \ + \ \varepsilon^{A}_t\, , \nonumber \\[2mm]
&&B^I_{t+1} \ = \ b B^I_{t} \ + \  \eta A_{t} \ + \ \varepsilon^{B^I}_t\, ,
\end{eqnarray}
Here, the coupling term $\eta$ introduces a linear dependence of the target on the source.  The parameters are set to $a = 0.5$, $b = 0.2$, with $\varepsilon_t^A \sim \mathcal{N}(0,1)$  and $\varepsilon_t^{B^I} \sim \mathcal{N}(0,0.25)$. The coupling coefficient $\eta$ is varied to investigate how the information transfer responds to changes in the 
interaction strength.

The estimated effective Rényi transfer entropy (ERTE) in both directions, \( T^{\text{eff}}_{\alpha, A \rightarrow B^I}(1,1) \) and \( T^{\text{eff}}_{\alpha, B^I \rightarrow A}(1,1) \), is presented in Fig.\ref{fig_rte_process1}. The Markov orders $r = 1$ and $l = 1$ are determined from the definition of the process. 
For each value of $\eta = 0.2, 0.5$, and $0.8$, the system is simulated 30 times, 
with each simulation comprising $N = 10^5$ time steps. During each simulation, the entropy 
terms $\hat{\mathscr{I}}_1$, $\hat{\mathscr{I}}_2$, $\hat{\mathscr{I}}_3$, 
$\hat{\mathscr{I}}_4$, $\hat{\mathscr{I}}_{12}$, and $\hat{\mathscr{I}}_{34}$
(see Fig.~\ref{fig_re_process1}) are computed and subsequently used to derive the ERTE, 
with the last two terms also computed on a shuffled source.

To summarize the variability of the estimates, the results are aggregated across the 30 simulations using a percentile-based visualization. The central curve for each value of $\alpha$ represents the median ERTE. Surrounding this median, semi-transparent bands indicate intervals between selected empirical percentiles: 1st–99th, 5th–95th, and 25th–75th. Bands closer to the median are plotted with greater opacity, reflecting higher concentration of the distribution and thus greater confidence in the estimate’s stability.

The $k$-NN parameter is set to $k = 3$ for $\alpha \in (0, 1.23)$, and increased to $k = 50$ for $\alpha \in (1.08, 5)$ to improve robustness in regimes where larger $\alpha$ values emphasize high-probability events. The gray vertical band indicates this transient interval. The two other vertical dashed lines are placed at $\alpha=0.5$ and $\alpha=1$, the Shannon case.

The blue, red, and green quantile bands (from bottom to top) correspond to ERTE estimates in the causal direction for coupling strengths \( \eta = 0.2 \), \( 0.5 \), and \( 0.8 \), respectively. These results closely match the theoretical values indicated by dashed lines. As both systems are Gaussian, the theoretical RTE is independent of $\alpha$ and the result dependents on a covariance matrix and dimensionality, see Eq.~(\ref{eq_RE_normal}) for details. Transfer entropy in the reverse direction is approximately zero across all values of $\eta$ (three gray quantile bands), which is consistent with the unidirectional structure of the model.

To validate the three reliability conditions introduced in Section~\ref{rel_conds}, 
Fig.~\ref{fig_re_process1} shows the estimated values of the entropy terms 
$\hat{\mathscr{I}}_1$, $\hat{\mathscr{I}}_2$, $\hat{\mathscr{I}}_3$, and 
$\hat{\mathscr{I}}_4$ underlying the ERTE from $A$ to $B^I$, along with their 
corresponding analytical values for a coupling of $\eta = 0.5$. The estimates are 
also presented with quantile bands; however, the deviations are too small to be visible 
on the scale of the $y$-axis. The vertical gray band indicates the interval over which 
the estimator parameter $k$ is switched from 3 to 50. For $\alpha < 1$, where $k = 3$, 
the estimates of the REs $\mathscr{I}_1$, $\mathscr{I}_2$, $\mathscr{I}_3$, and 
$\mathscr{I}_4$, with respective dimensions $2, 1, 3, 2$, deviate from the analytical 
values for $\alpha < 0.5$, consistent with the results shown in Figs.~\ref{fig_est_norm_size} and~\ref{fig_est_norm_knn}. Nevertheless, entropy estimates are non-negative, and panel (a) confirms that the Shannon transfer entropy, i.e. at $\alpha=1$, is non-negative as well, thus all three reliability conditions are fulfilled. REs underlying the opposite direction also satisfy reliability conditions, see Fig.~\ref{fig_re_process1_op}.

To illustrate cases where the reliability conditions are more difficult to satisfy, we present a slightly modified version of the Process~1 in Appendix~\ref{Appendix_NegativeRE}. The modification involves reducing the variance of $\varepsilon^B$, which increases the correlation strength and results in an almost deterministic coupling between the source and target processes. Under these conditions, the previously used memory and $k$-NN parameters no longer satisfy the reliability criteria. We demonstrate that correcting the negativity in the estimated REs requires increasing both the memory length and the number of nearest neighbors. 

To illustrate the importance of bias subtraction, we present the estimated RTE values 
(without the shuffled correction) in Fig.~\ref{SIfig_pr1_eff_big} for $k=50$ and 
Fig.~\ref{SIfig_pr1_eff_small} for $k=3$, each covering the full range of $\alpha$. 
For $\alpha < 1$, the RTE estimates diverge from the theoretical values due to 
finite-sample effects. This systematic bias is effectively corrected by subtracting the 
RTE computed from a shuffled source series, resulting in the effective RTE, which 
closely matches the analytical predictions across the entire $\alpha$ range for both 
values of $k$. Notably, the RTE estimates are accurate from $\alpha = 0.5$ for $k=3$, 
whereas for $k=50$, reliable estimates are obtained only starting around $\alpha \approx 1$.  This behavior is consistent with previous observations, as shown in Fig.~\ref{fig_est_norm_knn}.


\begin{figure*}[t] 
    \justifying
    \begin{subcaptionbox}{\justifying \footnotesize
    {\em Effective Rényi transfer entropy in both directions between $C$ and $B^{II}$.} The $x$-axis shows the Rényi parameter $\alpha$, while the $y$-axis presents estimated Rényi transfer entropies between two processes in bits. A value of $k=3$ is used for $\alpha<2.1$ and $k=50$ on the rest of the interval; the transition is marked by a vertical gray band. The memory parameters $r=l=1$ are determined by the process definition. The turquoise, red, and yellow quantile bands -- 1st–99th, 5th–95th, and 25th–75th with a line indicating median over 30 runs --    (from bottom to top) correspond to estimated effective transfer entropies in the causal direction, i.e. from $C$ to $B^{II}$ for coupling strengths $\kappa = 0.2$, $0.5$, and $0.8$, respectively. Greater coupling strength leads to higher information transfer for $\alpha >0.7$, but for smaller values this dependency is reversed. This subtle transition highlights the sensitivity of low-$\alpha$ Rényi entropy to tail events and the increased variability of the estimator in this regime, which is exacerbated by the heavy-tailed nature of the distribution introduced by the cubic transformation. Gray lines near zero depict transfer entropy in the opposite direction ($B^{II}$ to $C$), where no causality is expected. The estimates again show increased variability for small $\alpha$ values, primarily due to the influence of heavy-tailed distributions.  
    \label{fig_rte_process2}}
        {\includegraphics[width=0.48\textwidth]{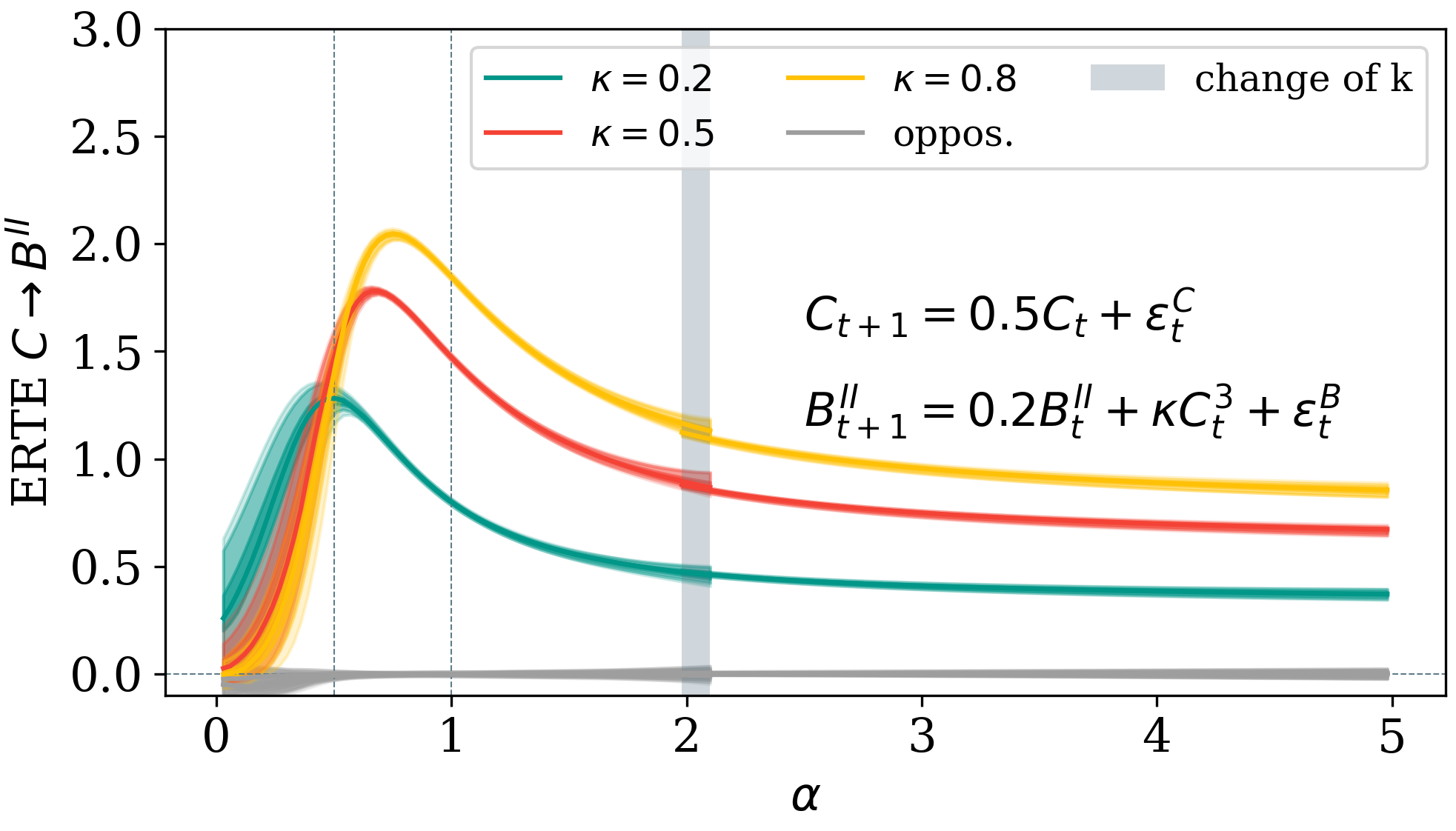}}
    \end{subcaptionbox}
    \hfill
    \begin{subcaptionbox}{\justifying \footnotesize
    {\em Rényi entropies $\mathscr{I}_1, \mathscr{I}_2, \mathscr{I}_3, \mathscr{I}_4, \mathscr{I}_{12}$ and $\mathscr{I}_{34}$ underlying the RTE $C \rightarrow B^{II}$ for coupling parameter $\eta = 0.5$.} The $x$-axis represents the Rényi parameter $\alpha$, and the $y$-axis shows estimated values of the corresponding Rényi entropies measured in bits. The quantile bands show estimated results over 30 runs. The entropies $\mathscr{I}_1$, $\mathscr{I}_2$, $\mathscr{I}_3$, and $\mathscr{I}_4$ (with dimensions 2,1,3,2) are represented by orange, magenta, blue, and green colors, respectively. For small values of $\alpha$, the estimates exhibit high variability, which further increases with dimensionality. In particular, $\mathscr{I}_{3}$, computed in three dimensions, shows the greatest variance. Conditional entropies $\mathscr{I}_{12}=\mathscr{I}_1 - \mathscr{I}_2$ and $\mathscr{I}_{34}=\mathscr{I}_3 - \mathscr{I}_4$ are shown in cyan and yellow colors. All reliability conditions for the estimates are satisfied, i.e $\hat{\mathscr{I}}_1, \hat{\mathscr{I}}_2, \hat{\mathscr{I}}_3, \hat{\mathscr{I}}_4, \hat{\mathscr{I}}_{12}$ and $\hat{\mathscr{I}}_{34}$, are positive.  Additionally, as $\hat{\mathscr{I}}_{12} >\hat{\mathscr{I}}_{34}$ for $\alpha=1$ Shannon transfer entropy is positive as well. This means that the resulting RTE and ERTE are interpretable. Transfer entropy becomes negative only for $\alpha <0.3$, but this bias --- caused by the finite sample size --- is effectively corrected by the effective RTE, see Fig.~\ref{figSI_effrte_process2}.
    \label{fig_re_process2}}
        {\includegraphics[width=0.48\textwidth]{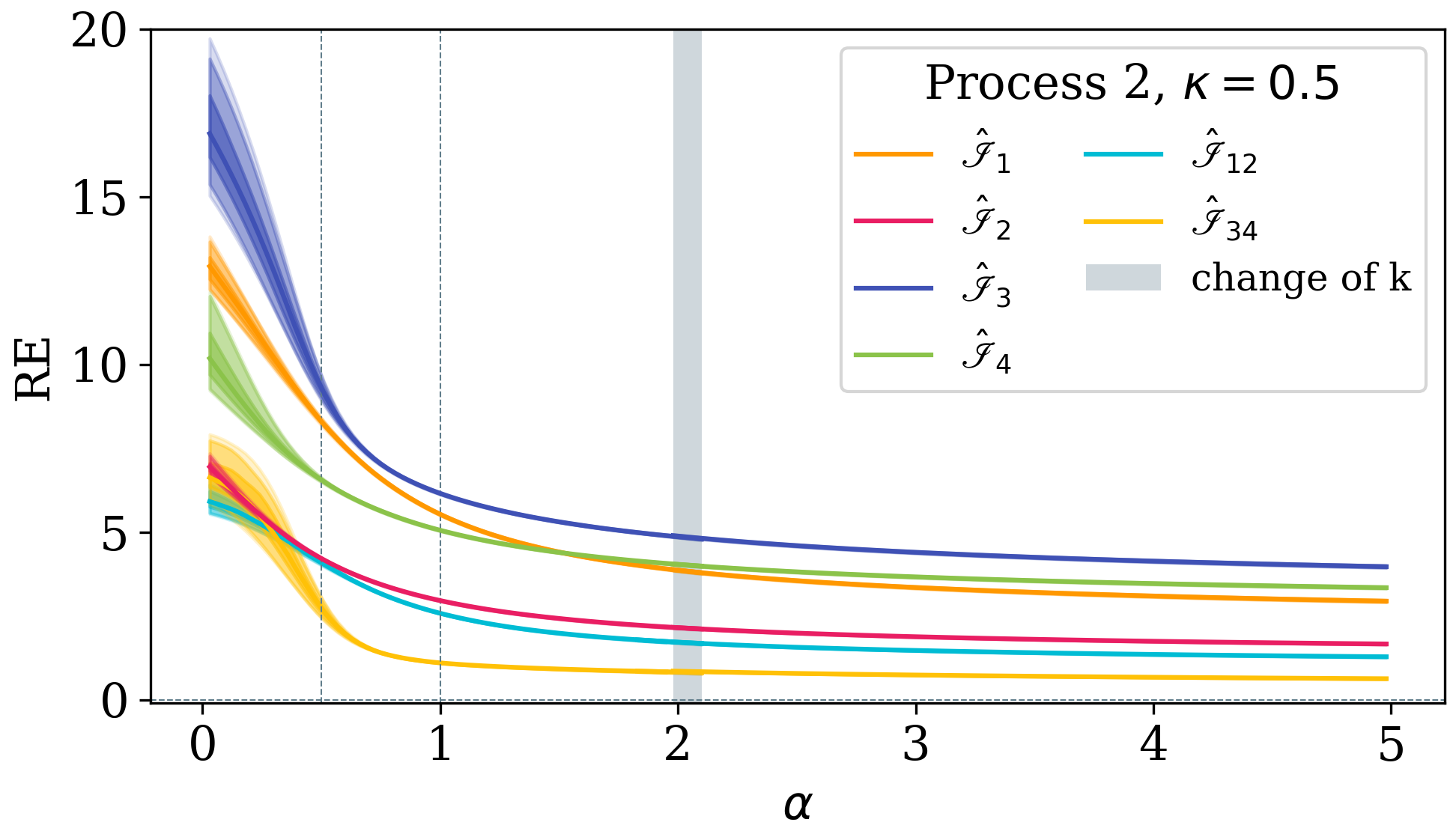}}
    \end{subcaptionbox}
    \caption{\small Process 2: Estimated Rényi Entropies and Effective Transfer Entropies for nonlinear Single-Source Coupling}
    \label{fig:both}
\end{figure*}

\subsection{Process 2: nonlinear coupling with one source}

The second process models a nonlinear interaction in which the target variable~$B^{II}$ depends cubically on the source process~$C$. The system evolves according to
\begin{eqnarray}
&& C_{t+1} \ = \ c C_t \ + \ \varepsilon^C_t \,, \nonumber \\[2mm]
&& B^{II}_{t+1}  \ = \  b B^{II}_t \ + \  \kappa C_t^3 \ + \ \varepsilon^{B^{II}}_t \,,
\end{eqnarray}
where the cubic term~$\kappa C_t^3$ introduces a nonlinear coupling from~$C$ to~$B^{II}$. For consistency across simulations, all parameters are fixed: $c = 0.5$, $b = 0.2$, $\varepsilon^C_t \sim \mathcal{N}(0,1)$, and $\varepsilon^{B^{II}}_t \sim \mathcal{N}(0,0.25)$. Given the model definition, the memory parameters are $l = r = 1$, and no information transfer is expected in the reverse direction (i.e., from~$B^{II}$ to~$C$).

Fig.~\ref{fig_rte_process2} shows the effective RTE in both directions, 
$T^{\text{eff}}_{\alpha, C \rightarrow B^{II}}(1,1)$ and 
$T^{\text{eff}}_{\alpha, B^{II} \rightarrow C}(1,1)$, 
for coupling strengths $\kappa = 0.2$, $0.5$, and $0.8$. The $x$-axis corresponds to the Rényi parameter~$\alpha$, while the $y$-axis displays the ERTE estimates in bits. For each coupling value, the system is simulated 30 times, with $N = 10^5$ samples per run. During each run, the estimates $\hat{\mathscr{I}}_1$, $\hat{\mathscr{I}}_2$, $\hat{\mathscr{I}}_3$, $\hat{\mathscr{I}}_4$, $\hat{\mathscr{I}}_{12}$, and $\hat{\mathscr{I}}_{34}$ are computed, from which the ERTE is subsequently derived.

Results are visualized using quantile-based summaries. For each $\alpha$, the central bold line denotes the median ERTE across the 30 realizations. Shaded regions around the median represent quantile intervals at increasing levels of dispersion: 1st–99th, 5th–95th, and 25th–75th percentiles, with increasing opacity indicating higher concentration of the distribution. This provides a clear visual representation of both the central tendency and variability of the estimates. The nearest-neighbor parameter is set to $k = 3$ for $\alpha \in (0, 2.1)$, and increased to $k = 50$ for $\alpha \in (1.8, 5)$ to ensure stability and accuracy of the estimates in higher $\alpha$ regimes.

In Fig.~\ref{fig_rte_process2}, the turquoise, red, and yellow lines (from bottom to top) represent the effective RTE from $C$ to $B^{II}$ for coupling strengths $\kappa = 0.2$, $0.5$, and $0.8$, respectively. As expected, the estimated information transfer increases with coupling strength, most clearly for $\alpha >0.7$. However, for smaller values of $\alpha$ (approximately up to $0.6$), an inversion occurs: the strongest transfer is observed for $\kappa=0.2$, and the weakest for $\kappa=0.8$, despite the values being closely aligned. This subtle transition reflects the sensitivity of low-$\alpha$ Rényi measures to nonlinear coupling in tail events. This region corresponds to an ill-defined Rényi entropy for the underlying process $B^{II}$. Due to the cubic dependence on $C$, the distribution of $B^{II}$ exhibits heavy tails, which—as shown in earlier sections—can result in undefined RE for small $\alpha$. Notably, ERTE for all of the three cases converges to zero as $\alpha \rightarrow 0$, indicating no information transfer

As in the first example, the large deviations observed here are primarily due to the use of a small $k$ in the estimator. A comparison between panels \textbf{(a)}, \textbf{(c)}, and \textbf{(e)}, which use $k=50$, and panels \textbf{(b)}, \textbf{(d)}, and \textbf{(f)}, which use $k=3$ in Fig.~\ref{figSI_effrte_process2}, confirms that the variance is noticeably lower for larger $k$.  We include results for small $k$-NN here because, as shown in Fig.~\ref{fig_est_norm_knn}, smaller $k$ values provide more accurate entropy estimates for small $\alpha$. Results in Fig.~\ref{figSI_effrte_process2} demonstrate the importance for the bias correction in the low-$\alpha$ regime, which further helps to reduce the variance. In this region, RTEs in both the causal and non-causal directions exhibit substantial bias, which increases with stronger coupling. To correct for this bias in both directions, the ERTE is calculated using a shuffled version of the process $A$, since it contributes most significantly to the estimation error.



To further demonstrate that the reliability conditions are satisfied, Fig.~\ref{fig_re_process2} presents the estimated entropy components $\hat{\mathscr{I}}_1$, $\hat{\mathscr{I}}_2$, $\hat{\mathscr{I}}_3$, $\hat{\mathscr{I}}_4$, $\hat{\mathscr{I}}_{12}$, and $\hat{\mathscr{I}}_{34}$, which underlie the calculation of ERTE from $C$ to $B^{II}$ for $\kappa = 0.5$. These components are shown in orange, magenta, blue, green, cyan, and yellow, respectively. As the figure indicates, all necessary reliability conditions are fulfilled.


\begin{figure}[] 
    {\includegraphics[width=0.48\textwidth]{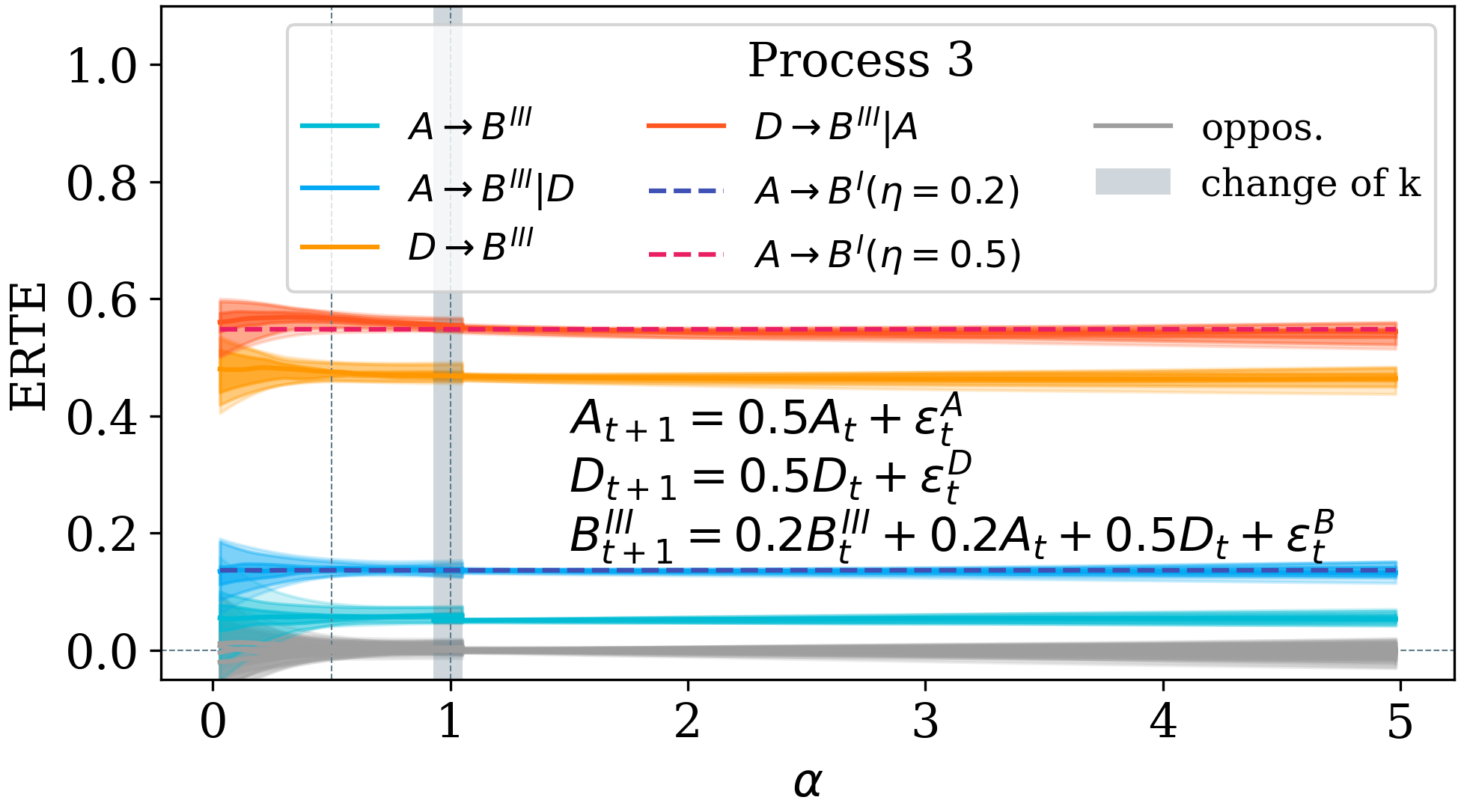}}
 
    \caption{\justifying \footnotesize
    {\em Process 3: Effective Rényi transfer entropy in both directions between $A$ or $D$ and $B^{III}$.} The $x$-axis shows the Rényi parameter $\alpha$, while the $y$-axis presents both estimated and analytical Rényi transfer entropy values between two processes in bits. A value of $k=3$ is used for $\alpha<1.08$ and $k=50$ on the rest of the interval; the transition is marked by a vertical gray band. The memory parameters $r=l=1$ are determined by the process definition. Quantile bands -- 1st–99th, 5th–95th, and 25th–75th with a line indicating median over 30 runs -- correspond to estimated effective transfer entropies in the causal direction. Since two source processes are involved, the conditional RTE is also computed. The cyan and orange quantile bands, centered around approximately $0.05$ and $0.47$ bits, represent the apparent RTEs from $A$ ($\eta_1=0.2$) and $D$ ($\eta_2=0.5$), respectively. Both estimates fall short of the theoretical values reported in Fig.~\ref{fig_rte_process1} by approximately $0.08$ bits. However, when conditioning on the other source process, the estimates align closely with the expected results: the light blue and red quantile bands, centered around $0.13$ and $0.55$ bits, respectively, overlap the theoretical predictions indicated by dashed lines. As expected, transfer entropy in the reverse direction is estimated to be zero.
    \label{fig_rte_process3} }
\end{figure}

\begin{figure}[] 
    {\includegraphics[width=0.48\textwidth]{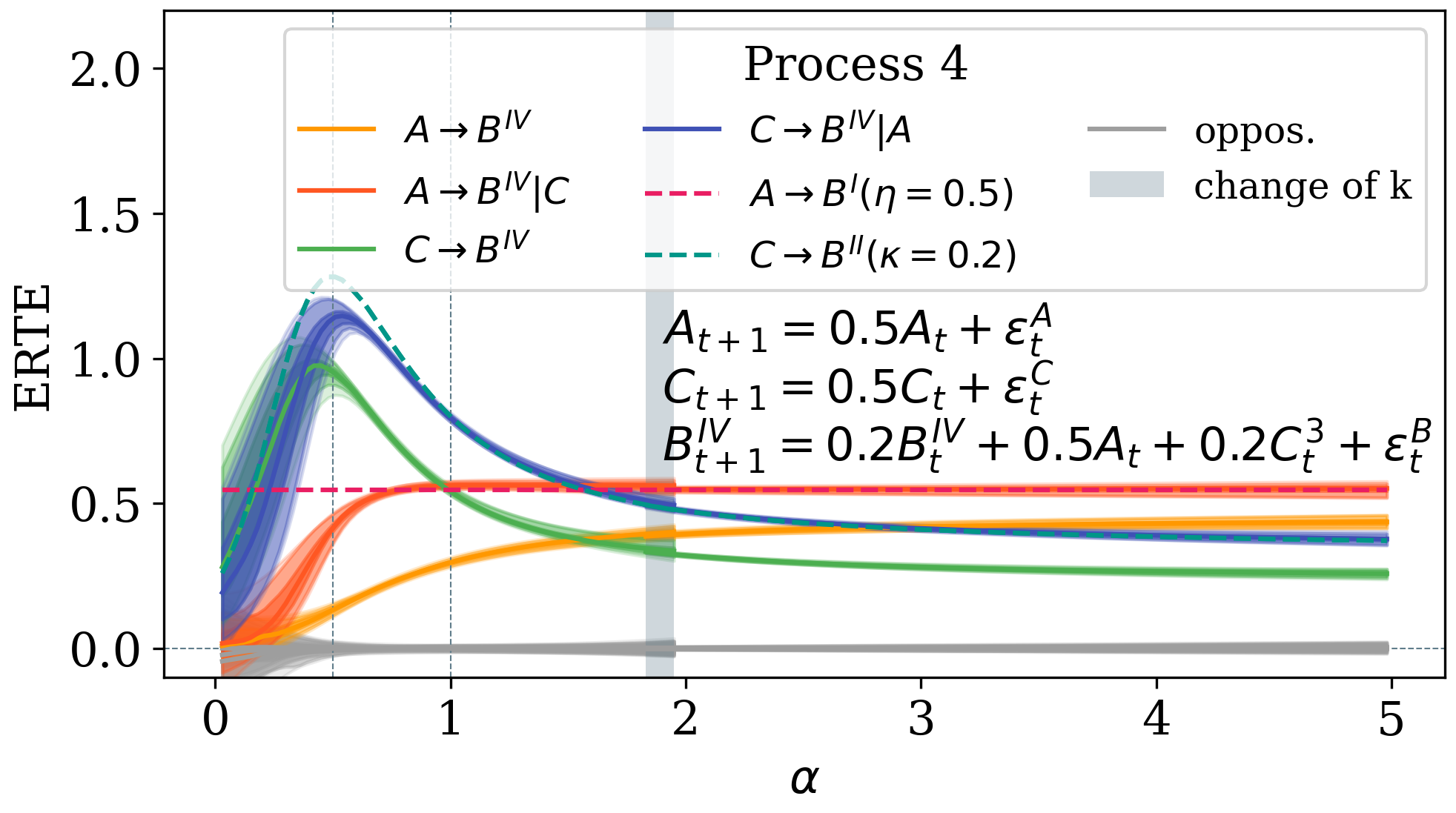}}
 
    \caption{\justifying \footnotesize
       {\em Process 4: Effective Rényi transfer entropy in both directions between $A$ or $C$ and $B^{IV}$.} The $x$-axis shows the Rényi parameter $\alpha$, while the $y$-axis presents estimated Rényi transfer entropy values between two processes in bits. A value of $k=3$ is used for $\alpha<1.98$ and $k=50$ on the rest of the interval; the transition is marked by a vertical gray band. The memory parameters $r=l=1$ are determined by the process definition. Quantile bands -- 1st–99th, 5th–95th, and 25th–75th with a line indicating median over 30 runs -- correspond to estimated effective transfer entropies in the causal direction. Since two source processes are involved, the conditional RTE is also computed. The estimated conditional effective RTEs, $T^{\text{eff}}_{\alpha, A \rightarrow B^{IV}|C}$ and $T^{\text{eff}}_{\alpha, C \rightarrow B^{IV}|A}$, shown in red and dark blue, respectively, align well with the expected values derived from earlier processes, as indicated by the corresponding dashed reference lines for $\alpha>0.7$. The divergence observed at smaller $\alpha$ values is attributed to the ill-defined nature of Rényi entropy under heavy-tailed distributions. In contrast, the apparent RTEs— $T^{\text{eff}}_{\alpha, A \rightarrow B^{IV}}$ and $T^{\text{eff}}_{\alpha, C \rightarrow B^{IV}}$, represented by orange and green curves—exhibit lower levels of information transfer, reflecting the absence of conditioning. Estimates in the reverse direction remain close to zero, as expected, though increased variability is observed for small $\alpha$. These findings demonstrate that, in the presence of multiple sources, complete RTE can be reliably extracted for $\alpha>0.7$. \label{fig_rte_process4} }
\end{figure}

\subsection{Process 3: linear coupling with two sources}

The third process extends the first by introducing an additional source variable~$D$, which is linearly coupled to the target process~$B^{III}$ via a second coupling coefficient~$\eta_2$, namely
\begin{eqnarray}
     &&A_{t+1}\ = \ aA_{t} \ + \ \varepsilon^A_t\, , \nonumber \\[2mm]
     &&D_{t+1} \ = \ aD_{t} \ + \  \varepsilon^D_t\, , \nonumber \\[2mm]
     &&B^{III}_{t+1} \ = \ bB^{III}_{t} \ + \ \eta_1 A_{t} \ + \  \eta_2 D_{t} \ + \  \varepsilon^{B^{III}}_t\, .
     \label{pr3_eq}
\end{eqnarray}
All model parameters, including noise terms 
$\varepsilon^A_t, \varepsilon^D_t \sim \mathcal{N}(0,1)$, and $\varepsilon^{B^{III}}_t \sim \mathcal{N}(0,0.25)$, are fixed as in previous experiments. Only the coupling strengths 
$\eta_1$ and $\eta_2$ are varied. The process memory parameters are by design
$r=l=1$ and, no transfer of information is expected in the reverse direction.

Results for $\eta_1=0.2$ and $\eta_2=0.5$ are presented in Fig.~\ref{fig_rte_process3}, using the same simulation protocol as in the preceding cases. Given the presence of multiple source variables, we also estimate the conditional Rényi transfer entropy, $T_{\alpha}^{CR}$, to account for shared information between sources. The conditional transfer entropy from $A$ to $B^{III}$ coupled with $\eta_1=0.2$, $T^{\text{eff}}_{\alpha, A \rightarrow B^{III}|D}(1,1,1)$, is shown as a light blue line with quantile bands. These results closely match those from the first process with the same coupling strength $\eta=0.2$ indicated by a dashed dark blue line at $0.13$ bits. In contrast, the apparent transfer entropy, $T^{\text{eff}}_{\alpha, A \rightarrow B^{III}}(1,1)$, shown in cyan, underestimates the true transfer and yields approximately $0.05$ bits, missing about $0.08$ bits of causal influence due to the absence of conditioning on $D$. This demonstrates that, for the given coupling setup, the complete RTE from $A$ to $B^{III}$ is accurately captured by the conditional RTE with memory parameters $r = l = s = 1$, i.e., $\bar{T}^R_{\alpha,A\rightarrow B^{III}} = T^{CR}_{\alpha,A\rightarrow B^{III}|D}(1,1,1)$.

A similar effect is observed for the second source. The conditional transfer entropy from $D$ to $B^{III}$ with $\eta_2=0.5$, $T^{\text{eff}}_{\alpha, D \rightarrow B^{III}|A}(1,1,1)$, shown as the red line with quantile bands, aligns with the theoretical result from Process 1 for $\eta=0.5$ (red dashed line at $0.55$ bits). The corresponding apparent transfer entropy, $T^{\text{eff}}_{\alpha, D \rightarrow B^{III}}(1,1)$, shown in orange, again underestimates the true value by approximately $0.08$ bits. This again shows that the complete RTE from $D$ to $B^{III}$ is correctly represented by the conditional RTE with memory parameters $r = l = s = 1$, i.e., $\bar{T}^R_{\alpha,D\rightarrow B^{III}} = T^{CR}_{\alpha,D\rightarrow B^{III}|A}(1,1,1)$.

Since the process involves linear dynamics and Gaussian noise, the joint distribution of $B^{III}$ remains Gaussian. As a result, the true transfer entropy is theoretically constant across all values of $\alpha$, see Eq.\ref{eq_RE_normal}. Minor deviations for $\alpha<0.5$ are attributed to estimation noise in the baseline computed from shuffled time series. Nonetheless, the median estimates converge reliably to the expected theoretical values. Fig.\ref{figSI_effrte_process3} presents RTE and ERTE estimates computed with $k=3$ and $k=50$ across the entire $\alpha$ interval. As observed in the first process, deviations are more pronounced when using $k=3$. 

\subsection{Process 4: nonlinear coupling with two sources}

The final process combines features of the previous systems by incorporating both linear and nonlinear couplings from two source processes. Specifically, the target process $B^{IV}$ is influenced linearly by 
$A$ and nonlinearly (cubic) by 
$C$, as described by the following equations:
\begin{eqnarray}
&&A_{t+1} \ = \  aA_{t} \ + \  \varepsilon^A_t\, , \nonumber \\[2mm]
&&C_{t+1} \ = \  cC_{t} \ + \  \varepsilon^C_t\, , \nonumber \\[2mm]
&&B^{IV}_{t+1}  \ = \ bB^{IV}_{t} \ + \  \eta A_{t} \ + \  \kappa C^3_{t} \ + \ \varepsilon^{B^{IV}}_t\, .
\end{eqnarray}
All model coefficients and the simulation protocol are consistent with those used in the previous experiments. Results for this configuration are presented in Fig.~\ref{fig_rte_process4}. The conditional effective Rényi transfer entropy from $A$ (with linear coupling strength $\eta=0.5$) and from 
$C$ (with cubic coupling strength $\kappa=0.2$) to $B^{IV}$, $T^{\text{eff}}_{\alpha, A \rightarrow B^{IV}|C}(1,1,1)$ and $T^{\text{eff}}_{\alpha, C \rightarrow B^{IV}|A}(1,1,1)$, are shown as red and dark blue quantile lines, respectively. These estimates are in agreement with the expected values obtained from the earlier processes, as indicated by the corresponding dashed reference lines, particularly for $\alpha>0.7$. This again demonstrates, that the complete RTEs in this coupling type can be approximated by conditional RTE, i.e. $\bar{T}^R_{\alpha,A\rightarrow B^{IV}} = T^{CR}_{\alpha,A\rightarrow B^{IV}|C}(1,1,1)$ and $\bar{T}^R_{\alpha,C\rightarrow B^{IV}} = T^{CR}_{\alpha,C\rightarrow B^{IV}|A}(1,1,1)$.  The deviations observed for smaller $\alpha$ values arise from the ill-defined Rényi entropy in the presence of heavy-tailed distributions.

When no conditioning is applied, the resulting apparent transfer entropies, 
$T^{\text{eff}}_{\alpha, A \rightarrow B^{IV}}(1,1)$ and 
$T^{\text{eff}}_{\alpha, C \rightarrow B^{IV}}(1,1)$, shown in orange and green respectively, underestimate the total information transfer. In the reverse direction, the ERTE—both with and without conditioning—remains zero (see Fig.~\ref{figSI_effrte_process4}).

\subsection{Relative Explanation Added}
%

To evaluate explanatory power of obtained transfer entropies, we use the Relative Explanation Added (REA) measure, introduced in~\cite{marschinski2002analysing}:
\begin{equation}
\label{eq_rea}
\mathcal{R}_{\alpha, Y\rightarrow X|Z}(r, l, s)  \ \equiv \  \frac{T^R_{\alpha, Y \rightarrow X \mid Z}(r, l, s)}{H_{\alpha}(x_{t+1} \mid x_t^{(r)}, z_t^{(s)})}\, ,
\end{equation}
where the denominator is equal to what we also denoted by $\mathscr{I}_{12}$. $\mathcal{R}$ quantifies the proportion of uncertainty in the target process $X$ that is reduced by incorporating information from a source process $Y$, beyond what is already explained by the past of $X$ and any additional sources $Z$, if present. Given, that $\mathscr{I}_{12}$ and $\mathscr{I}_{34}$ are non-negative (in accordance with the reliability conditions), a value $\mathcal{R} \in \left[0,1\right]$ indicates that the source reduces uncertainty in the target, with $\mathcal{R}\times100\%$ expressing the relative reduction. Values $\mathcal{R}<0$ imply that incorporating the source increases uncertainty.

Estimation of $\mathcal{R}$ also requires addressing biases introduced by finite sample sizes. While the effective Rényi transfer entropy, $T^{\text{eff}}_{\alpha}$, accounts for such biases in the numerator, a correction is still needed in the denominator—specifically for the term $\hat{H}_{\alpha}(x_{t+1} \mid x_t^{(r)}, z_t^{(s)})$. To estimate this bias, we assume that $x_{t+1}$ is statistically independent of the shuffled past states $x_t^{(r), \rm{sh}}$ and $z_t^{(s),\rm{sh}}$. Under this assumption, the conditional entropy should equal the marginal entropy, i.e., $\hat{H}_{\alpha}(x_{t+1} \mid x_t^{(r), \rm{sh}}, z_t^{(s), \rm{sh}})=\hat{H}_{\alpha}(x_{t+1})$. Any deviation from this equality arises purely from estimator bias. We define this bias as:

\begin{equation}
    H_{12}^{\rm{bias}} \ \equiv \  \hat{H}_{\alpha}\big(x_{t+1} \mid x_t^{(r),\rm{sh}}, z_t^{(s),\rm{sh}}\big) \ - \ \hat{H}_{\alpha}(x_{t+1})\, ,
\end{equation}
Incorporating this correction into the relative explanation added yields:

\begin{eqnarray}
\label{eq_rea}
&&\mbox{\hspace{-8mm}}\widehat{\mathcal{R}}_{\alpha, Y\rightarrow X|Z}(r, l, s)  \nonumber \\[2mm] &&\mbox{\hspace{9mm}} \equiv \  \frac{T^{\text{eff}}_{\alpha, Y \rightarrow X \mid Z}(r, l, s)}{\hat{H}_{\alpha}(x_{t+1} \mid x_t^{(r)}, z_t^{(s)}) \ - \  H_{12}^{\rm{bias}}}\, .
\end{eqnarray}

Table~\ref{tab_rea} presents the relative explanation added values, expressed as percentages, for all examined processes in the direction of coupling. For each process pair, $\mathcal{R}$ is computed at $\alpha = 0.2, 0.5, 0.8, 1, 2, 3, 4$, and $5$ using $k = 50$, with results averaged over 30 independent simulation runs. For comparison, values obtained using $k = 3$ at low $\alpha$ are shown in parentheses.

With the exception of one coupling type ($C \xrightarrow{\kappa = 0.2} B$), a notable growing $\mathcal{R}$ trend can be seen in the table. This observation is important: while ERTE estimates alone might suggest stronger or equal information transfer in the tails of the distribution, the $\mathcal{R}$ reveals that this impression is misleading. In fact, the underlying uncertainty is higher in the tails, which inflates absolute ERTE values. $\mathcal{R}$, being a relative measure, corrects for this and shows that information transfer is actually weaker in the tails compared to regions near the mean. 

In the case of weak nonlinear (cubic) coupling, specifically for the interaction $C \xrightarrow{\kappa = 0.2} B$ (see rows 4, 13, and 14), the $\mathcal{R}$ increases up to $\alpha = 1$, after which it slightly declines or remains constant. This behavior suggests that the influence of the weak cubic term $\kappa C^3$ is attenuated by the noise component $\varepsilon^B$ or process $A$ at higher $\alpha$ values, which emphasize dominant, high-probability events. However, its contribution remains more pronounced in the tail regions, consistent with the elevated $\mathcal{R}$ observed at lower $\alpha$. Overall, the trends of observations in Table \ref{tab_rea} are a consequence of the experimental design, where coupling contributions are comparable across all events. Alternative experimental setups could be constructed in which the coupling strength is modulated to predominantly affect either low- or high-probability events, e.g. in \cite{paluvs2024causes}.

REA values are also consistent across processes that share the same source and coupling type. The symbols $\diamond, \star$ and $\bullet$ in the table mark cases where complete transfer entropy estimates are in agreement across different processes, indicating correctly identified information flow (on intervals where the estimator is reliable). For instance, the $\star$ symbol marks flows driven by source process $A$ (or $D\equiv A$) with linear coupling $\eta=0.5$. As shown in Figs.~\ref{fig_rte_process1}, \ref{fig_rte_process3}, and \ref{fig_rte_process4}, this configuration yields approximately $0.55$ bits of transfer entropy. Correspondingly, the $\mathcal{R}$ values for all three $\star$-labeled cases agree for $\alpha\geq 0.8$, confirming consistency in both absolute and relative information transfer measures.

\begin{table}[b]
\caption{\label{tab_rea} \justifying \footnotesize
{\em Relative Explanation Added for All Coupling Types.} The values in the table represent the mean $\mathcal{R}$ across 30 simulation runs for each process, computed using Eq.~(\ref{eq_rea}) with $k=50$ and expressed as percentages (for $k=3$ results for small $\alpha$ are in brackets). Columns correspond to different values of the Rényi parameter, 
$\alpha = 0.2, 0.5, 0.8, 1, 2, 3, 4$ and $5$. Each row reports the $\mathcal{R}$ for a specific source–target process pair. For example, the first row shows that process 
$A$ reduces uncertainty about $B^I$ by approximately $10\%$ for $\alpha=0.5$ and $18\%$ for $\alpha=5$.
The symbols $\diamond, \star$ and $\bullet$ indicate cases where the complete transfer entropy estimates are consistent across different processes, suggesting correctly extracted information flows. The $\mathcal{R}$ values for matching symbols agree within averaging error bounds of $1–3\%$. Deviations occur primarily for small values of $\alpha$, where the estimator becomes less reliable. However, for $\alpha\geq 1$, the results are highly consistent across comparable coupling scenarios.}

\begin{tabular}{lccccccccc}
\hline
& &$0.2$ & $0.5$ & $0.8$ & $1$ & $2$ & $3$ & $4$ & $5$ \\
\hline
\hline
$A\xrightarrow{\eta=0.2}B^I$ & $\diamond$  &8(8)  & 9(10) & 11(11) & 11 & 14 & 16 & 17 & 18 \\

$A\xrightarrow{\eta=0.5}B^I$& $\star$ &26(26) & 30(30) & 33(33) & 35 & 40 & 43 & 45 & 47 \\

$A\xrightarrow{\eta=0.8}B^I$& &37(38) & 43(43) & 46(47) & 48 & 53 & 57 & 59 & 60 \\

\hline
$C\xrightarrow{\kappa=0.2}B^{II}$& $\bullet$ &9(15) & 32(43) & 41(46) & 41 & 36 & 36 & 37 & 38 \\

$C\xrightarrow{\kappa=0.5}B^{II}$& &2(4)  & 21(36) & 46(56) & 51 & 51 & 51 & 51 & 52 \\

$C\xrightarrow{\kappa=0.8}B^{II}$& &0(2)  & 15(30) & 43(58) & 52 & 56 & 56 & 57 & 58 \\
\hline

$A\xrightarrow{\eta_1=0.2}B^{III}$& &2(2)  & 3(3)  & 3(3)  & 3  & 4  & 4  & 4  & 4  \\

$A\xrightarrow{\eta_1=0.2}B^{III}|D$& $\diamond$ &7(8)  & 9(10) & 11(11) & 12 & 14 & 16 & 17 & 18 \\

$D\xrightarrow{\eta_2=0.5}B^{III}$& &21(21) & 25(25) & 27(27) & 29 & 33 & 35 & 37 & 38 \\

$D\xrightarrow{\eta_2=0.5}B^{III}|A$& $\star$ &25(25) & 30(30) & 33(33) & 35 & 39 & 42 & 44 & 46 \\
\hline

$A\xrightarrow{\eta=0.5}B^{IV}$& &1(1)  & 4(4)  & 10(10) & 14 & 23 & 27 & 29 & 31 \\

$A\xrightarrow{\eta=0.5}B^{IV}|C$& $\star$ &1(2)  & 13(19) & 28(32) & 32 & 39 & 43 & 45 & 47 \\

$C\xrightarrow{\kappa=0.2}B^{IV}$& &8(14)  & 22(30) & 25(28) & 23 & 19 & 18 & 18 & 18 \\

$C\xrightarrow{\kappa=0.2}B^{IV}|A$& $\bullet$ &4(11)  & 25(38) & 37(44) & 38 & 36 & 36 & 37 & 38 \\
\hline

\end{tabular}

\end{table}

\section{Discussion}\label{sec_discussion}

In this paper, we evaluate the performance of the $k$-NN estimator of RE, as originally proposed by Leonenko et al. \cite{leonenko2008class}, and extend its application to RTE. Our analysis proceeds in two stages. First, we assess the estimator's accuracy using synthetic data drawn from multivariate normal and multivariate $t$-distributions with varying tail heaviness. Second, we demonstrate its applicability to a set of simulated coupled stochastic processes. Our results show that the accuracy of RE estimation depended on sample size $N$, dimensionality $d$ (via memory), and the choice of $k$-NN. Estimation performance also varied across $\alpha$-parameters, depending on the combination of the aforementioned parameters.

For the multivariate normal distribution --- with identical marginals across dimensions --- the estimates closely match analytical values for $\alpha > 1$ with as few as $N = 2000$ samples. As $\alpha$ decreases, larger sample sizes are required: reliable estimates are obtained for $\alpha \geq 0.75$ with $N > 10^4$, for $0.5\leq \alpha < 0.75$ with $N > 10^5$, and for $0.4\leq \alpha < 0.5$ with $N = 10^6$. For smaller values of $\alpha$, even larger sample sizes would be necessary to improve accuracy. However, as the estimator is defined for PDFs with bounded support, precision for the normal distribution with unbounded support would not be reached for very small $\alpha$s in any case. These findings indicate that using small $\alpha$ requires careful consideration of sample size, as limited data can reduce estimation accuracy, but also setting lower limit for $\alpha$. Nevertheless, in the context of RTE estimation, the impact of finite-sample bias can be mitigated by computing the effective RTE or alternatively using surrogate data approach \cite{paluvs2007nonlinearity, lancaster2018surrogate} in order to establish the statistical significance of the RTE estimate.

Another critical factor influencing the accuracy of RE and RTE estimation is the dimensionality, $d$, of the input data. For homogeneously distributed multivariate data, the estimator's precision decreases as $d$ increases, although results remain reasonable for $d < 6$. In the context of RTE, the dimensionality of each entropy term (i.e., $\mathscr{I}_1$ through $\mathscr{I}_4$ is determined by the memory or Markov order parameters $r,l$ and $s_i$, which defines the length of the time windows used in the estimation. These parameters are typically chosen based on prior knowledge of the system or determined empirically using Markov order tests.

It is important to note that excessively large memory parameters can introduce redundancy due to strong autocorrelations in time series data, potentially degrading estimator performance. In such cases, data preprocessing techniques --- such as subsetting or time shifting --- may be beneficial. Where appropriate, dimensionality reduction methods, such as principal component analysis (PCA), can be employed to extract the most informative features, thereby reducing complexity while preserving the underlying structure of the process under study \cite{vejmelka2015non, runge2015identifying}.

Another key parameter influencing the estimation process is the number of nearest neighbors, $k$, serving as the estimator's parameter. Conceptually, this parameter governs how local probability density is estimated within different regions of the phase space. In sparsely populated regions, smaller values of $k$ yield more localized and thus more accurate estimates, while in densely populated areas, larger $k$ values are preferable to avoid overfitting to noise.

This sensitivity is particularly important when estimating RE across different values of the Rényi parameter $\alpha$. For $\alpha<1$, RE places greater emphasis on rare or low-probability events—effectively amplifying the importance of tail behavior. In this regime, smaller $k$ values are more suitable, as they better capture localized variations in sparse regions. Conversely, for $\alpha \geq 1$ larger $k$ values are recommended to ensure stable estimates of high-probability regions that dominate the entropy calculation.

Moreover, the choice of $k$ is subject to a mathematical constraint derived from the domain of the gamma function used in the estimator. Specifically, the estimator is well-defined only when $\Gamma(k+1-\alpha)$ finite. This imposes a lower bound on permissible values of $k$, especially for large $\alpha$, and should be carefully considered when selecting estimator parameters.

\subsection{Insights from Coupled Stochastic Processes}

To evaluate the estimator's ability to detect directional information transfer, we apply it to a variety of simulated coupled stochastic processes with known causal structure. These settings include both linear and nonlinear interactions, as well as varying coupling strengths. Our findings indicate that, when memory parameters are correctly specified, the complete RTE reliably captures information flows, even in the presence of multiple sources. Apparent RTE, i.e. when additional causal sources are not accounted for, tends to underestimate the true transfer, confirming the necessity of conditional RTE in presence of multiple sources.

Our findings suggest that the relative explanation added, $\mathcal{R}$, increases with $\alpha$ across most coupling types. However, in example with nonlinear cubic coupling $\kappa=0.2$, $\mathcal{R}$ peaks around $\alpha=0.8$ and further declines as $\alpha$ increases, indicating that tail events can carry proportionally more information. This transition illustrates how different $\alpha$ regimes emphasize distinct dynamical features, providing a richer view of causal interactions than the Shannon-based TE alone.


\subsection{Challenges with Heavy-Tailed Distributions and Small $\alpha$}

Our results also highlight the limitations of RTE estimation for small values of $\alpha$ in the context of heavy-tailed distributions. Since Rényi entropy with $\alpha < 1$ places increasing weight on low-probability events, the estimator becomes more sensitive to tail behavior. This is particularly problematic when the underlying distribution has unbounded support, such as the multivariate Student's $t$-distribution. As the tail heaviness increases (i.e., as the degrees of freedom $\nu$ decrease), the range of $\alpha$ for which Rényi entropy remains well-defined narrows. We find that for heavy-tailed distributions, deviations from analytical values begin to appear close to $\alpha = 1$, and grow more pronounced as $\alpha$ decreases. 

Nonetheless, even for the extreme case of the Cauchy distribution (corresponding to $\nu = 1$), Rényi entropy is still well-defined for $\alpha > 0.75$. This provides a usable interval in which tail-sensitive information measures like RTE can still be meaningfully applied. Thus, although caution is warranted in interpreting RTE for small $\alpha$ under heavy-tailed conditions, the estimator remains applicable in a range that captures significant tail behavior, especially when sample sizes are sufficiently large and the $k$ parameter is appropriately tuned.

\subsection{Practical Steps for Reliable RTE Estimation}

This study is based entirely on simulated data, where the underlying distributions and parameters are fully specified. In contrast, estimating RE and RTE from empirical joint distributions—especially those involving heterogeneous correlated marginals—presents significant challenges.

Unlike Shannon transfer entropy, which is non-negative by definition, RTE can assume negative values. This characteristic makes it essential to distinguish between negativity that reflects meaningful properties of the system under study and negativity that arises from estimation bias or methodological limitations. Additionally, negative component entropies can lead to false-positive RTE. To address this ambiguities, we introduce three reliability conditions that should be satisfied for the resulting RTE estimates to be considered interpretable.

These conditions require that all estimated entropy components $\mathscr{I}_1, \mathscr{I}_2, \mathscr{I}_3, \mathscr{I}_4, \mathscr{I}_{12}, \mathscr{I}_{34}$ must be non-negative and the corresponding Shannon transfer entropy must also be non-negative. These criteria are grounded in fundamental properties of discrete Rényi entropy, conditional Rényi entropy, and Shannon-based transfer entropy, and serve as a minimal diagnostic check to ensure interpretability.

Given the many factors influencing estimation accuracy, we propose a set of practical steps to help calibrate parameters in a way that increases the likelihood of satisfying the reliability conditions. These recommendations are derived directly from the systematic analysis of simulated processes presented in this work, where ground truth is known and estimator behavior can be thoroughly evaluated. While these steps do not guarantee success, they offer a structured approach to guide parameter selection and to evaluate whether results are trustworthy or further data preprocessing is necessary.

\begin{itemize}
\item Start with minimal memory: set $r=1$, $l=1$ and $s_i=1$ for all additional sources.
\item For small values of $\alpha$, use small values of $k$ (e.g., $k=1,2,3,4,5$) to better capture tail events emphasized by Rényi entropy. 
\item For large $\alpha$, increase $k$ to stabilize the estimator in high-probability regions.
\item If RE is negative for $\alpha\geq1$, increase $k$ significantly, e.g., $k=100,500,...$. This reduces the influence of extremely short distances, which can cause the argument inside the logarithm to become smaller than 1 and yield negative values.
\item If RE is negative consider increasing memory parameters, as RE tends to increase with dimensionality (see Appendix \ref{Appendix_NegativeRE}). Change of memory parameter of a source process affects $\mathscr{I}_3$ and $\mathscr{I}_4$, while change of memory parameter of a target process has impact on all component entropies.
\item Estimation biases due to the finite sample size are pronounced by high dimensionality or heavy-tailed distributions. When computing effective RTE, it is essential to shuffle the process that contributes most to these biases.

\item Note: If above steps do not succeed, consider preprocessing the data (e.g., filtering, normalization, shifting time series) or incorporating additional relevant sources to better account for shared information structure.

\end{itemize}
These guidelines are intended to support practical use of the $k$-NN estimator for RTE, particularly in empirical applications where model parameters and distributions are unknown. They also highlight the importance of diagnostic validation and careful parameter tuning in order to draw meaningful conclusions from information-theoretic measures in complex systems.

\section{Conclusions}\label{sec_conclusion}

In this work, we investigated bivariate information flow using the RTE estimated via the $k$-nearest-neighbor method. We first analyzed how three key factors --- the sample size $N$, the data dimensionality $d$, and the nearest-neighbor parameter $k$ --- affect the estimation of the Rényi entropy across a range of Rényi parameter values $\alpha$. We then applied the estimator to quantify information flow in four simulated coupled stochastic processes that exhibit progressively increasing complexity of influence.

Our analysis shows that estimating RE becomes most challenging at small values of $\alpha$. This difficulty arises for two reasons. First, low $\alpha$ values place greater weight on the distribution's tails, which encode extreme events that are typically undersampled and therefore degrade estimator precision. Second, when the underlying distributions have unbounded support, although this effect is negligible in most empirical applications. Overall, our results indicate that estimation at small $\alpha$ is more reliable for small $k$, whereas for $\alpha > 1$, larger $k$ values yield improved accuracy. As expected, finite-sample biases also affect the estimates, but it can be reduced by increasing  the sample size $N$ or, when this is not possible, by applying a shuffle-based correction. To facilitate consistent and interpretable RTE estimation, we formulated three reliability conditions derived from fundamental properties of RE.

This study establishes a first step into previously unexplored territory: the estimation of Rényi transfer entropy for simulated coupled processes across the wide continuum of the Rényi parameter $\alpha$.  Using four simple examples, we observed variability in how information flow contributes to extreme versus frequent events. The character of this information transfer depends on both the coupling strength and the functional form of the interactions. Moreover, when a target process is influenced by multiple source processes, the resulting interactions become more complex, and all sources must be included in the calculations.

Our findings highlight the subtlety involved in extracting detailed information flows with the versatile $k$-NN method. With careful parameter tuning that balances estimator precision against preservation of the underlying data structure, we demonstrate that the RTE framework can accurately capture causal interactions in both linear and nonlinear coupled systems. Furthermore, the framework shows considerable potential for robustly identifying the causal mechanisms driving extreme events.

Finally, results obtained suggest several directions for future research. One avenue involves developing improved estimation methods, particularly for degenerate processes with fractal dimensions of configuration space. Another direction concerns constructing more complex coupled systems with multiple sources or intricate coupling mechanisms based on known underlying dynamics. Such systems can be designed to selectively influence specific aspects of the distribution—such as the tails or the mean --- providing insight into how information is transferred, shared, or interacts in complex systems. Finally, our findings provide a foundation for extending Rényi-based information-theoretic analyses to empirical applications in diverse fields, including climate science, medicine, and finance.

\section{Author contributions}
Z.T., P.J., H.L.: Writing – original draft, Methodology, Conceptualization. P.J., M.P.: Writing – review \& editing, Supervision. Z.T., H.L.: Software, Investigation, Formal analysis.

\section{Declaration of generative AI and AI-assisted technologies in the manuscript preparation process}

During the preparation of this work the author(s) used ChatGPT in order to enhance the clarity and readability of the text, correct grammatical issues, and assist in identifying relevant literature. After using this tool/service, the author(s) reviewed and edited the content as needed and take(s) full responsibility for the content of the published article.

\section{Code and Data availability}

All code used for simulations and data processing is available on GitHub at:\\[1mm]
\url{https://github.com/zlatataa/OnPracticalEstimationofRenyiTransferEntropy}

\begin{acknowledgments}
%
We thank Stefan Thurner for useful discussions. P.J., H.L., Z.T. and 
M.P. were supported by the Czech Science Foundation Grant (GA\v{C}R), Grant No.
25-18105S. M.P. was also supported by  the Johannes Amos Comenius Programme (P JAC),
project No. $CZ.02.01.01/00/22\_008/0004605$, Natural and anthropogenic georisks.
\end{acknowledgments}



\appendix

\section{Glossary}\label{Appendix_Notation}

\begin{itemize}
\item \textbf{General Symbols}
    \begin{description}
    \item[$\alpha \in \mathbb{R}_0^+$] Rényi parameter
    \item[$r, l, s \in \mathbb{N}$] Markov orders (memory parameters)
    \item[$N \in \mathbb{N}$] Sample size
    \item[$d \in \mathbb{N}$] Dimension
    \item[$k \in \mathbb{N}$] Nearest-neighbour order
    \item[$\rho_{j,i} \in \mathbb{R}$] Euclidean distance between data points $j$ and $i$
    \item[$\mathcal{P_S}$] Student's t-distribution
    \item[$\nu$] Degrees of freedom in Student's t-distribution\\[-1mm]
    \end{description}

\item \textbf{Stochastic Processes}
    \begin{description}
    \item[$X \equiv \{x_1, x_2, \ldots, x_N\}$] Target process
    \item[$Y \equiv \{y_1, y_2, \ldots, y_N\}$] Source process
    \item[$Z \equiv \{z_1, z_2, \ldots, z_N\}$] Conditioning process
    \item[$y_t^{(l)}$] Window of $l$ past observations of $Y$, i.e.\\  $y_t, y_{t-1}, \ldots, y_{t-l+1}$
    \item[$Y_{\mathrm{sh}}$] Shuffled version of $Y$
    \item[$x_t^{(r), \mathrm{sh}}$] Shuffled history of $X$
    \item[$z_t^{(s), \mathrm{sh}}$] Shuffled history of $Z$\\[-1mm]
    \end{description}

\item \textbf{Information-Theoretic Quantities}
    \begin{description}
    \item[$T^R_{\alpha,Y\rightarrow X}(r,l)$] Rényi Transfer Entropy from $Y$ to $X$
    \item[$T^{CR}_{\alpha,Y\rightarrow X|Z}(r,l,s) $] Conditional Transfer Entropy
    \item[$\bar{T}^{R}_{\alpha,Y\rightarrow X}$] Complete Transfer Entropy
    \item[$I_{\alpha}(x_{t+1}:y_t^{(l)}|x_t^{(r)})$] Mutual Information
    \item[$H_{\alpha}(x_{t+1}|x_t^{(r)})$] Conditional Rényi Entropy
    \item[$\hat{H}_\alpha(N,k,\alpha,d)$] Estimate of Rényi Entropy
    \item[$T^{\mathrm{eff}}_{\alpha,Y\rightarrow X}(r,l)$] Effective Rényi Transfer Entropy 
    \item[$\mathcal{R}_{\alpha, Y\rightarrow X|Z}(r, l, s)$] Relative Explanation Added 
    \item[$H_{12}^{\mathrm{bias}}$] Estimation bias of $\hat{\mathscr{I}}_{12}$\\[-1mm]
    \end{description}

\item \textbf{RTE Components}
    \begin{description}
    \item[$\mathscr{I}_1$] $H_\alpha(x_{t+1}, x_t^{(r)}, z_t^{(s)})$
    \item[$\mathscr{I}_2$] $H_\alpha(x_t^{(r)}, z_t^{(s)})$
    \item[$\mathscr{I}_3$] $H_\alpha(x_{t+1}, x_t^{(r)}, z_t^{(s)}, y_t^{(l)})$
    \item[$\mathscr{I}_4$] $H_\alpha(x_t^{(r)}, z_t^{(s)}, y_t^{(l)})$
    \item[$\mathscr{I}_{12}$] $\mathscr{I}_1 - \mathscr{I}_2$
    \item[$\mathscr{I}_{34}$] $\mathscr{I}_3 - \mathscr{I}_4$\\[-1mm]
    \end{description}

\item \textbf{Simulated Processes and Parameters}
    \begin{description}
    \item[$A, C, D$] Source (driving) AR(1) processes
    \item[$B^I, B^{II}, B^{III}, B^{IV}$] Target (driven) processes
    \item[$\eta, \eta_1, \eta_2, \kappa \in \{0.2, 0.5, 0.8\}$] Coupling coefficients
    \item[$a = c = 0.5, \; b = 0.2$] AR parameters
    \item[$\varepsilon_t^A \sim \mathcal{N}(0,1)$] Noise in $A$
    \item[$\varepsilon_t^{B^I} \sim \mathcal{N}(0,0.25)$] Noise in $B^I$
    \item[$\varepsilon_t^{B^{II}} \sim \mathcal{N}(0,0.25)$] Noise in $B^{II}$
    \item[$\varepsilon_t^C \sim \mathcal{N}(0,1)$] Noise in $C$
    \item[$\varepsilon_t^D \sim \mathcal{N}(0,1)$] Noise in $D$
    \item[$\varepsilon_t^{B^{III}} \sim \mathcal{N}(0,0.25)$] Noise in $B^{III}$
    \end{description}
\end{itemize}

\section{
Some relevant concepts from information theory --- Shannon's entropy case \label{Appendix_RE} }


Consider a discrete random variable $X$ taking values in a finite set $\mathfrak{X}$, with a probability distribution function (PDF) $\mathcal{P}_X$. For simplicity, we write for the realizations $p(x)$ instead of the more precise notation $p_X(x)$. The \emph{Shannon entropy} $H(X)$ of $X$ is defined as~\cite{Shannon:1948}:
\begin{eqnarray}
H(X) \  \equiv \ H(\mathcal{P}_X) \ = \ -\sum_{x \in \mathfrak{X}} p(x) \log_2 p(x)\, .
\end{eqnarray}
Now, consider another random variable $Y$ taking values in a set $\Upsilon$, with marginal PDF $p(y)$, and a joint PDF $p(x, y)$ defined over $\mathfrak{X} \times \Upsilon $. The \emph{joint entropy} $H(X, Y)$ is given by
\begin{eqnarray}
H(X, Y) \ = \ -\sum_{x \in \mathfrak{X}} \sum_{y \in \Upsilon} p(x, y) \log_2 p(x, y)\, .
\end{eqnarray}
This definition generalizes naturally to the joint entropy of $n$ discrete variables.

The \emph{conditional entropy} $H(Y|X)$, which quantifies the remaining uncertainty in $Y$ given $X$, is defined as
\begin{eqnarray}
H(Y|X) \ = \ -\sum_{x \in \mathfrak{X}} \sum_{y \in \Upsilon} p(x, y) \log_2 p(y|x)\, ,
\end{eqnarray}
where the conditional probability $p(y|x) = {p(x, y)}/{p(x)}$, assuming $p(x) \neq 0$.

The amount of shared information between $X$ and $Y$ is measured by the \emph{mutual information}, defined as
\begin{eqnarray}
I(X; Y) \ &=& \  \sum_{x \in \mathfrak{X}} \sum_{y \in \Upsilon} p(x, y) \log_2 \frac{p(x, y)}{p(x) p(y)}\nonumber \\[2mm]
&=& \ H(X) \ + \ H(Y) \ - \  H(X,Y) \nonumber \\[2mm]
&=& \ H(X)  \ - \ H(X|Y)\, .
\end{eqnarray}
One can use Jensen's inequality for concave functions [in this case $\log_2(\cdots)$] to show that $I(X; Y) \geq 0$. The latter is also known as Gibbs' inequality.

The \emph{conditional mutual information} (CMI) of $X$ and $Y$ given a third variable $Z$ is
\begin{eqnarray}
I(X; Y|Z) \ = \  \sum_{x, y, z} p(x, y, z) \log_2 \frac{p(x, y|z)}{p(x|z) p(y|z)}\, .~~~~
\end{eqnarray}
It is clear from this definition that when $Z$ is statistically independent of both $X$ and $Y$, then
\begin{eqnarray}
I(X; Y|Z) \ = \  I(X; Y)\, .
\end{eqnarray}
A simple algebraic rearrangement leads to the important identity
\begin{eqnarray}
I(X; Y|Z) \ &=& \  H(X, Z) \ + \  H(Y, Z) \nonumber \\[2mm] &-& \  H(X, Y, Z) \ - \  H(Z)\, .
\label{A.7.cv}
\end{eqnarray}

The identity~(\ref{A.7.cv}) for CMI can be extended to the framework of \emph{Rényi entropy}. RE was introduced by A.~Rényi in his seminal series of papers~\cite{renyi1961measures,Renyi:1976a}. RE is an information-theoretic measure that constitutes a one-parameter family of entropy measures, which generalizes the classical Shannon entropy.
%
%
%
RE defined as
\begin{eqnarray}
H_\alpha(X) \ \equiv\ H_\alpha(\mathcal{P}_X) \ = \  \frac{1}{1 - \alpha} \log \left( \sum_{x \in \mathfrak{X}}  p^{\alpha}(x) \right),  ~~~~~
\end{eqnarray}
with $\alpha \in \mathbb{R}_0^+$, which reduces to Shannon entropy in the limit $\alpha \to 1$, i.e., $ \lim_{\alpha \to 1} H_\alpha(X) = H(X)$.

In this paper, we define CMI and its Rényi generalization for one-dimensional (scalar) variables. This form has proven sufficient to uncover unidirectional causal relationships in our numerical experiments presented in the main body of the paper. For extensions to higher-dimensional forms of CMI, and their application to multivariate time series, see e.g.~\cite{jizba2012renyi,JLT:22}.

Building on this, the mutual information of order $\alpha$ and the conditional Rényi entropy of order $\alpha$ are defined respectively as:
\begin{eqnarray}
I_{\alpha}(X ; Y) \ &=& \ \frac{1}{1 - \alpha} \log_2 \frac{\sum_{x \in \mathfrak{X}, y \in \Upsilon} p^{\alpha}(x) p^{\alpha}(y)}{\sum_{x \in \mathfrak{X}, y \in \Upsilon} p^{\alpha}(x, y)}\, , \\
H_{\alpha}(X|Y) \ &=& \  \frac{1}{1 - \alpha} \log_2 \frac{\sum_{x \in \mathfrak{X}, y \in \Upsilon} p^{\alpha}(x, y)}{\sum_{y \in \Upsilon} p^{\alpha}(y)} \nonumber \\[2mm] &=& \ \frac{1}{1 - \alpha} \log_2 \frac{\sum_{x \in \mathfrak{X}, y \in \Upsilon} p^{\alpha}(x|y) p^{\alpha}(y)}{\sum_{y \in\Upsilon} p^{\alpha}(y)}\nonumber  \\[2mm]
&\equiv& \ \frac{1}{1 - \alpha} \log_2 \Big( \sum_{x \in \mathfrak{X}, y \in \Upsilon} p^{\alpha}(x|y) \rho_{\alpha}(y) \Big)\, , ~~~~~~~~
\end{eqnarray}
where 
\begin{eqnarray}
\rho_{\alpha}(y) \ \equiv \ \frac{p^{\alpha}(y)}{\sum_{y' \in \Upsilon} p^{\alpha}(y')}\, , 
\end{eqnarray}
is known as the {\em escort distribution}. This distribution plays a key role in imparting the so-called zooming property to Rényi-based information measures, including Rényi transfer entropy.

The conditional Rényi entropy as defined above satisfies several important properties:\\[-1mm]
\begin{itemize}
\item $H_\alpha(X|Y) \geq 0$ for all $\alpha \in \mathbb{R}_0^+$;
\item $H_\alpha(X|Y) = H_\alpha(X, Y) - H_\alpha(Y)$;
\item The inequality $H_\alpha(X|Y) \leq H_\alpha(X)$ can be guaranteed only when $X$ or $Y$ follow a uniform distribution.
\end{itemize}
In particular, from the last point we see that the {\em mutual information of order} $\alpha$ between $X$ and $Y$, i.e.
\begin{eqnarray}
I_{\alpha}(X;Y) \ = \ H_{\alpha}(X) \ - \  H_{\alpha}(X|Y)\, ,
\end{eqnarray}
might be in some cases negative.

\begin{widetext}
\subsection{RE of multivariate $t$-distribution}

Here we derive RE of multivariate Student $t$-distribution:
\begin{equation}
    \label{eq_student}
    \mathcal{P} _{S}(\textbf{x}) \ = \ \frac{\Gamma\left(\frac{\nu+n}{2}\right)}{\Gamma(\frac{\nu}{2})(\nu \pi)^{\frac{n}{2}} \sqrt{|
    \Sigma|}}\left[1 \ + \ \frac{1}{\nu}(\textbf{x}-\boldsymbol\mu)^T\Sigma^{-1}(\textbf{x}-\boldsymbol\mu)\right]^{-\frac{\nu+n}{2}}.
\end{equation}
$\Sigma$ is a PD $n \times n$ scaling matrix and $\boldsymbol x, \boldsymbol\mu$ are real $n$-dimensional vectors. We put $\boldsymbol\mu = \boldsymbol 0$, $a \equiv \frac{\nu + n}{2}$, and to calculate the integral we use the \textit{Schwinger's trick}, which is widely used in quantum field theory. It is based on integral definition of Gamma function, that can be rewritten as
\begin{equation}
    \frac{1}{A(z)^n} \ = \ \frac{1}{\Gamma(n)}\int_0^{+\infty} du \textrm{ }u^{n-1} \exp(-uA(z))\, .
    \label{schwinger}
\end{equation}
Thus, we have
\begin{eqnarray}
    \label{re_integral_stud}
    \int_{\mathbb{R}} \frac{d^n x}{\left(1+\frac{1}{\nu}\boldsymbol x^T\Sigma^{-1}\boldsymbol x\right)^{\alpha a}} \ &=& \ \frac{1}{\Gamma(\alpha a )} \int_{\mathbb{R}}d^n x \int_{\mathbb{R}^+}du \textrm{ } u^{\alpha a -1}\exp(-u[1+\nu^{-1}\boldsymbol x^T \Sigma^{-1}\boldsymbol x])\nonumber \\[2mm]
    &=& \ \frac{1}{\Gamma(\alpha a )} \int_{\mathbb{R}^+}du \textrm{ } u^{\alpha a -1} e^{-u} \left(\frac{\nu \pi}{u} \right)^{\frac{n}{2}}\sqrt{|\Sigma|}\nonumber \\[2mm]
    &=& \ \frac{\sqrt{|\Sigma|}}{\Gamma(\alpha a )} (\nu \pi)^{\frac{n}{2}} \int_{\mathbb{R}^+}du \textrm{ } u^{\alpha a -1-\frac{n}{2}} e^{-u} \nonumber \\[2mm]
    &=& \ \frac{\sqrt{|\Sigma|}}{\Gamma(\alpha a )} (\nu \pi)^{\frac{n}{2}}\Gamma(\alpha a -{n}/{2})\, ,
\end{eqnarray}
where $\nu$ and $\alpha$ have to satisfy $Re(\alpha\frac{\nu+n}{2}-\frac{n}{2})>0$. Putting this result and (\ref{eq_student}) together gives explicit definition for RE of multivariate $t$-distribution
\begin{equation}
    H_{\alpha}[\mathcal{P} _S] \ = \ \frac{1}{1-\alpha}\log_2\left[\frac{\Gamma^{\alpha}(\frac{\nu+n}{2})\Gamma(\alpha\frac{\nu+n}{2}-\frac{n}{2})}{\Gamma^{\alpha}(\frac{\nu}{2})\Gamma(\alpha\frac{\nu+n}{2})}|\Sigma|^{\frac{1}{2}(1-\alpha)}(\nu \pi)^{\frac{1}{2}n(1-\alpha)}\right].
    \label{eq_RE_student}
\end{equation}
For $\nu=1$ Student's t-distribution is equal to the Cauchy distribution, and for $\nu \to \infty$ it converges to multivariate Normal distribution, RE of which is defined as:
\begin{eqnarray}
    \label{eq_RE_normal}
H_\alpha[\mathcal{P}_{\mathcal{N}}] \ = \ \log_2\left[(2\pi)^{\frac{n}{2}}|\tilde{\Sigma}|^{\frac{1}{2}}\right] \ + \ \frac{1}{1-\alpha}\log_2\left[\alpha^{-\frac{n}{2}}\right],
\end{eqnarray}
where $\tilde{\Sigma}$ is a positive definite covariance matrix. It can be observed that the term involving $\alpha$ depends solely on the dimensionality, $n$. In the computation of RTE, these terms cancel out, and therefore RTE on Gaussian processes is $\alpha$-independent.


\clearpage

\newpage

\section{Negativity of estimates}\label{Appendix_NegativeRE}

Here, we examine a scenario, named Process 1.b, involving a strongly correlated source and target process. As noted earlier, such configurations require careful treatment due to the risk of estimator bias and interpretability issues. The system is defined using the same structure as in Process 1:

\begin{eqnarray}
&&A_{t+1} \ = \ a A_{t} \ + \ \varepsilon^{A}_t\, , \nonumber \\[2mm]
&&B^{I.b}_{t+1} \ = \ b B^{I.b}_{t} \ + \  \eta A_{t} \ + \ \varepsilon^{B^{I.b}}_t\, ,
\end{eqnarray}

but with modified noise properties. Specifically, we reduce the variance of the noise in the target process to 
\( \varepsilon_t^{B^{I}} \sim \mathcal{N}(0,0.0001) \), while keeping the remaining parameters unchanged: 
 \( a = 0.5 \), \( b = 0.2 \), \( \varepsilon_t^A \sim \mathcal{N}(0,1) \). We focus on the case with coupling strength $\eta = 0.5$.
Due to the extremely low noise level in the target process, $B^I$ becomes nearly a deterministic function of the source, resulting in a degenerate structure. Consequently, the joint distribution’s covariance matrix is no longer positive definite, making an analytical solution \ref{eq_RE_normal} inapplicable.

Figure~\ref{fig_re_process1B} shows the estimated Rényi entropies $\hat{\mathscr{I}}_1$, $\hat{\mathscr{I}}_2$, $\hat{\mathscr{I}}_3$, and $\hat{\mathscr{I}}_4$, plotted in orange, magenta, dark blue, and light green, respectively, for Process 1.B. These estimates are obtained on $N = 10^5$ samples, with memory parameters $r = l = 1$, and are computed using the $k$-nearest neighbor parameter $k = 3$ (solid lines) and $k = 50$ (dashed lines). The corresponding effective Rényi transfer entropy (ERTE) in the causal direction is shown in cyan, also differentiated by line style for the two values of $k$.

As shown, not all reliability conditions are satisfied—most notably, $\hat{\mathscr{I}}_3$ is negative. In contrast, the reverse direction (Fig.~\ref{fig_re_process1B_op}) yields more reliable results for $r = l = 1$: all reliability conditions are fulfilled, and as expected, the ERTE is approximately zero, indicating no causal influence in this direction.

To satisfy the reliability conditions in the causal direction, the nearest neighbor parameter, $k$, needs to be increased. High correlation between the source and target processes causes points to cluster tightly in phase space, leading to small arguments in the logarithmic term and negative entropy estimates. Additionally, increasing the memory parameter helps reduce these correlations. Figure~\ref{figSI_RE_process1B_H34} demonstrates how the estimates of $\hat{\mathscr{I}}_3$ and $\hat{\mathscr{I}}_4$ become positive as both the memory parameter $l$ (for the source process $A$) and $k$ increase. As $\hat{\mathscr{I}}_1$ and $\hat{\mathscr{I}}_2$ are independent of the source, they remain unchanged. For comparison, the figure includes estimates for $k = 50$ and $l = 1$ (as shown in Fig.~\ref{fig_erte_process1B}), marked by the thinnest lines. Increasing $k$ to 1000 while keeping $l = 1$ yields a positive $\hat{\mathscr{I}}_3$ (dashed blue line, second from the bottom), although it still remains below $\hat{\mathscr{I}}_4$, resulting in a negative $\hat{\mathscr{I}}_{34}$. Further increases in both $k$ and $l$ are necessary to achieve $\hat{\mathscr{I}}_{34} \geq 0$. The condition is eventually satisfied with $k = 7000$ and $l = 5$, fulfilling all reliability criteria.

Figure~\ref{fig_erte_process1B} shows the ERTE values calculated using the REs from Fig.\ref{figSI_RE_process1B_H34}. All ERTE estimates are non-negative and tend to decrease as $k$ and $l$ increase. The corresponding relative explanation added values are shown in Fig.\ref{fig_rea_process1B}. Unlike ERTE, the REA metric serves as a direct diagnostic for false positives, as values exceeding 1 (indicated by the horizontal dotted line) violate interpretability. Valid ERTEs are thus obtained for $k = 1000$ and $l = 4$ for $\alpha \leq 1$, and for $k = 7000$ and $l = 4$ across the entire $\alpha$ range shown.

The observed trend suggests that further increases in $k$ or $l$ would continue to reduce ERTE, potentially driving it toward zero or even negative values. This indicates that, for specific parameter configurations, it is possible to capture directional entropy transfer suggestive of a causal relationship. However, we interpret these findings as reflecting apparent transfer entropy, shaped by the estimator’s sensitivity to parameter selection rather than representing the underlying causal mechanism in full. To accurately capture complete transfer entropy—more closely aligned with true information flow—a deeper investigation of the process, including its intrinsic dimension, structure would be required, which lies beyond the scope of this study.

\begin{figure*}[h] 
    \begin{subcaptionbox}{\justifying \footnotesize \textit{Effective Rényi transfer entropy and component entropies for the causal direction in Process 1.b.} The $x$-axis shows the Rényi parameter $\alpha$, while the $y$-axis presents both the estimated component Rényi entropies and the resulting effective Rényi transfer entropy in causal direction, $A \rightarrow B^{I.b}$. Estimates are based on $N = 10^5$ samples, with memory parameters $r = l = 1$, using $k = 3$ (solid lines) and $k = 50$ (dashed lines). From bottom to top, the lines represent: $\hat{\mathscr{I}}_3$ (dark blue), $\hat{\mathscr{I}}_2$ (magenta), $\hat{\mathscr{I}}_1$ (orange), $\hat{\mathscr{I}}_4$ (light green), and ERTE (cyan). The ERTE shown is a false positive, as the reliability conditions are violated due to the negative value of $\hat{H}3$.
    \label{fig_re_process1B} }
        {\includegraphics[width=0.48\textwidth]{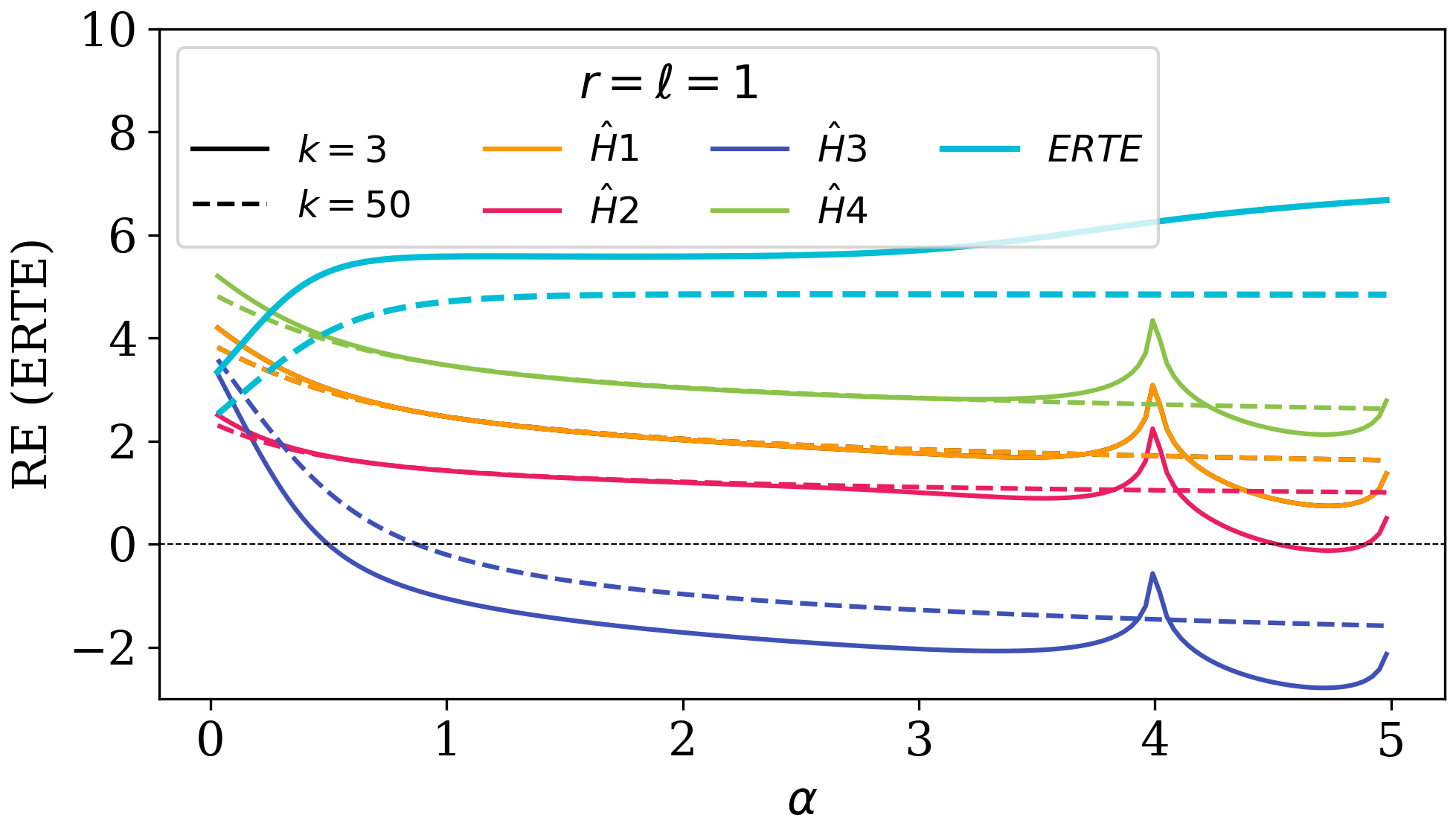}}
    \end{subcaptionbox}
    \hfill
    \begin{subcaptionbox}{\justifying \footnotesize \textit{Effective Rényi transfer entropy and component entropies for the non-causal direction in Process 1.b.} The $x$-axis shows the Rényi parameter $\alpha$, while the $y$-axis presents both the estimated component Rényi entropies and the resulting effective Rényi transfer entropy in non-causal direction, $B^{I.b}\rightarrow A$. Estimates are based on $N = 10^5$ samples, with memory parameters $r = l = 1$, using $k = 3$ (solid lines) and $k = 50$ (dashed lines). From bottom to top, the lines represent: ERTE (cyan around zero), $\hat{\mathscr{I}}_2$ (magenta), $\hat{\mathscr{I}}_4$ (light green), $\hat{\mathscr{I}}_1$ (orange),  $\hat{\mathscr{I}}_3$ (dark blue). All reliability conditions are satisfied, and, as expected, ERTE is zero. \label{fig_re_process1B_op}
    }
        {\includegraphics[width=0.48\textwidth]{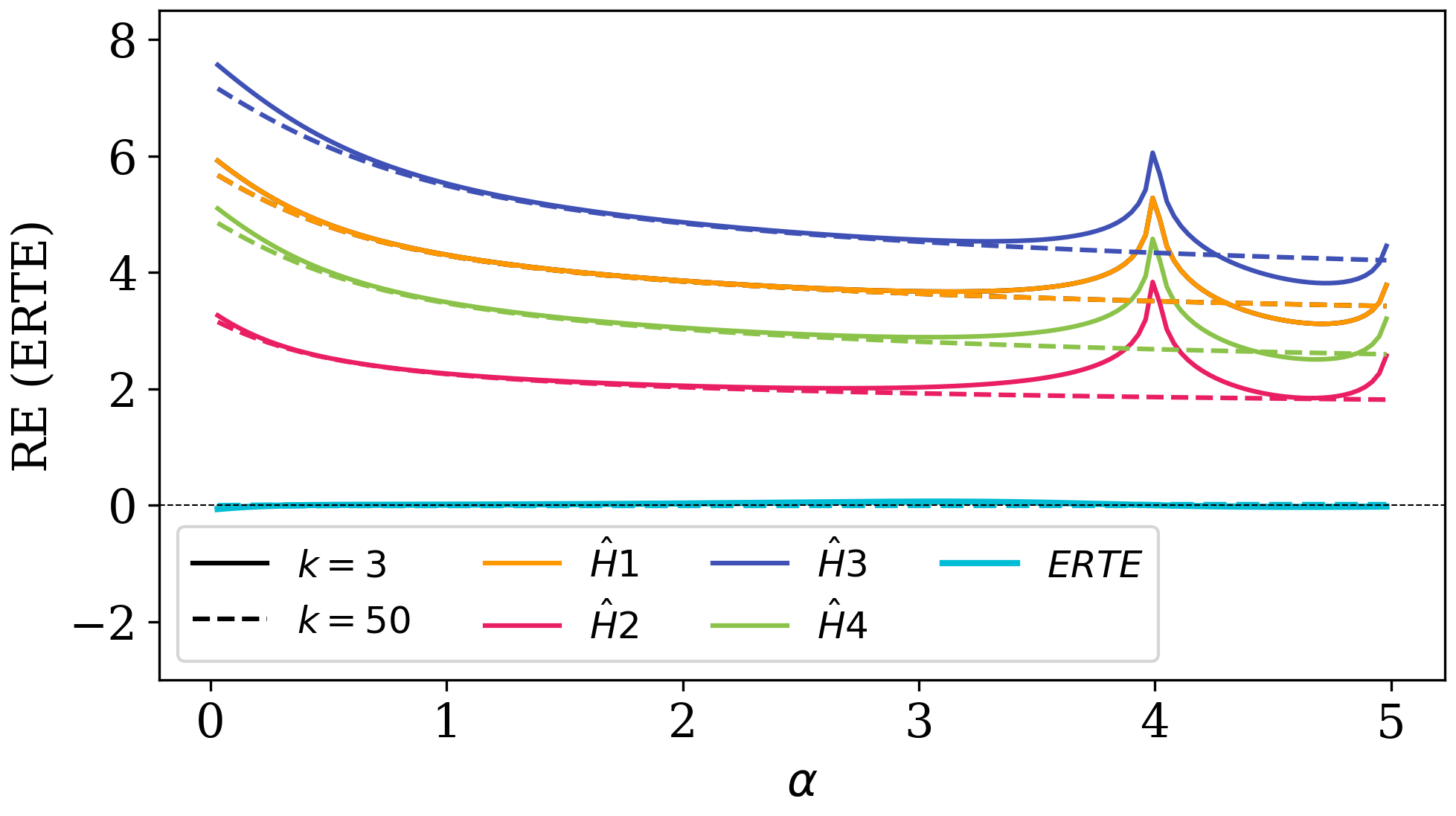}}
    \end{subcaptionbox}
    \caption{\small Process 1.b: Estimated Rényi Entropies and Effective Transfer Entropies for Linear Single-Source Coupling with Strong Correlations}
    \label{figSI_H14_process1B}
\end{figure*}

\begin{figure}[h] 
    {\includegraphics[width=0.48\textwidth]{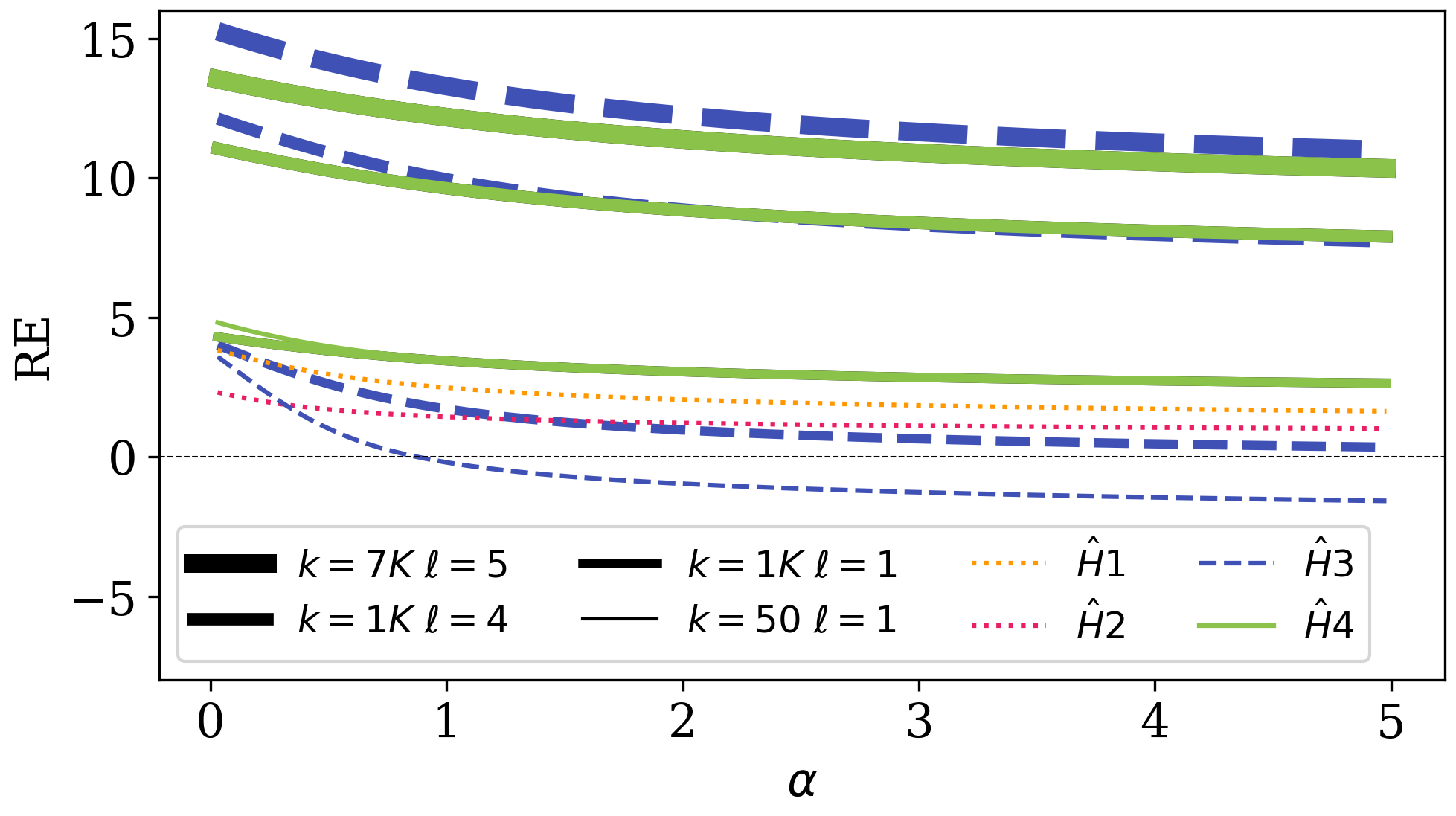}}
    \caption{\justifying \footnotesize \textit{Estimates of Rényi entropies $\mathscr{I}_3$ and $\mathscr{I}_4$ under varying $k$-NN and memory parameters for the causal direction in Process 1.b.} The $x$-axis shows the Rényi parameter $\alpha$, and the $y$-axis shows the estimated component Rényi entropies $\mathscr{I}_1$ through $\mathscr{I}_4$ for the causal direction $A \rightarrow B^{I.b}$. All estimates are based on $N = 10^5$ samples. The target memory parameter is fixed at $r=1$, while the source memory parameter $l$ and the $k$-nearest neighbor parameter are varied to assess their influence. The entropy estimates are shown as follows: $\hat{\mathscr{I}}_1$ (dotted orange), $\hat{\mathscr{I}}_2$ (dotted magenta), $\hat{\mathscr{I}}_3$ (dashed dark blue), and $\hat{\mathscr{I}}_4$ (solid light green). As $l$ and $k$ increase, $\hat{\mathscr{I}}_3$ and $\hat{\mathscr{I}}_4$ grow due to increasing dimensionality and larger distances in phase space. To satisfy the reliability conditions, $\hat{\mathscr{I}}_3$ must be positive and $\hat{\mathscr{I}}_{34} = \hat{\mathscr{I}}_3 - \hat{\mathscr{I}}_4$ must also be non-negative—that is, the light green line ($\hat{\mathscr{I}}_4$) must remain below the dashed dark blue line ($\hat{\mathscr{I}}_3$). These conditions are met for $k = 1000$, $l = 4$ for $\alpha \leq 1$, and for $k = 7000$, $l = 5$ across the entire range of $\alpha$ values. In contrast, $\hat{\mathscr{I}}_1$ and $\hat{\mathscr{I}}_2$ remain relatively stable, as they are unaffected by changes in $l$ and converge for large $k$, with minor deviations at low $\alpha$. These results demonstrate that the reliability conditions can be satisfied for specific parameter combinations; however, since such combinations are not unique and the resulting transfer entropy depends strongly on parameter tuning, the observed values should be interpreted as \textit{apparent} rather than complete transfer entropy. }
    \label{figSI_RE_process1B_H34}
\end{figure}

\begin{figure*}[h] 
    \begin{subcaptionbox}{\justifying \footnotesize \textit{Effective Rényi transfer entropies under varying $k$-NN and memory parameters for the causal direction in Process 1.b.} The $x$-axis denotes the Rényi parameter $\alpha$, and the $y$-axis shows the estimated effective Rényi transfer entropies derived from the component entropies presented in Fig.~\ref{figSI_RE_process1B_H34}. All estimates are based on $N = 10^5$ samples. While all ERTE values shown are non-negative, some are derived from intermediate Rényi entropy or conditional Rényi entropy estimates that violate the reliability conditions—such as negativity of individual components—which undermines their validity. The only ERTE curve that satisfies all reliability criteria across the full range of $\alpha$ values corresponds to the parameter combination $k = 7000$ and $l = 5$, indicated by the orange line.\label{fig_erte_process1B} }
        {\includegraphics[width=0.48\textwidth]{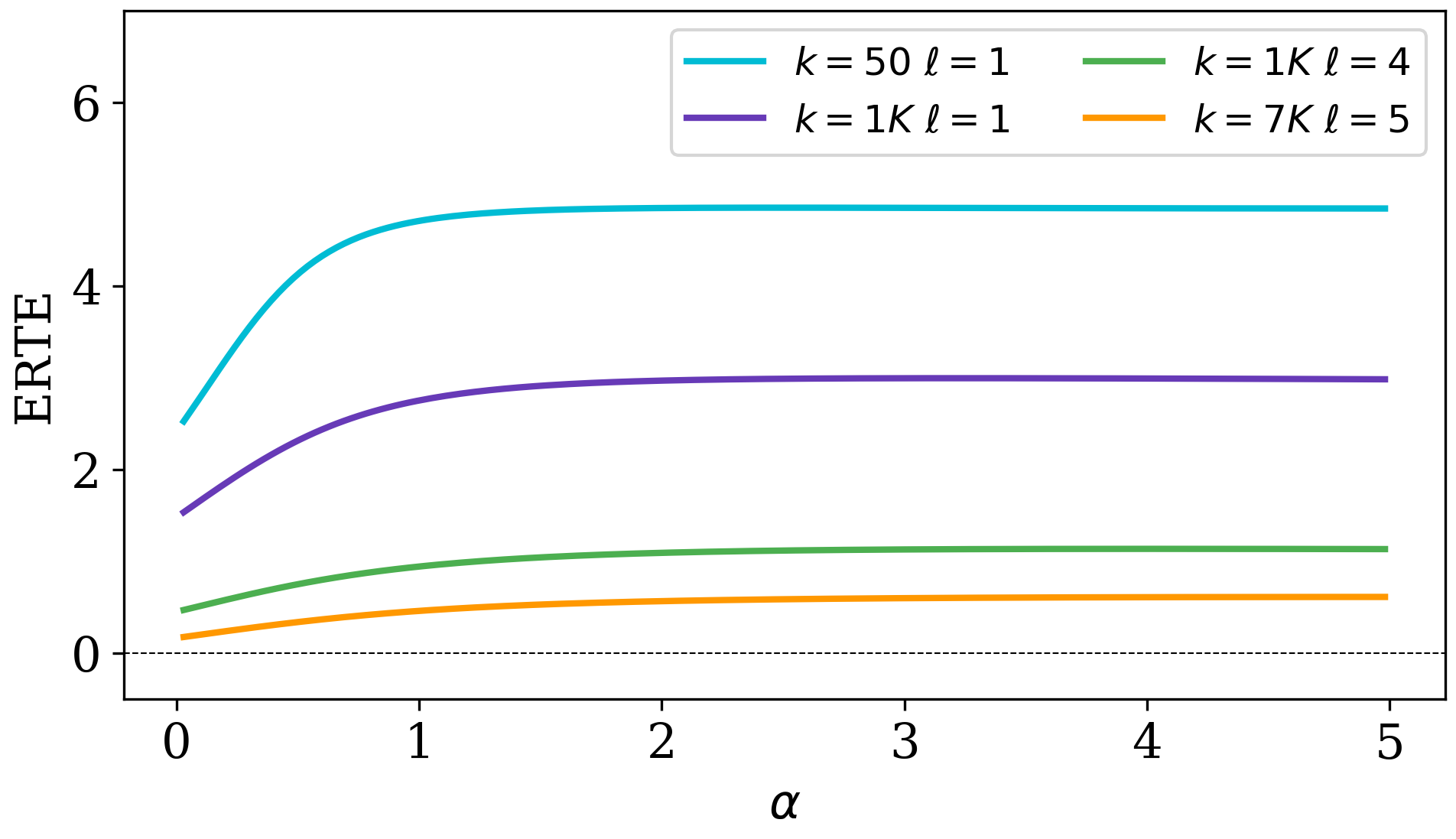}}
    \end{subcaptionbox}
    \hfill
    \begin{subcaptionbox}{\justifying \footnotesize \textit{Relative explanation added (REA) for the causal direction in Process 1.b.  under varying $k$-NN and memory parameters.} The $x$-axis denotes the Rényi parameter $\alpha$, and the $y$-axis displays the relative explanation added (REA), as defined in Eq.~\ref{eq_rea}. In contrast to ERTE, the REA metric offers a direct diagnostic for identifying false positives: values exceeding 1 (highlighted by the horizontal dotted line) indicate violations of interpretability. Based on this criterion, valid ERTEs are achieved for $k = 1000$, $l = 4$ when $\alpha \leq 1$, and for $k = 7000$, $l = 4$ across the entire range of $\alpha$ values presented.\label{fig_rea_process1B}
    }
        {\includegraphics[width=0.48\textwidth]{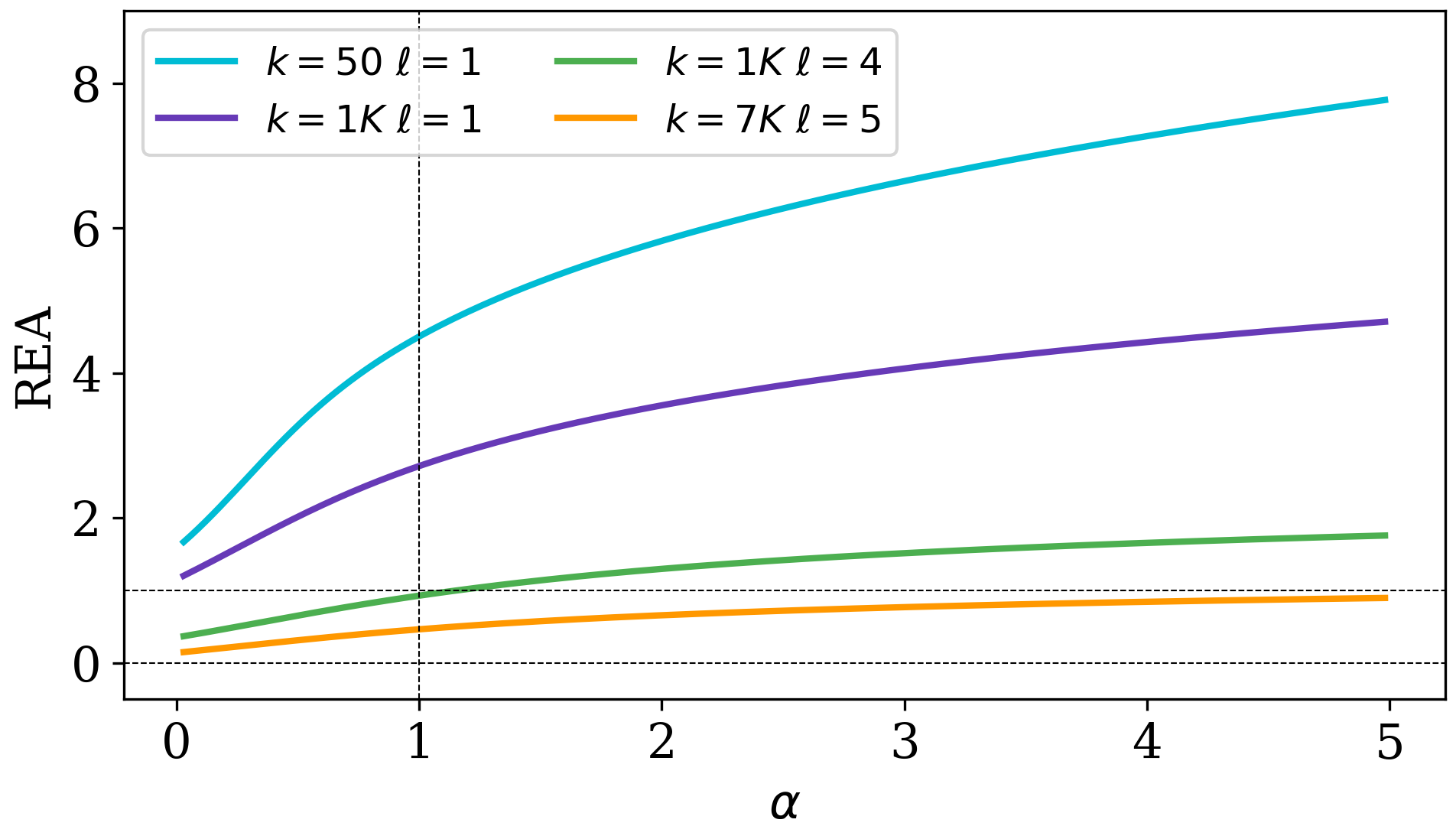}}
    \end{subcaptionbox}
    \caption{\small Process 1.b: Estimated Effective Transfer Entropies and Relative Explanations Added Under Varying $k$-NN and Memory Parameters}
    \label{figSI_process1B_rea_erte}
\end{figure*}

\clearpage

\section{Estimators properties on Cauchy distribution}\label{Appendix_Cauchy}
%

\begin{figure*}[h] 
    \begin{subcaptionbox}{\justifying \footnotesize \textit{Influence of the $k$-Nearest Neighbor Parameter on Estimation Accuracy for the 3-Dimensional Cauchy Distribution.}
    The $x$-axis represents the Rényi parameter $\alpha$, and the $y$-axis shows the estimated Rényi entropy for a 3-dimensional Cauchy distribution with an identity scaling matrix. Analytical values are marked by cyan crosses, while solid lines correspond to estimates obtained using $k = 1, 2, 3, 4, 5, 10, 50, 100,$ and $1000$. All estimates are based on samples of size $N = 10^5$. Due to the dependence of the estimator on $\Gamma(k+1-\alpha)$, certain combinations of $k$ and $\alpha$ result in divergence when the gamma function’s argument approaches zero or becomes negative. For $\alpha < 1$, the most accurate estimates are obtained with the smallest neighborhood size ($k = 1$), whereas for $\alpha \geq 1$, larger values of $k$ provide more stable estimates. These findings are consistent with the trends observed in Fig.~\ref{fig_est_norm_knn}, underscoring that the optimal choice of $k$ is $\alpha$-dependent.\label{fig_cauchy_neigh} }
        {\includegraphics[width=0.48\textwidth]{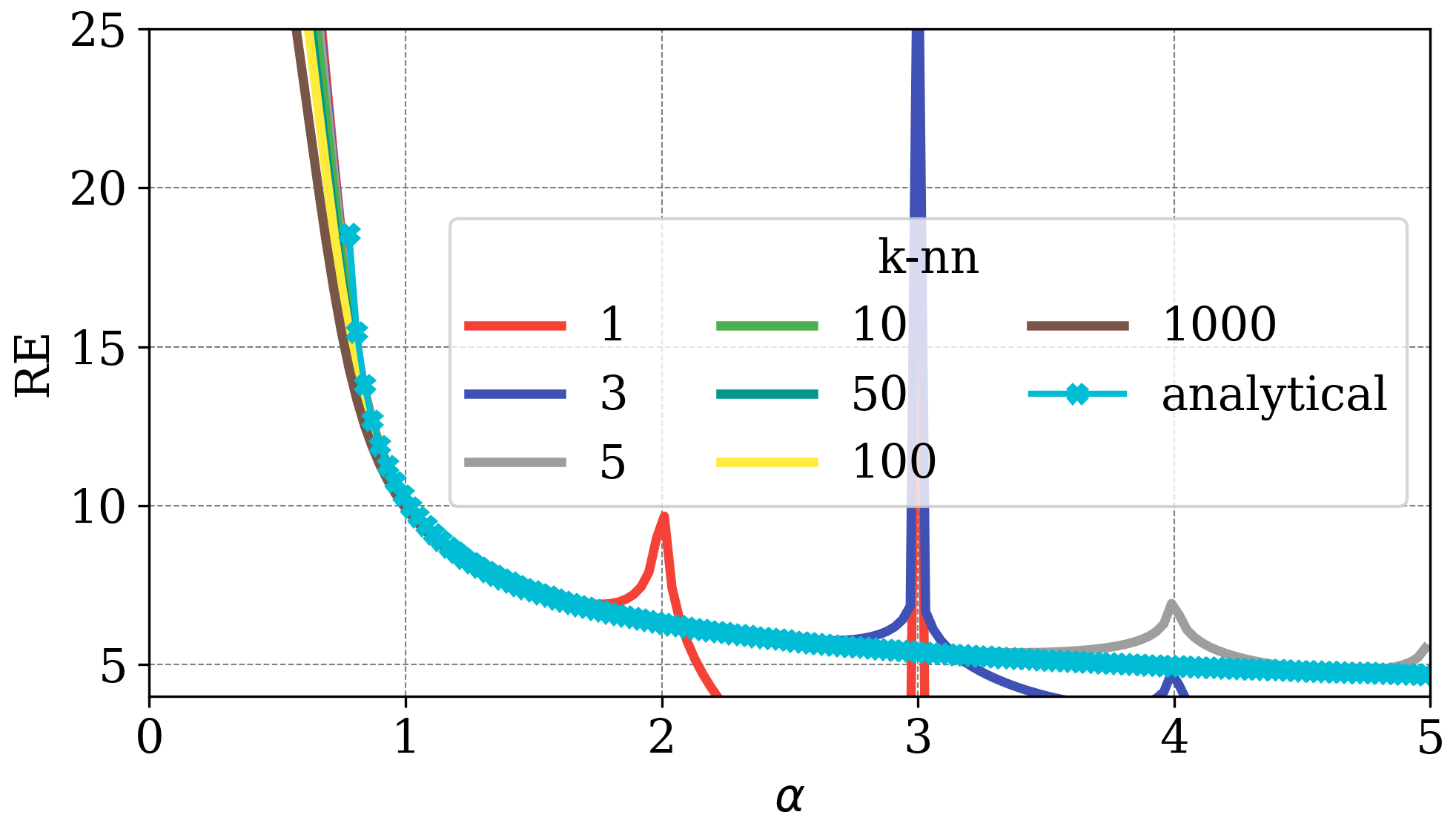}}
    \end{subcaptionbox}
    \hfill
    \begin{subcaptionbox}{\justifying \footnotesize \textit{Effect of Sample Size $N$ on the Estimation of Rényi Entropy for the Cauchy Distribution.}
    The $x$-axis represents the Rényi parameter $\alpha$, and the $y$-axis displays the estimated Rényi entropy for a 3-dimensional Cauchy distribution with an identity scale matrix. Analytical values are indicated by cyan crosses (note that Rényi entropy is not defined for $\alpha < 0.75$ in this case; see Eq.~\ref{re_integral_stud}). Solid lines correspond to estimates obtained using $k = 10$ for sample sizes $N = 100$, $10^3$, $2 \times 10^3$, $5 \times 10^3$, $7 \times 10^3$, $10^4$, $10^5$, and $10^6$, shown in red, blue, light gray, green, turquoise, yellow, brown, and dark gray, respectively. For $\alpha > 1$, accurate estimates are achieved even with moderate sample sizes (e.g., $N = 10^3$). However, estimates in the range $\alpha \in (0.75, 1)$—which emphasize low-probability events—are more sensitive to sample size and converge more slowly to theoretical values. Reliable estimates in this regime begin to emerge for $N \geq 10^5$.
    \label{fig_cauchy_size}}
        {\includegraphics[width=0.48\textwidth]{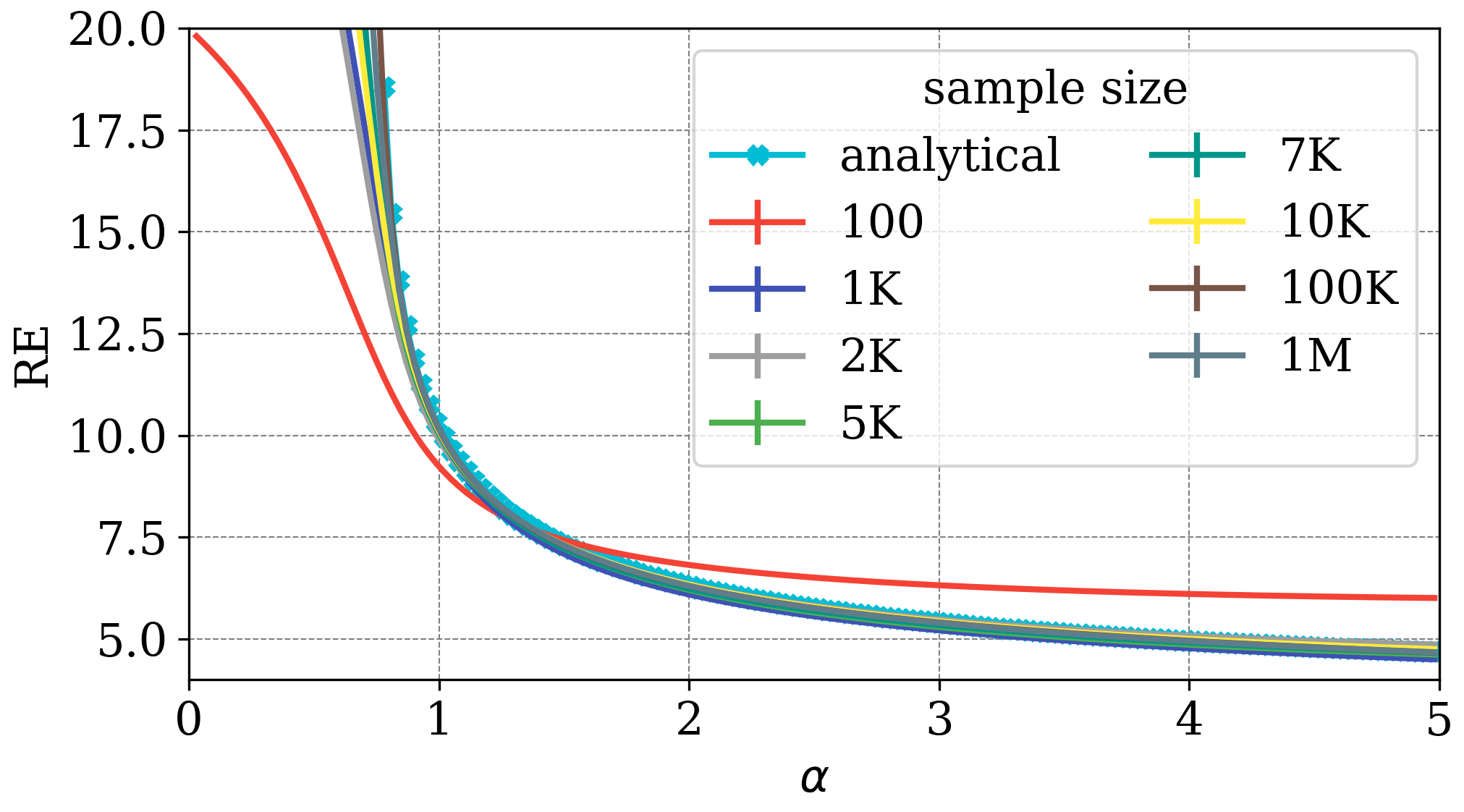}}
    \end{subcaptionbox}

    \begin{subcaptionbox}{\justifying \footnotesize \textit{Influence of Sample Dimensionality $d$ on Estimation Accuracy for the Cauchy Distribution.}
    The $x$-axis represents the Rényi parameter $\alpha$, and the $y$-axis shows the estimated (solid lines) and analytical (x-markers) Rényi entropy for $d$-dimensional Cauchy distributions with identity scaling matrices. As the dimensionality increases, the domain over which the Rényi entropy is defined becomes progressively narrower, particularly for small $\alpha$ values. Estimation accuracy declines with higher dimensions due to heavier tails and increased sparsity in high-dimensional space. Reliable estimates are generally obtained for $d < 10$, as illustrated in panel (d). }
    {\includegraphics[width=0.48\textwidth]{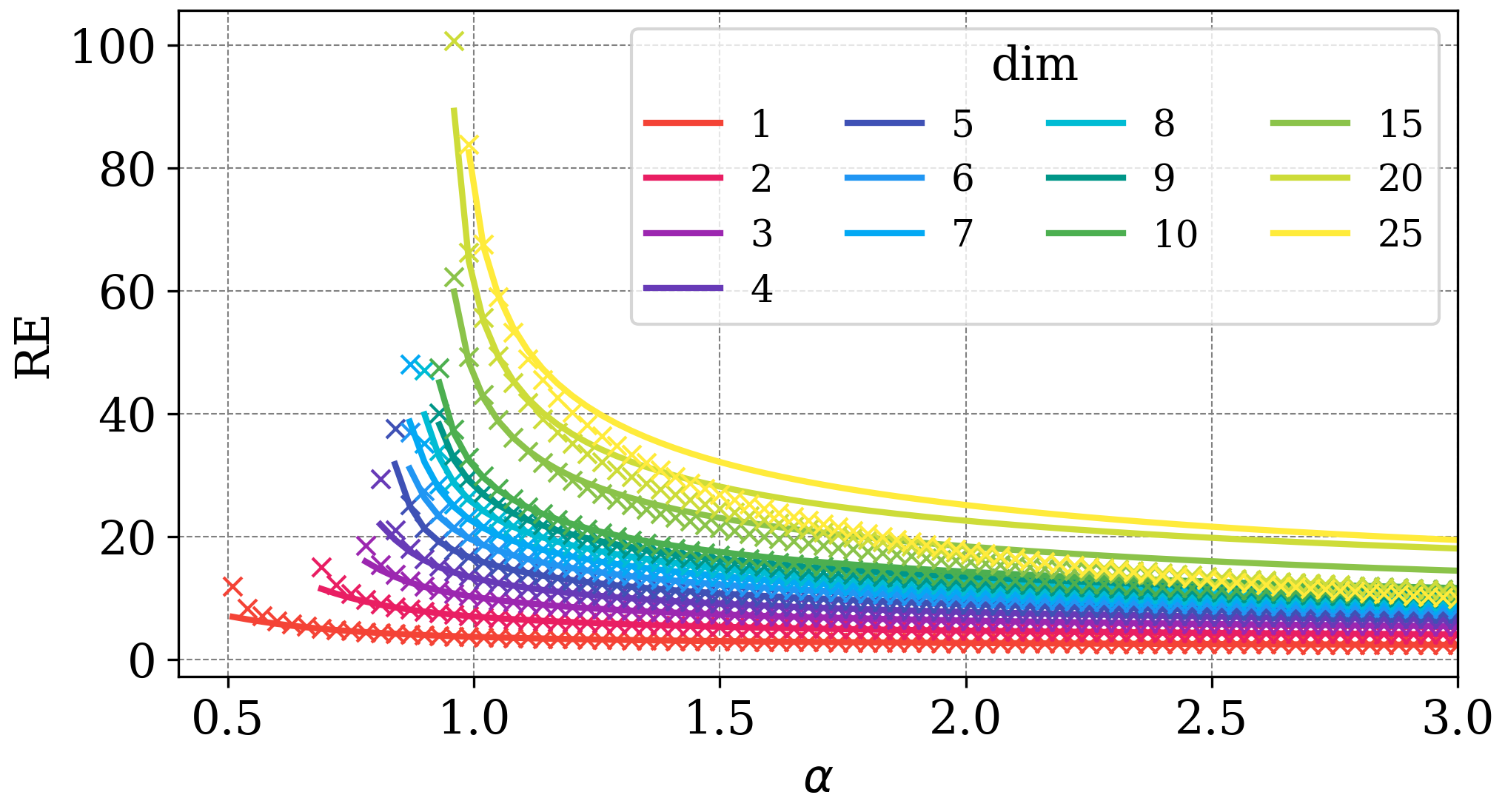}}
    \end{subcaptionbox}
    \hfill
    \begin{subcaptionbox}{ \justifying \footnotesize \textit{ Estimation Errors on Cauchy Samples with Increasing Dimensionality}
    The $x$-axis represents the Rényi parameter $\alpha$, and the $y$-axis shows the ratio of estimated to analytical Rényi entropy for $d$-dimensional Cauchy distributions with identity scaling matrices. Estimates are computed using samples of size $N = 10^5$ with $k = 10$. A ratio near 1 indicates high estimation accuracy. The results reveal that increasing dimensionality substantially degrades estimation performance, consistent with the trend observed for normal distributions in Fig.~\ref{fig_est_dims}.
    \label{fig_cauchy_dims2}}
        {\includegraphics[width=0.48\textwidth]{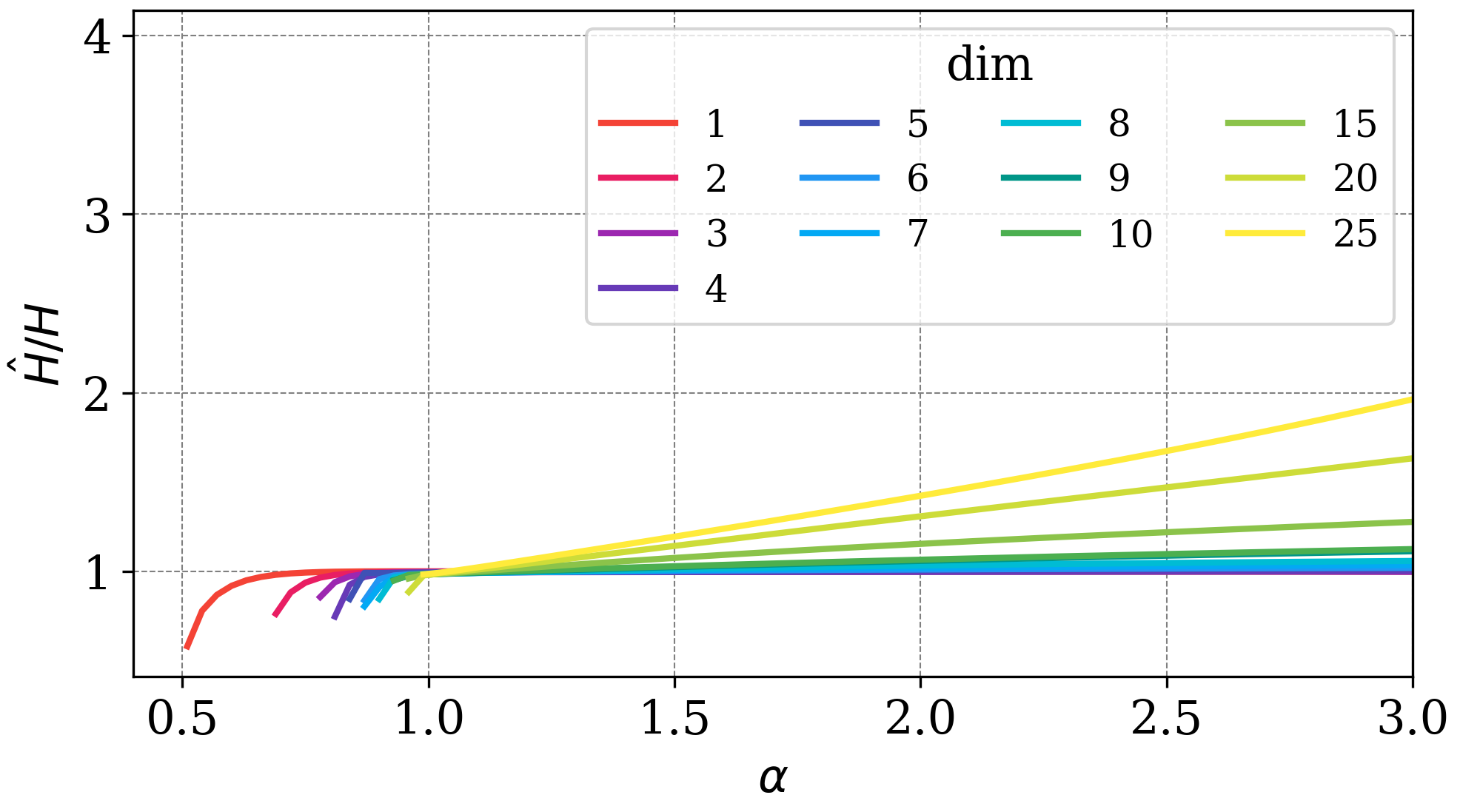}}
    \end{subcaptionbox}
    \caption{\small Performance of the $k$-NN Rényi Entropy Estimator on Cauchy-Distributed Data.}
\end{figure*}

\clearpage
\section{Processes 1-4, detailed visualizations}\label{Appendix_Cauchy}

\begin{figure*}[h] 
    \begin{subcaptionbox}{\justifying \footnotesize    {\em Rényi entropies $\mathscr{I}_1, \mathscr{I}_2, \mathscr{I}_3, \mathscr{I}_4, \mathscr{I}_{12}$ and $\mathscr{I}_{34}$ underlying the RTE in non-causal direction, i.e. $B^I \rightarrow A$, with coupling parameter $\eta = 0.5$.} The $x$-axis represents the Rényi parameter $\alpha$, and the $y$-axis shows estimated (solid lines) and analytical (dashed lines) values of the corresponding Rényi entropies measured in bits. The quantile bands show estimated results over 30 runs. The entropies $\mathscr{I}_1$, $\mathscr{I}_2$, $\mathscr{I}_3$, and $\mathscr{I}_4$ (with dimensions 2,1,3,2) are represented by orange, magenta, blue, and green colors, respectively. For small values of $\alpha$, approximately smaller than $0.5$, the estimates diverge from analytical predictions due to unbounded support of normal distribution. Conditional entropies $\mathscr{I}_{12}=\mathscr{I}_1 - \mathscr{I}_2$ and $\mathscr{I}_{34}=\mathscr{I}_3 - \mathscr{I}_4$ are shown in cyan and yellow colors. All reliability conditions for the estimates are satisfied, i.e $\hat{\mathscr{I}}_1, \hat{\mathscr{I}}_2, \hat{\mathscr{I}}_3, \hat{\mathscr{I}}_4, \hat{\mathscr{I}}_{12}$ and $\hat{\mathscr{I}}_{34}$, are positive.  Additionally, as $\hat{\mathscr{I}}_{12} \geq \hat{\mathscr{I}}_{34}$ for $\alpha=1$ Shannon transfer entropy is non-negative as well. This means that the resulting RTE and ERTE are interpretable. As $\hat{\mathscr{I}}_{12}$ and $\hat{\mathscr{I}}_{34}$ are almost equal on the entire range of $\alpha$, RTE is zero, as given by the definition of the process. Small bias for $\alpha<0.5$ is caused by the sample finiteness, and is corrected for in ERTE. \label{fig_re_process1_op} }
        {\includegraphics[width=0.48\textwidth]{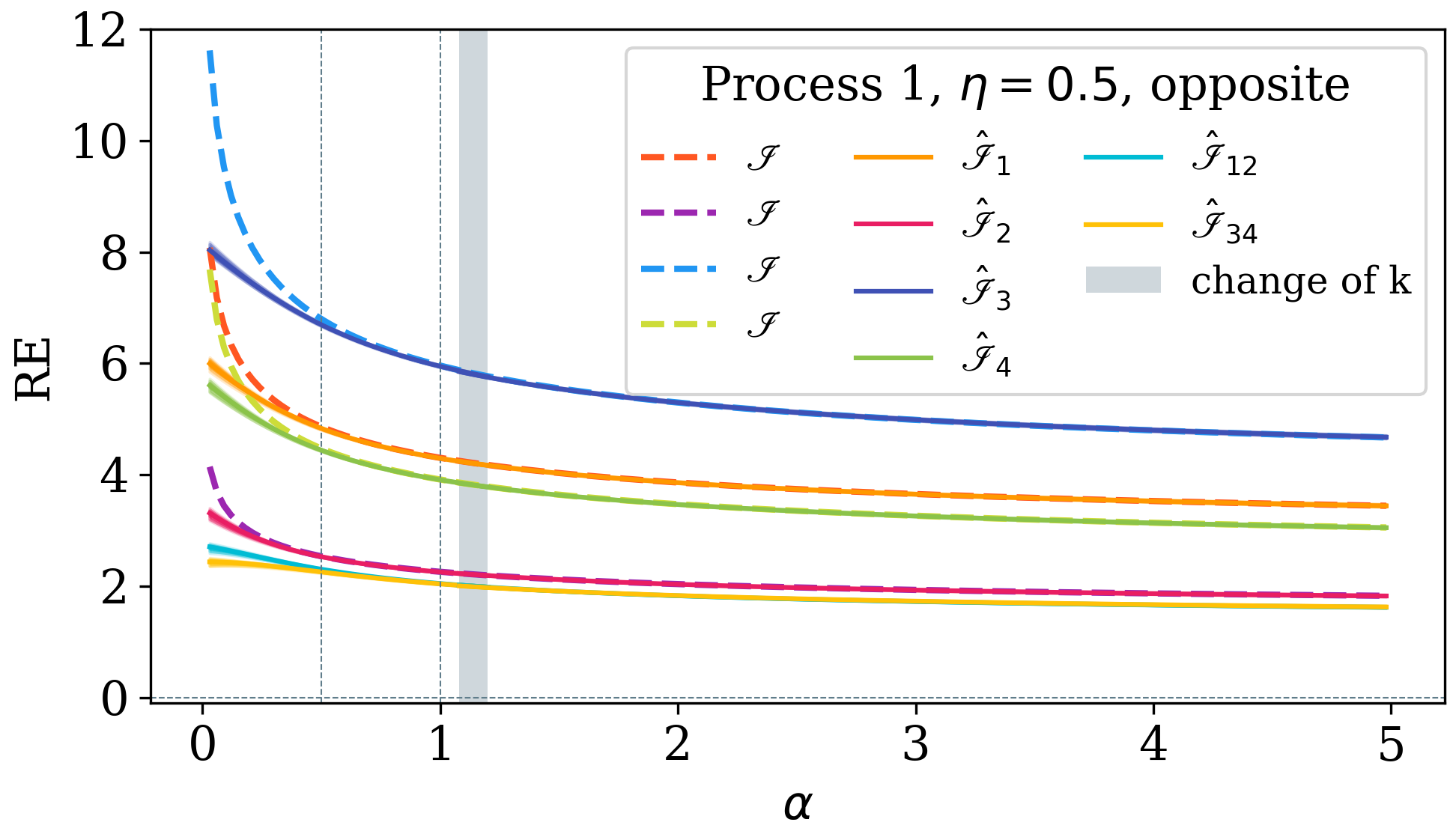}}
    \end{subcaptionbox}
    \hfill
    \begin{subcaptionbox}{\justifying \footnotesize{\em Rényi entropies $\mathscr{I}_1, \mathscr{I}_2, \mathscr{I}_3, \mathscr{I}_4, \mathscr{I}_{12}$ and $\mathscr{I}_{34}$ underlying the RTE in non-causal direction, i.e. $B^{II} \rightarrow C$, with coupling parameter $\eta = 0.5$.} The $x$-axis represents the Rényi parameter $\alpha$, and the $y$-axis shows estimated values of the corresponding Rényi entropies measured in bits. The quantile bands show estimated results over 30 runs. The entropies $\mathscr{I}_1$, $\mathscr{I}_2$, $\mathscr{I}_3$, and $\mathscr{I}_4$ (with dimensions 2, 1, 3, 2) are represented by orange, magenta, blue, and green colors, respectively. For small values of $\alpha$, the estimates exhibit high variability, which further increases with dimensionality. In particular, $\mathscr{I}_3$, computed in three dimensions, shows the greatest variance. Conditional entropies $\mathscr{I}_{12}=\mathscr{I}_1 - \mathscr{I}_2$ and $\mathscr{I}_{34}=\mathscr{I}_3 - \mathscr{I}_4$ are shown in cyan and yellow colors. All reliability conditions for the estimates are satisfied, i.e $\hat{\mathscr{I}}_1, \hat{\mathscr{I}}_2, \hat{\mathscr{I}}_3, \hat{\mathscr{I}}_4, \hat{\mathscr{I}}_{12}$ and $\hat{\mathscr{I}}_{34}$, are positive.  Additionally, as $\hat{\mathscr{I}}_{12} >\hat{\mathscr{I}}_{34}$ for $\alpha=1$ Shannon transfer entropy is positive as well. This means that the resulting RTE and ERTE are interpretable. Transfer entropy becomes negative only for $\alpha <0.5$, this bias—caused by the finite sample size—can be partly corrected by the effective RTE, see Fig.\ref{figSI_effrte_process2}. \label{fig_re_process2_op}
    }
        {\includegraphics[width=0.48\textwidth]{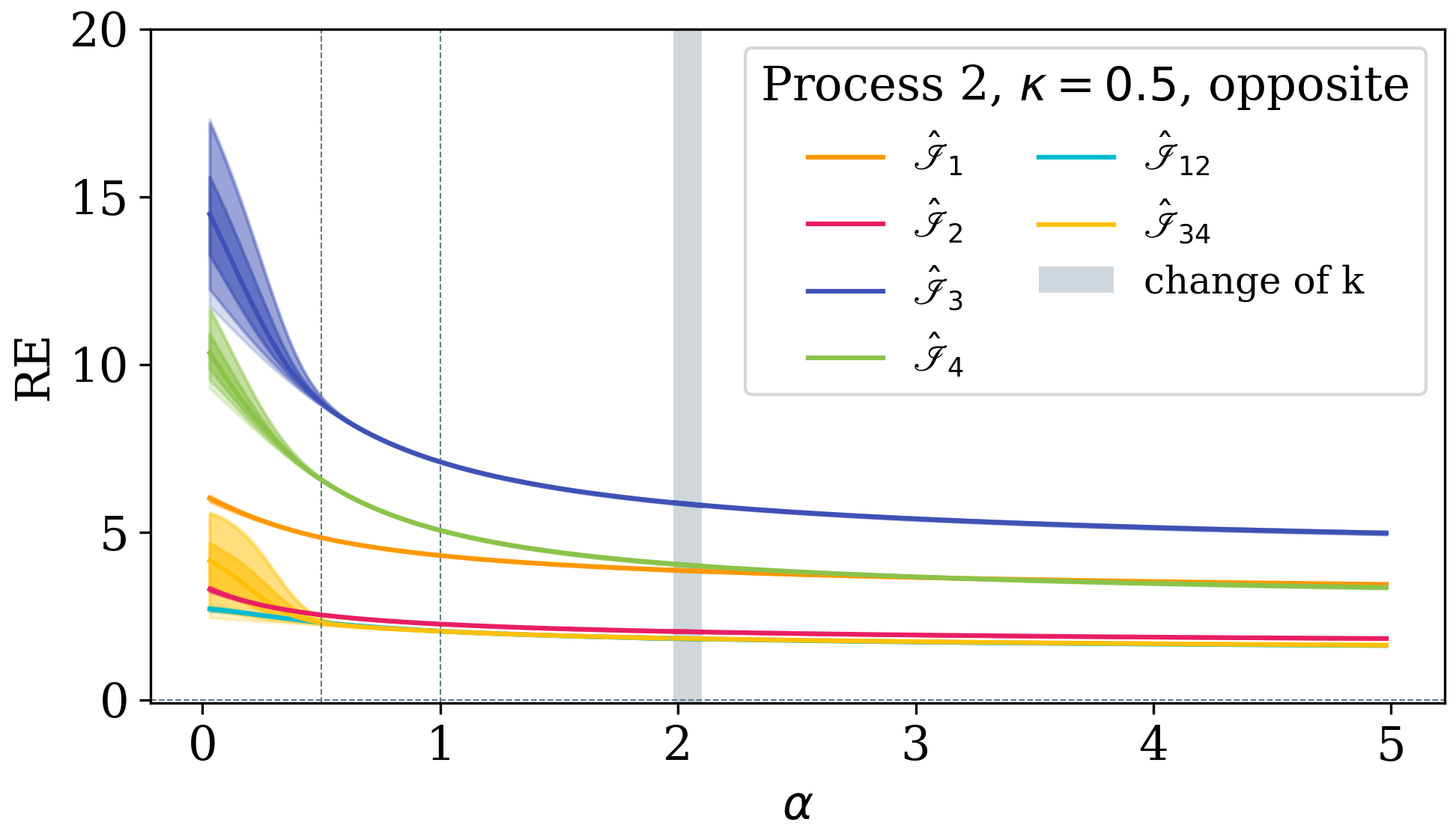}}
    \end{subcaptionbox}
    \caption{ \small {Rényi entropies, $\mathscr{I}_1-\mathscr{I}_4$, for opposite direction in processes 1 and 2 with coupling parameters $\eta=0.5$ and $\kappa=0.5$. } }
    \label{figSI_H14_oppos}
\end{figure*}

\begin{figure*}[h] 
    \begin{subcaptionbox}{ \justifying \footnotesize {RTE and ERTE in both directions for $k=50$.} 
    \label{SIfig_pr1_eff_big} }
        {\includegraphics[width=0.48\textwidth]{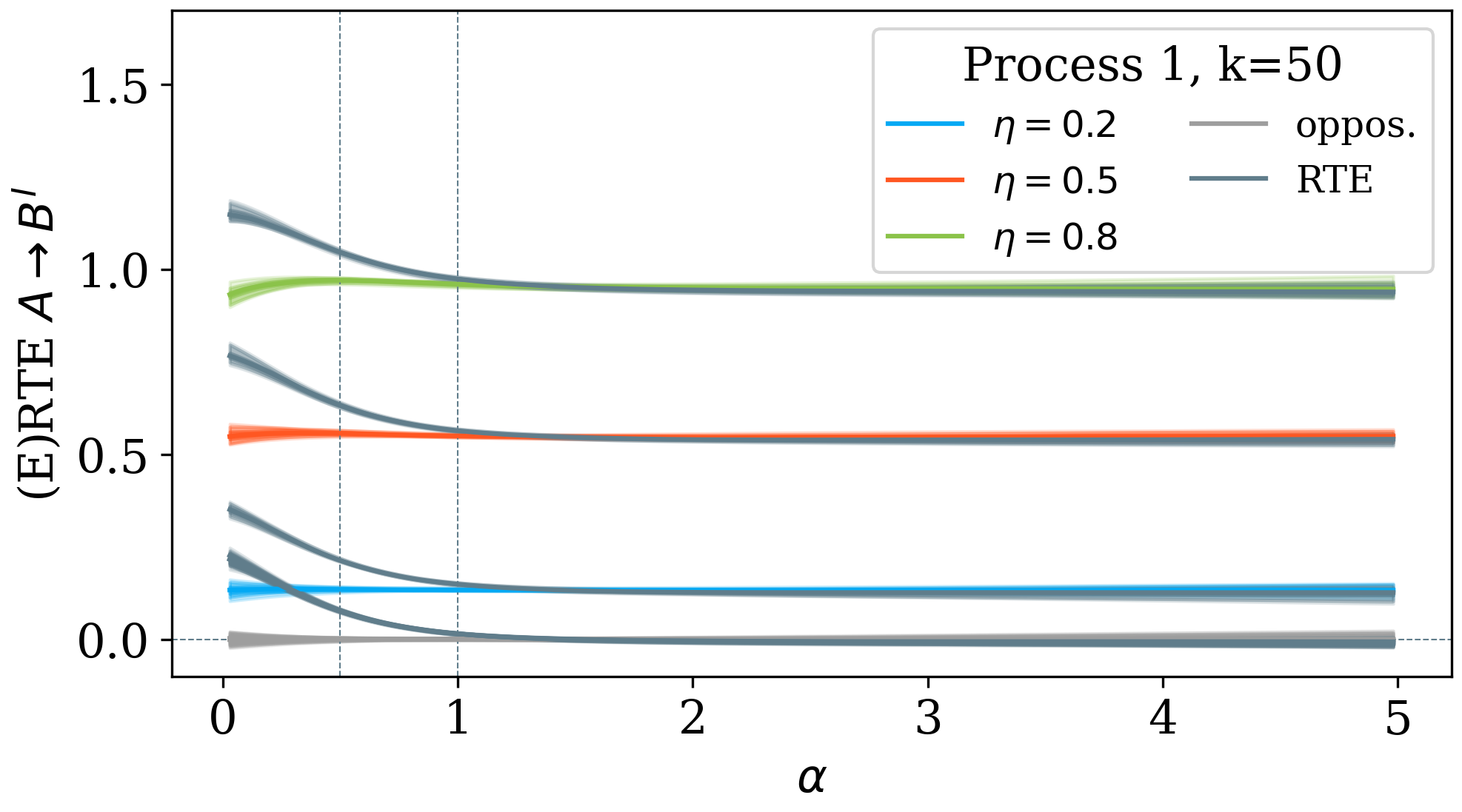}}
    \end{subcaptionbox}
    \hfill
    \begin{subcaptionbox}{ \justifying \footnotesize {RTE and ERTE in both directions for $k=3$.} 
    \label{SIfig_pr1_eff_small} 
    }
        {\includegraphics[width=0.48\textwidth]{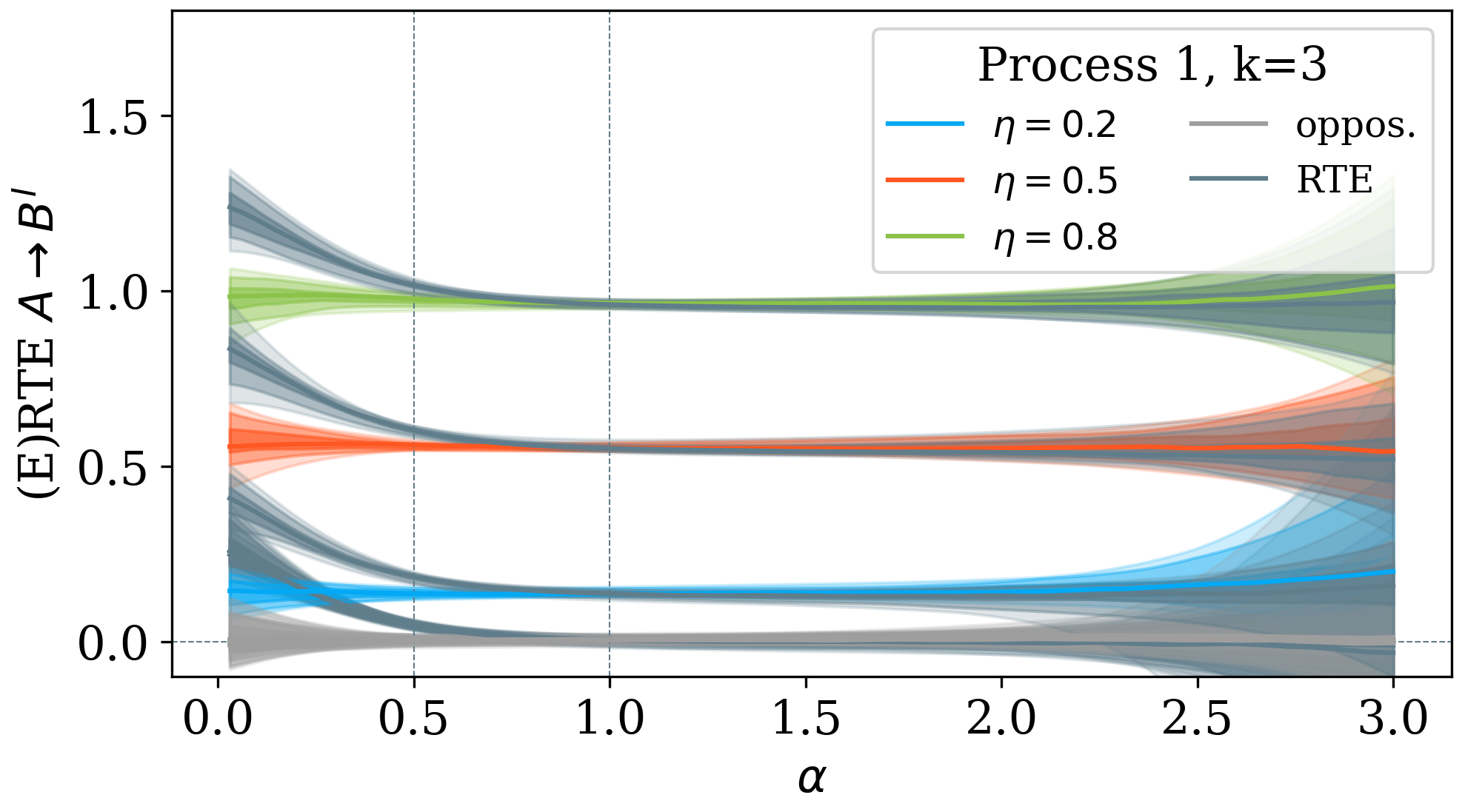}}
    \end{subcaptionbox}
    \caption{\justifying \footnotesize {\em RTE and ERTE of Process 1 in both directions for $k=50$ and $k=3$.} The $x$-axis in both panels shows the Rényi parameter $\alpha$, the $y$-axis shows estimated RTE and ERTE. The memory parameters are $r=l=1$ and sample size is $N=10^5$. The quantile bands -- 1st–99th, 5th–95th, and 25th–75th with a line indicating median  -- show results obtained over 30 runs. Blue, red, and green quantile bands represent effective Rényi transfer entropies in the respective coupling directions. Additionally, gray bands centered around zero correspond to ERTEs in the non-coupling direction, confirming the absence of a causal influence from $B$ to $A$. The dark gray bands indicate estimated RTEs in coupling direction that exhibit noticeable estimation bias for $\alpha < 1$. This highlights the importance of computing effective RTE using shuffled surrogate data to correct for such bias. Notably, even though less noisy, RTE estimates are accurate already from $\alpha=0.5$ for $k = 3$, while for $k = 50$, reliable estimates are obtained only starting approximately at 1. This again confirms, that small values of $k$ are better for estimation on tails. }
    \label{figSI_effrte_process1}
\end{figure*}

\begin{figure*}[h] 
    \begin{subcaptionbox}{\justifying \footnotesize { RTE and ERTE in both directions for $k=50$ and $\kappa=0.2$.} }
        {\includegraphics[width=0.48\textwidth]{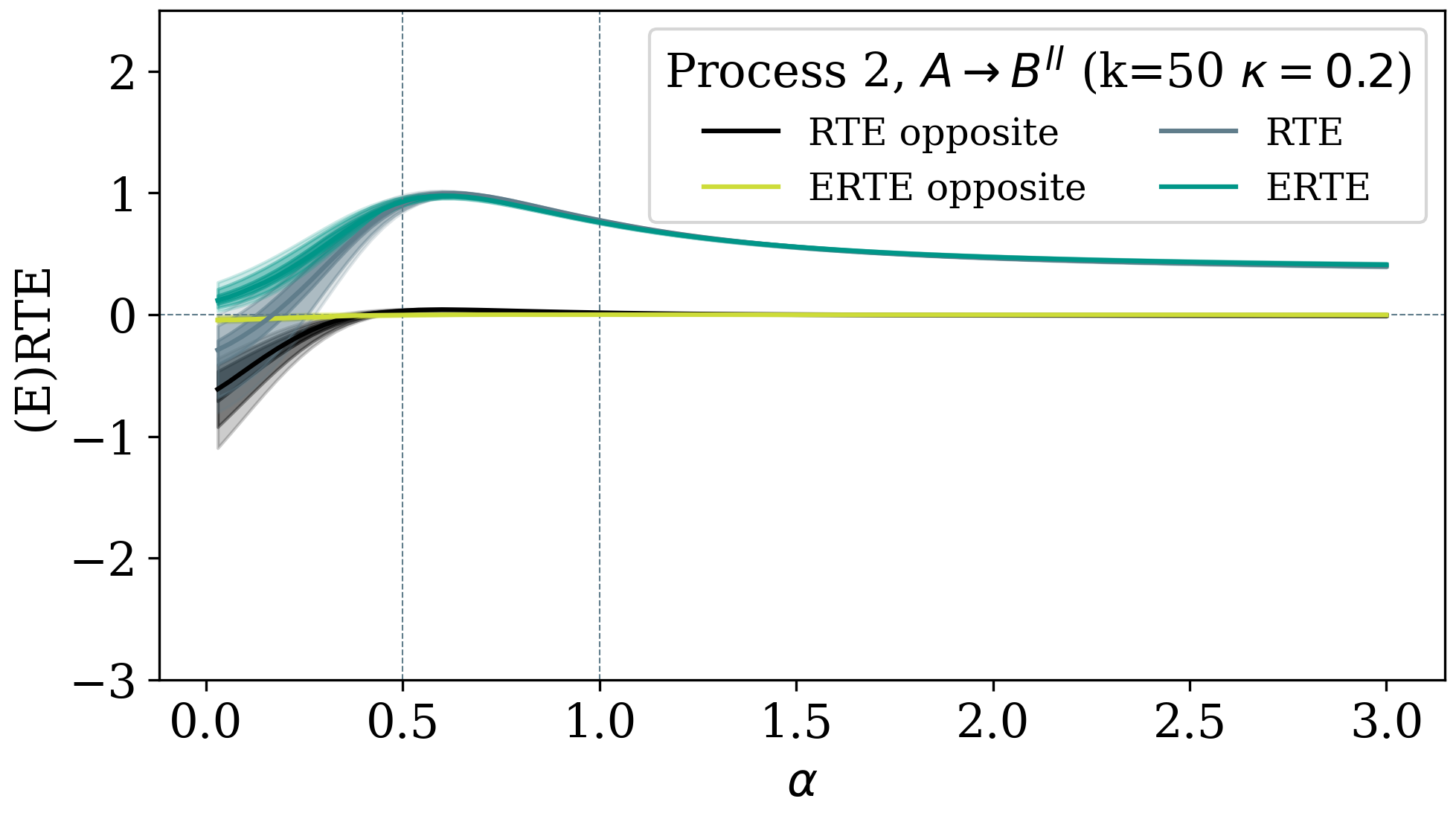}}
    \end{subcaptionbox}
    \hfill
    \begin{subcaptionbox}{\justifying \footnotesize { RTE and ERTE in both directions for $k=3$ and $\kappa=0.2$.}
    }
        {\includegraphics[width=0.48\textwidth]{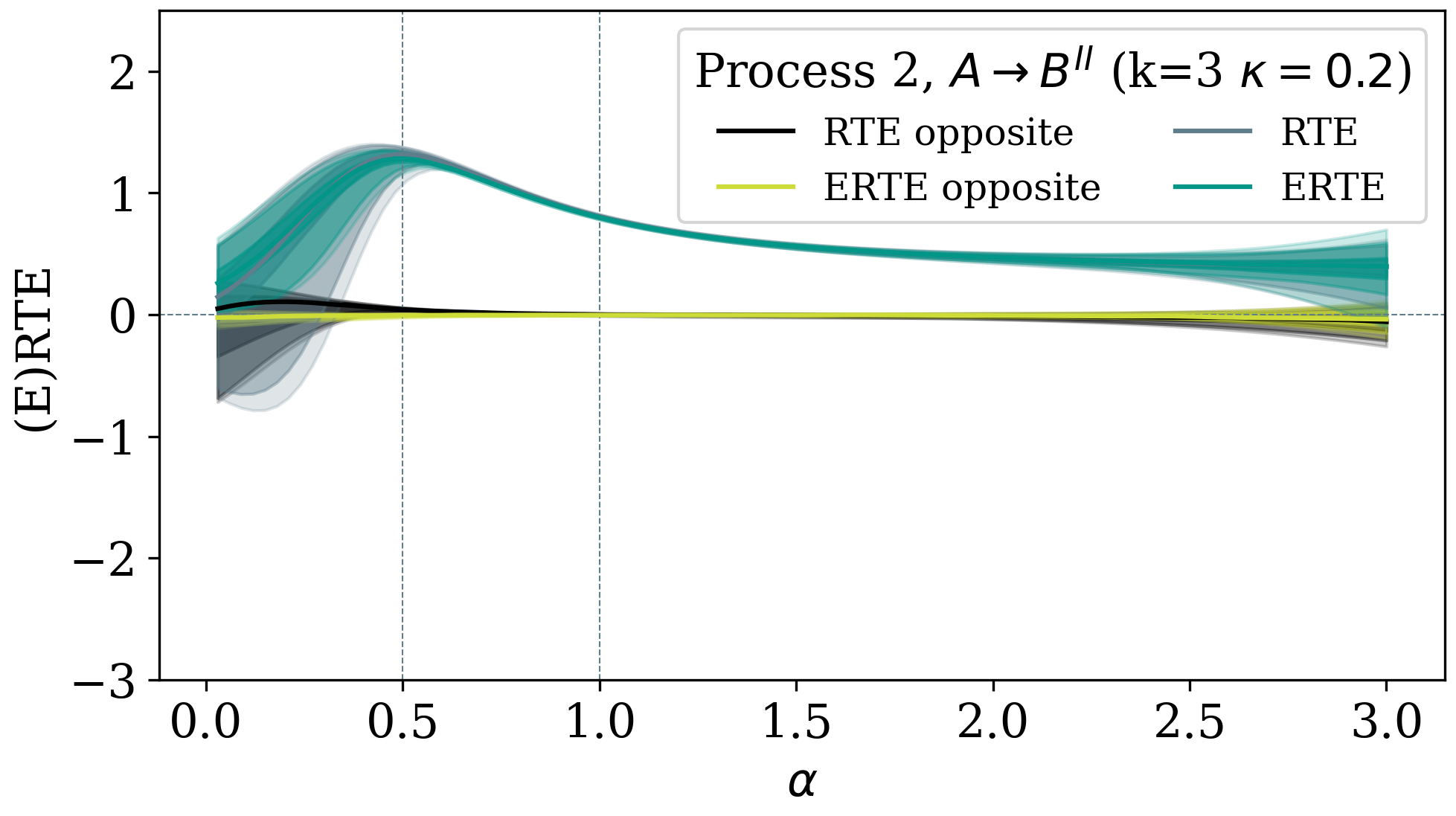}}
    \end{subcaptionbox}

    \begin{subcaptionbox}{\justifying \footnotesize { RTE and ERTE in both directions for $k=50$ and $\kappa=0.5$.} 
    }
        {\includegraphics[width=0.48\textwidth]{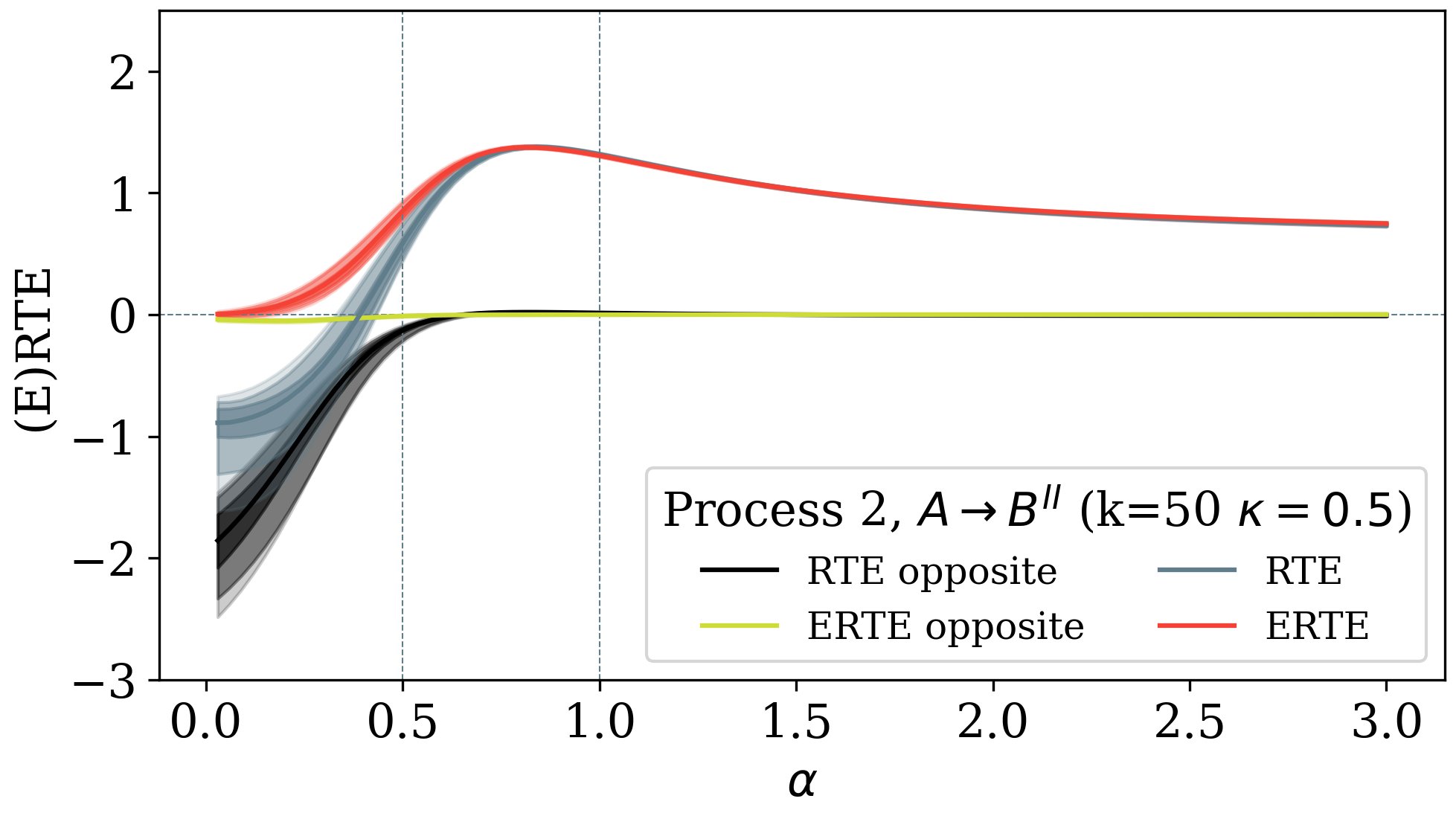}}
    \end{subcaptionbox}
    \hfill
    \begin{subcaptionbox}{\justifying \footnotesize { RTE and ERTE in both directions for $k=3$ and $\kappa=0.5$.} 
    }
        {\includegraphics[width=0.48\textwidth]{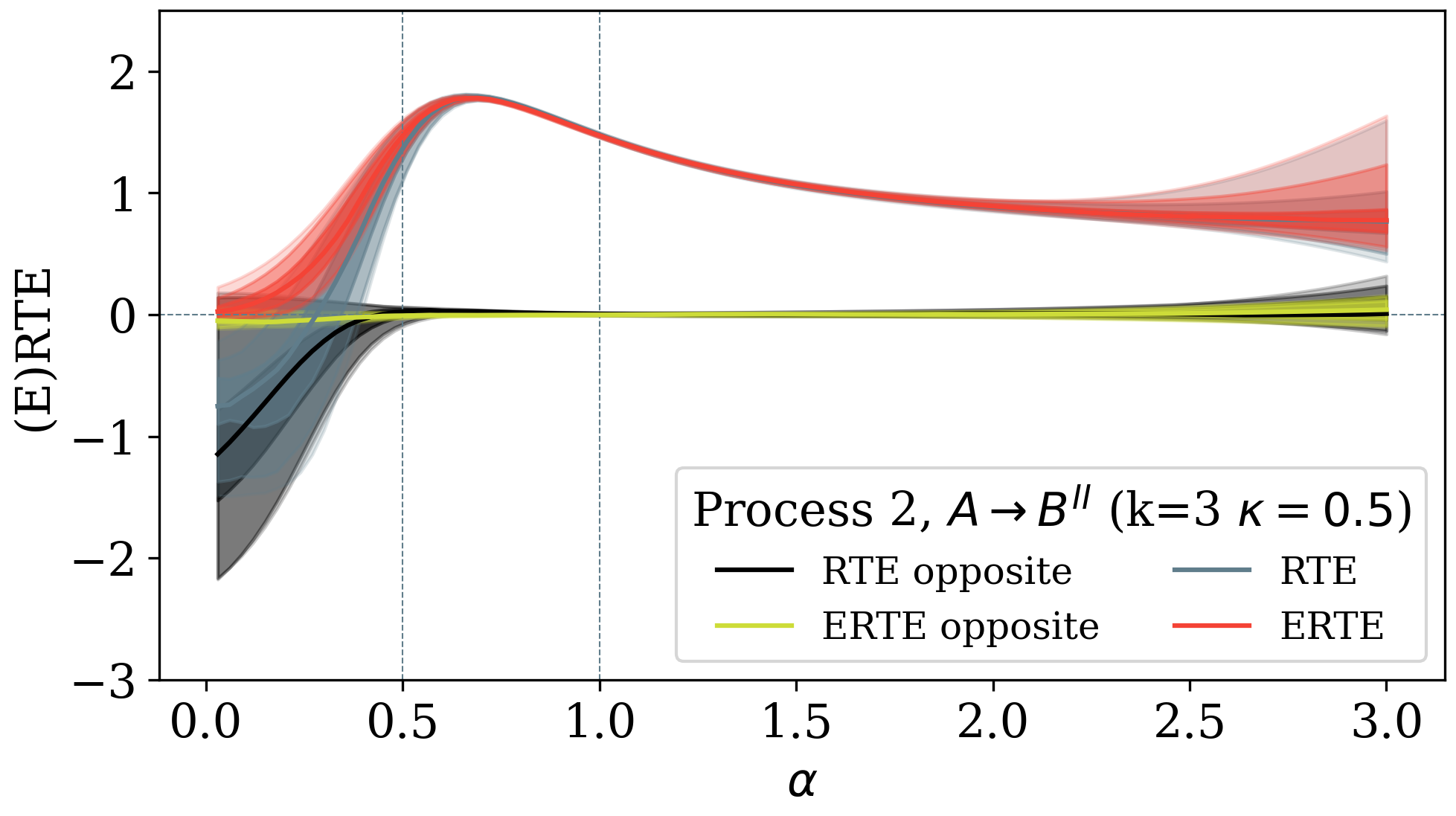}}
    \end{subcaptionbox}

    \begin{subcaptionbox}{\justifying \footnotesize {RTE and ERTE in both directions for $k=50$ and $\kappa=0.8$.}
    }
        {\includegraphics[width=0.48\textwidth]{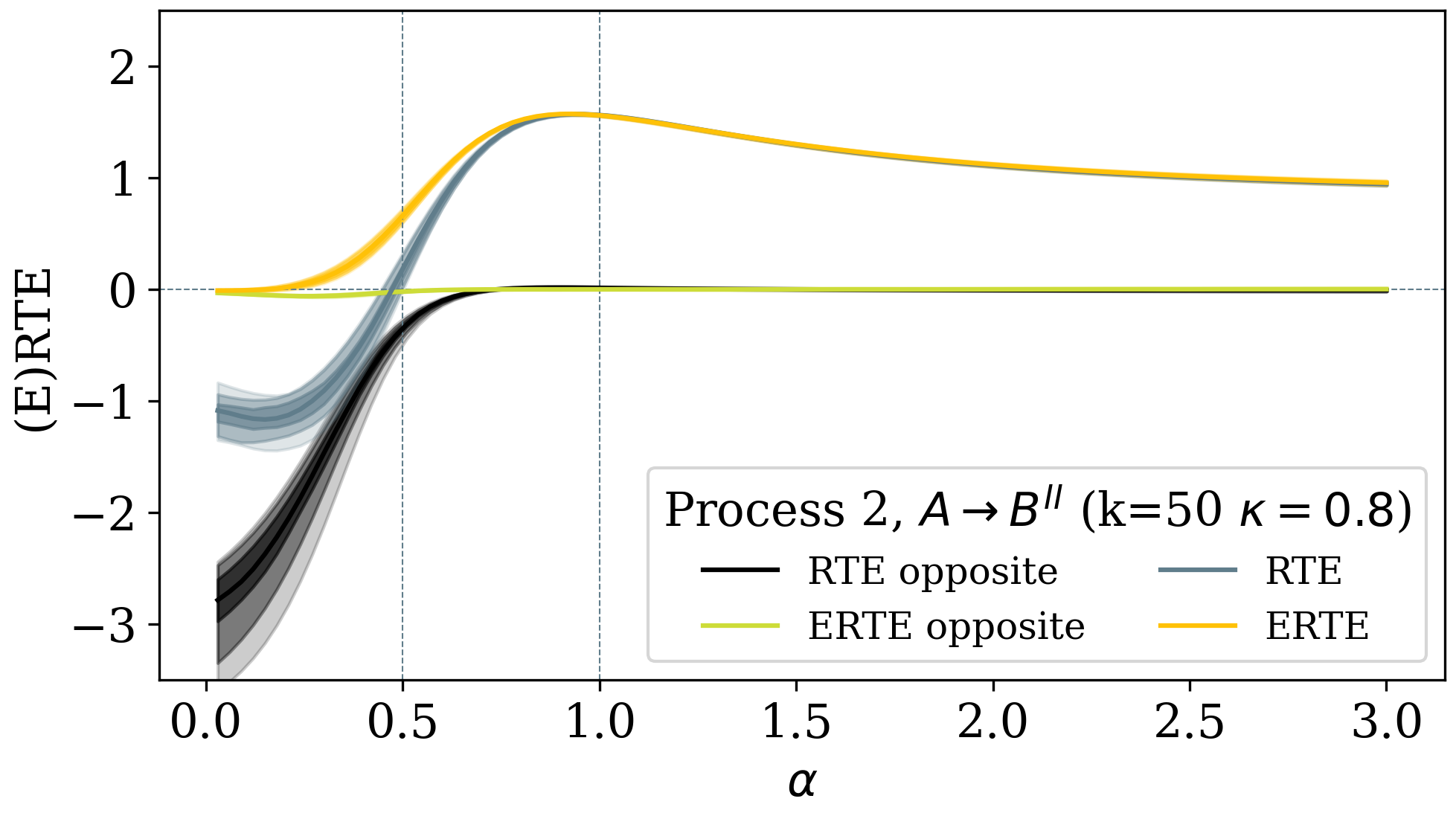}}
    \end{subcaptionbox}
    \hfill
    \begin{subcaptionbox}{\justifying \footnotesize {RTE and ERTE in both directions for $k=3$ and $\kappa=0.8$.}
    }
        {\includegraphics[width=0.48\textwidth]{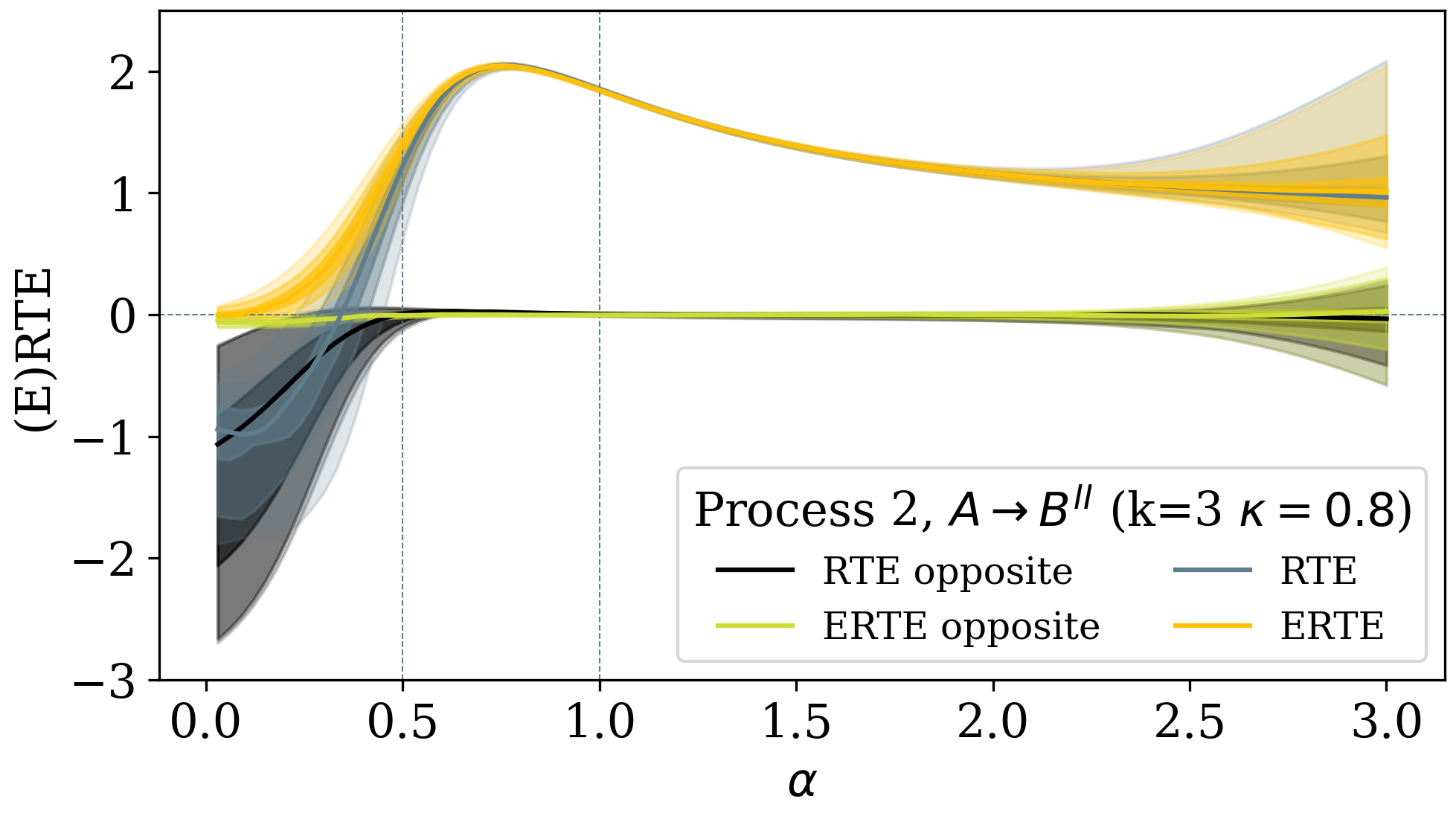}}
    \end{subcaptionbox}
    \caption{\justifying \footnotesize \textit{ RTE and ERTE of Process 2 in both directions for $k=50$ and $k=3$.} The $x$-axis in all panels shows the Rényi parameter $\alpha$, the $y$-axis shows estimated RTE and ERTE. The memory parameters are $r=l=1$ and sample size is $N=10^5$. The quantile bands -- 1st–99th, 5th–95th, and 25th–75th with a line indicating median  -- show results obtained over 30 runs. Across all panels, black bands represent RTE in the non-causal direction, dark gray bands indicate RTE in the causal direction, light yellow bands correspond to ERTE in the non-causal direction, and colored bands (yellow, red, and turquoise) show ERTE in the causal direction, as illustrated in Fig.~\ref{fig_rte_process2}. In each case, RTE estimates (black and gray) exhibit substantial bias for small values of $\alpha$ in both directions. This bias is primarily due to the heavy-tailed distribution of the process $B^{II}$ and is effectively corrected by subtracting RTE computed on the shuffled version of $B^{II}$.}
    \label{figSI_effrte_process2}
\end{figure*}

\begin{figure*}[h] 
    \begin{subcaptionbox}{\justifying \footnotesize {RTE and ERTE in both directions for $k=50$.}}
        {\includegraphics[width=0.48\textwidth]{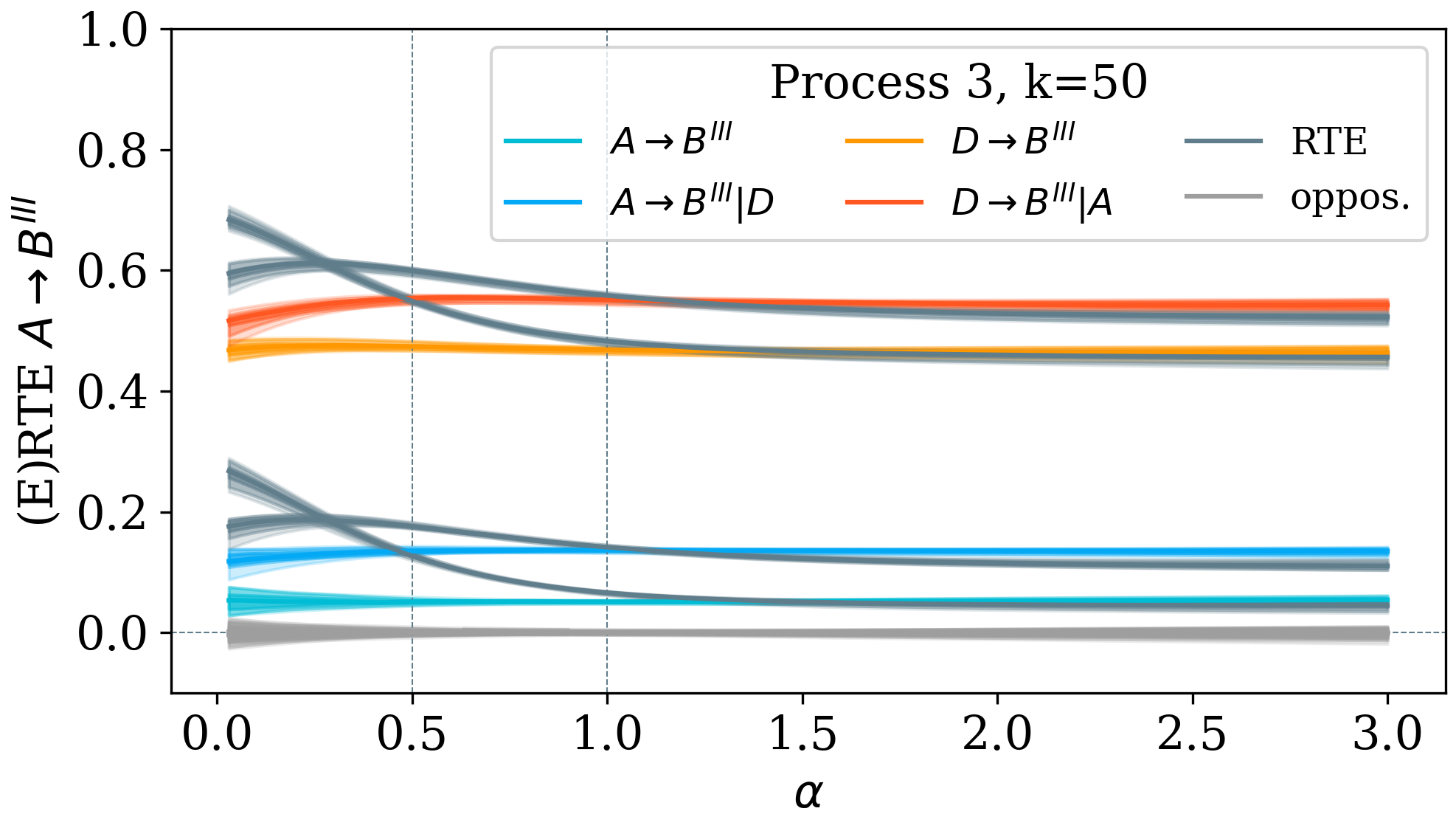}}
    \end{subcaptionbox}
    \hfill
    \begin{subcaptionbox}{\justifying \footnotesize {RTE and ERTE in both directions for $k=3$.}}
        {\includegraphics[width=0.48\textwidth]{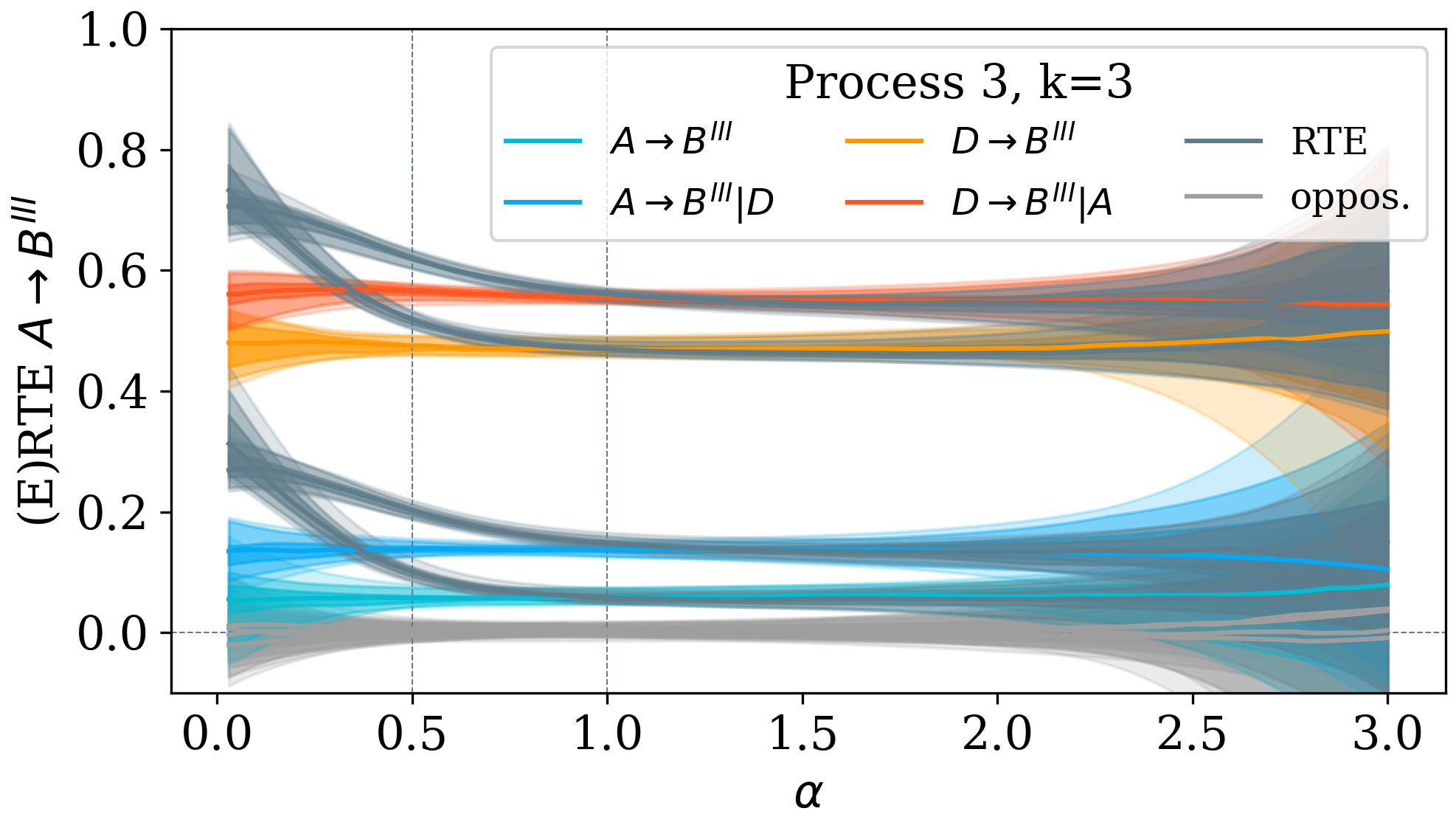}}
    \end{subcaptionbox}
    \caption{\justifying \footnotesize {\em RTE and ERTE of Process 3 in both directions for $k=50$ and $k=3$.} The $x$-axis in both panels shows the Rényi parameter $\alpha$, the $y$-axis shows estimated RTE and ERTE. The memory parameters are $r=l=1$ and sample size is $N=10^5$. The quantile bands -- 1st–99th, 5th–95th, and 25th–75th with a line indicating median  -- show results obtained over 30 runs. In both panels, dark gray bands represent estimated RTE in causal direction, light gray bands (at zero) represent ERTE in non-causal direction, and colored (cyan, light blue, orange and red) represent ERTE in causal direction from Fig.\ref{fig_rte_process3}. Estimated RTEs exhibit noticeable estimation bias for $\alpha < 1$. This highlights the importance of computing effective RTE using shuffled surrogate data to correct for such bias. Notably, even though less noisy, RTE estimates are accurate already from $\alpha\approx0.7$ for $k = 3$, while for $k = 50$, reliable estimates are obtained only starting approximately at 1. This again confirms, that small values of $k$ are better for estimation on tails. }
    \label{figSI_effrte_process3}
\end{figure*}

\begin{figure*}[h] 
    \begin{subcaptionbox}{\justifying \footnotesize{RTE and ERTE in both directions for $k=50$ and $\eta=0.5$. }}
        {\includegraphics[width=0.48\textwidth]{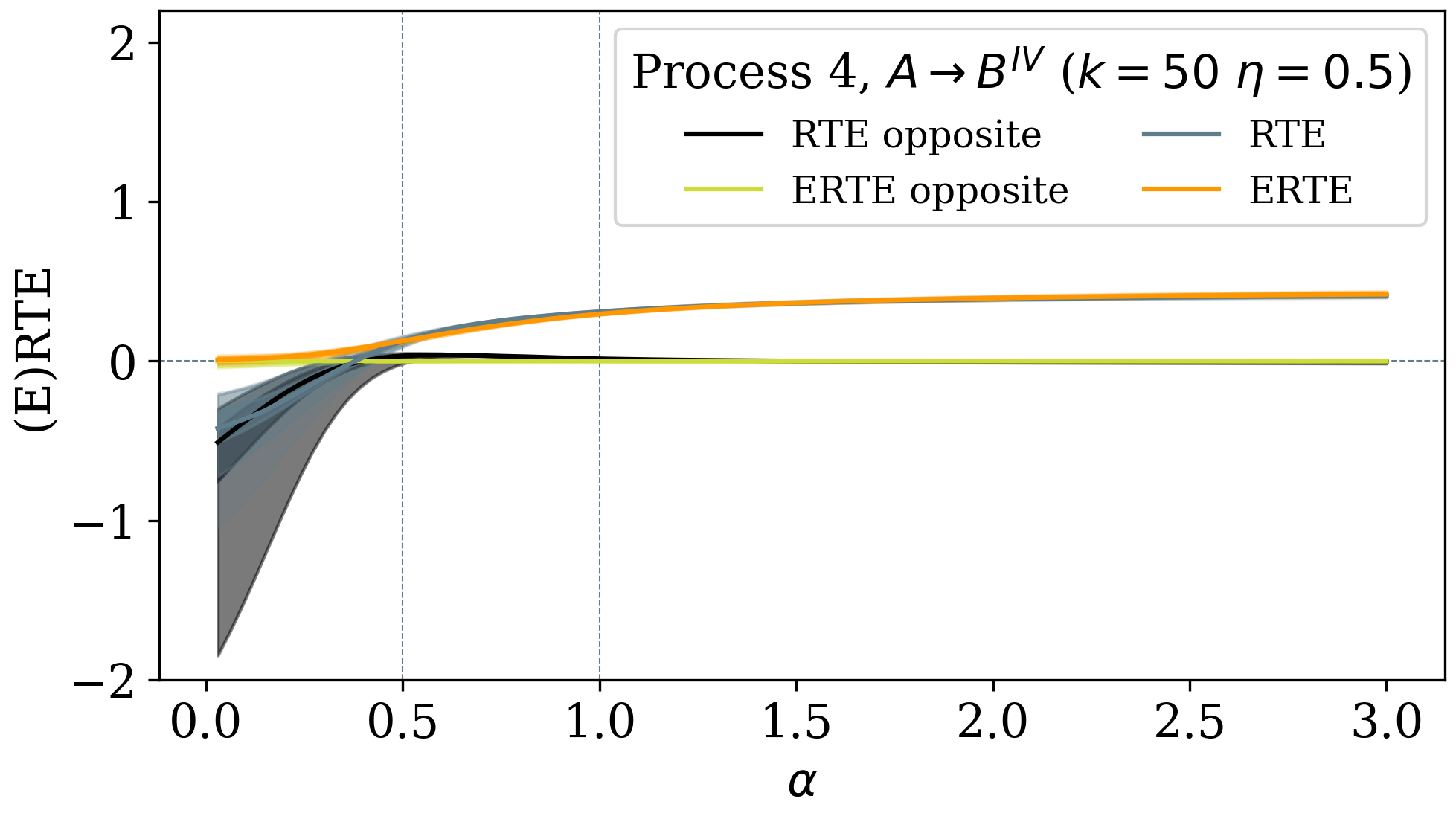}}
    \end{subcaptionbox}
    \hfill
    \begin{subcaptionbox}{\justifying \footnotesize{RTE and ERTE in both directions for $k=3$ and $\eta=0.5$.} 
    }
        {\includegraphics[width=0.48\textwidth]{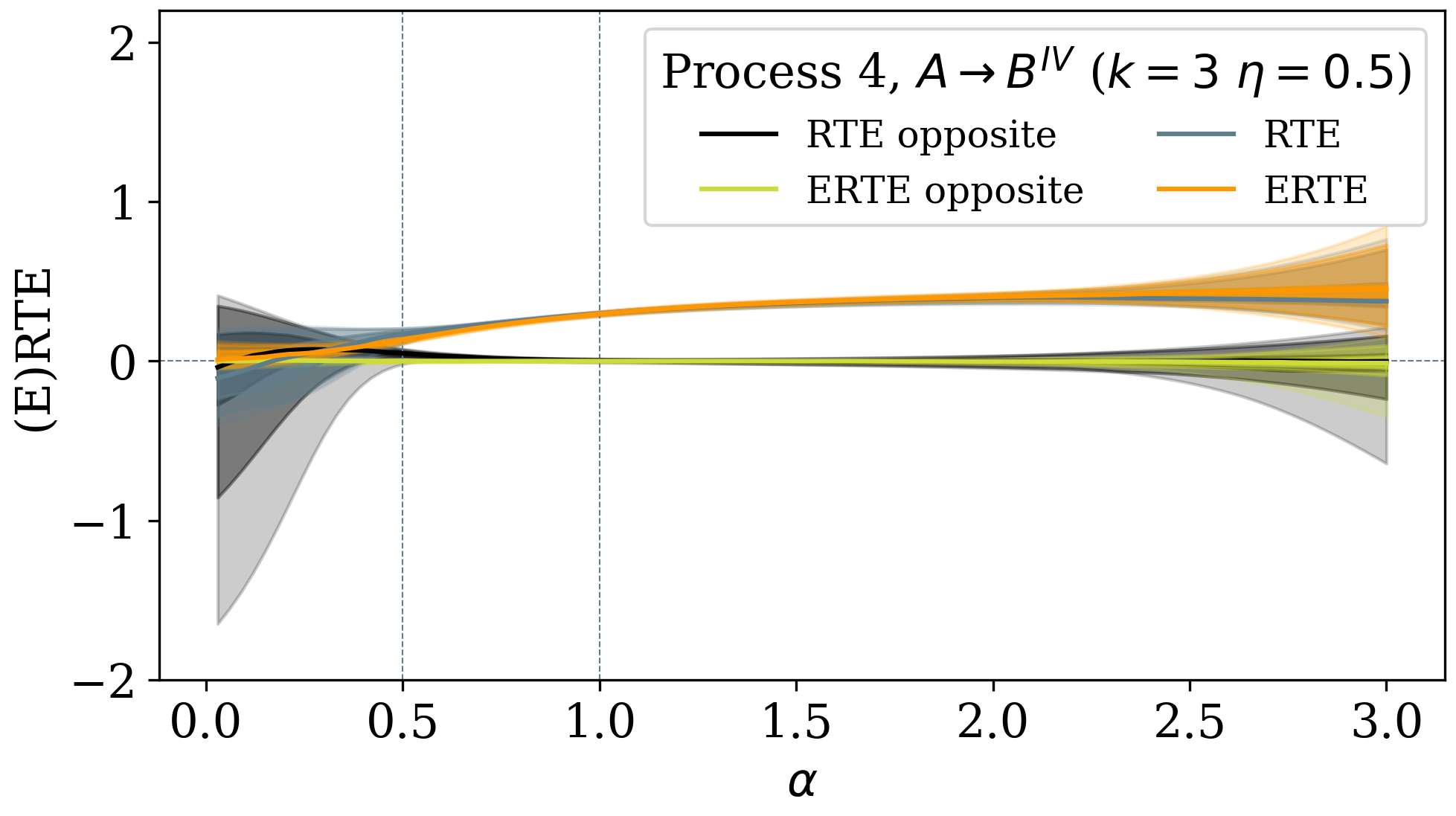}}
    \end{subcaptionbox}

    \begin{subcaptionbox}{\justifying \footnotesize{Conditional RTE and ERTE in both directions for $k=50$ and $\eta=0.5$.}}
        {\includegraphics[width=0.48\textwidth]{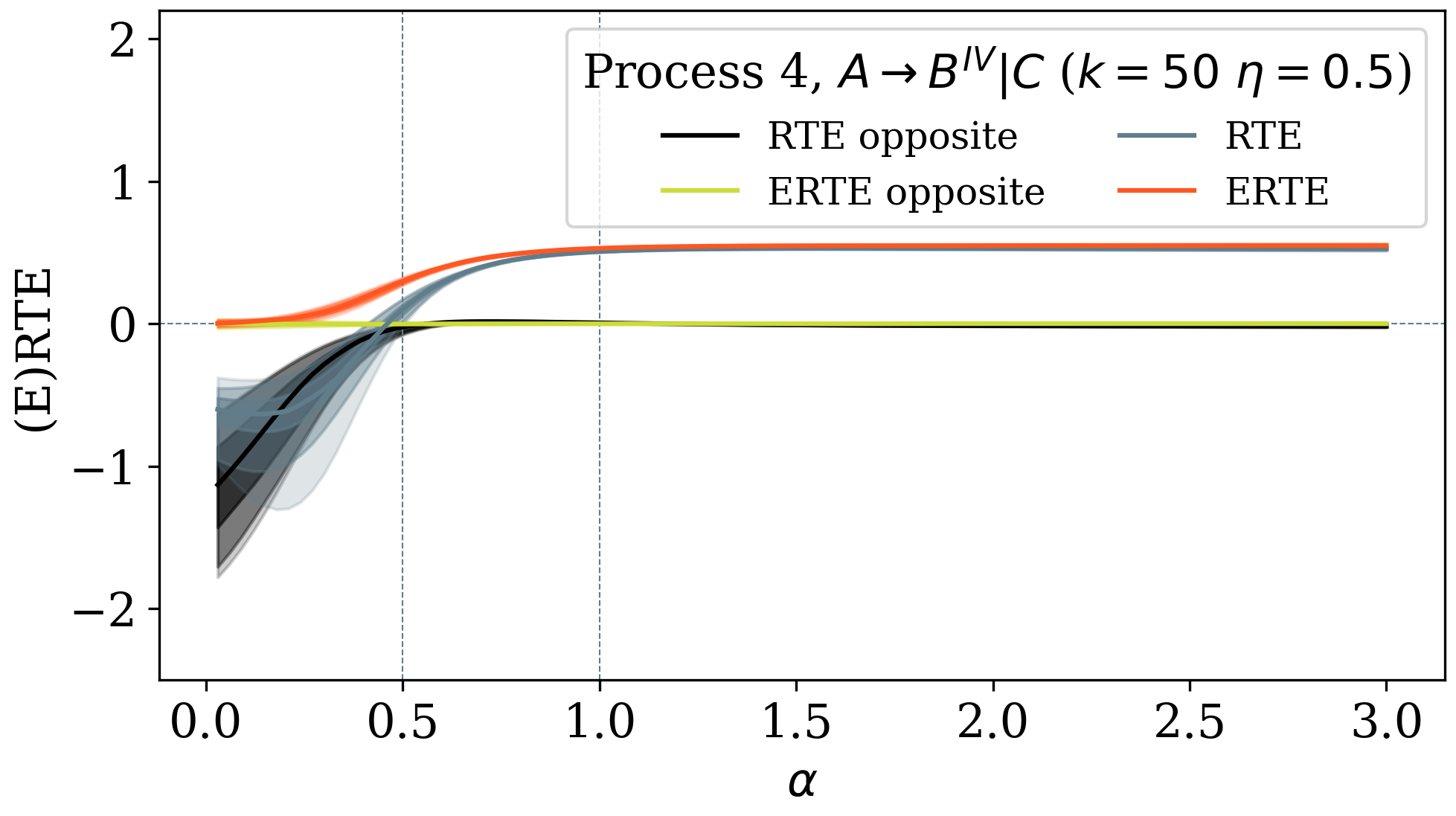}}
    \end{subcaptionbox}
    \hfill
    \begin{subcaptionbox}{\justifying \footnotesize{ Conditional RTE and ERTE in both directions for $k=3$ and $\eta=0.5$.} 
    }
        {\includegraphics[width=0.48\textwidth]{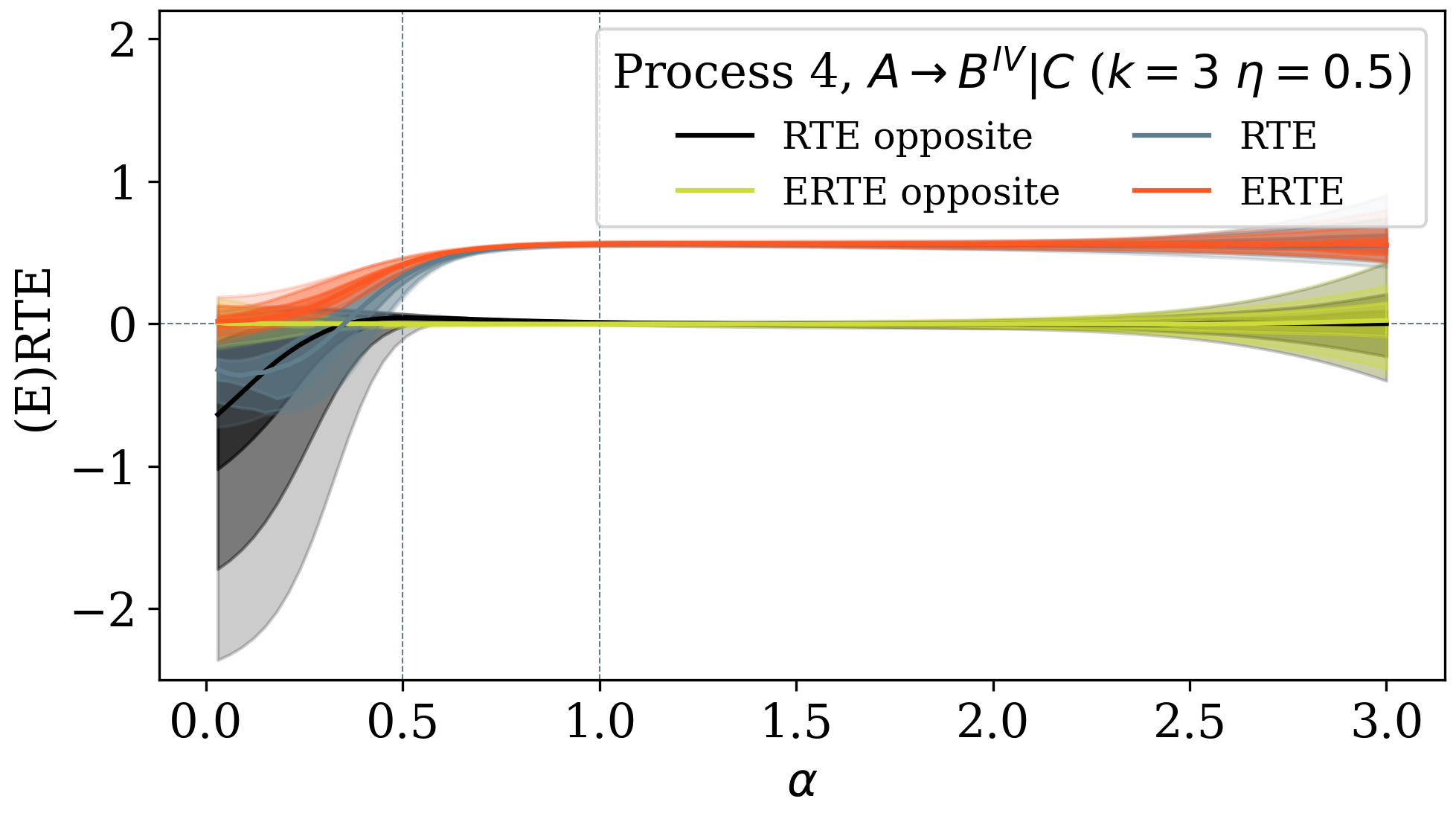}}
    \end{subcaptionbox}

    \begin{subcaptionbox}{\justifying \footnotesize{RTE and ERTE in both directions for $k=50$ and $\kappa=0.2$. }}
        {\includegraphics[width=0.48\textwidth]{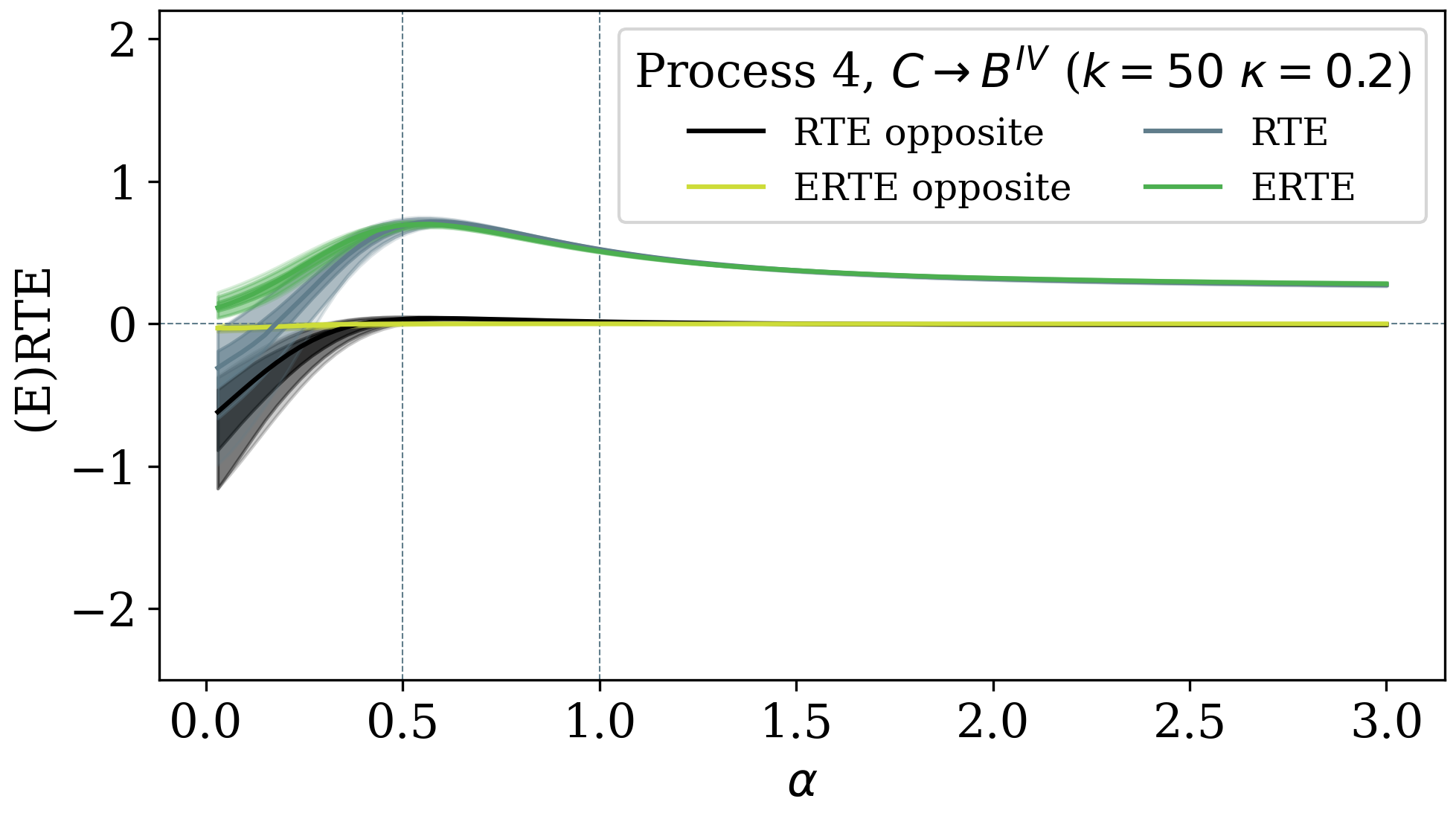}}
    \end{subcaptionbox}
    \hfill
    \begin{subcaptionbox}{\justifying \footnotesize{RTE and ERTE in both directions for $k=3$ and $\kappa=0.2$.} 
    }
        {\includegraphics[width=0.48\textwidth]{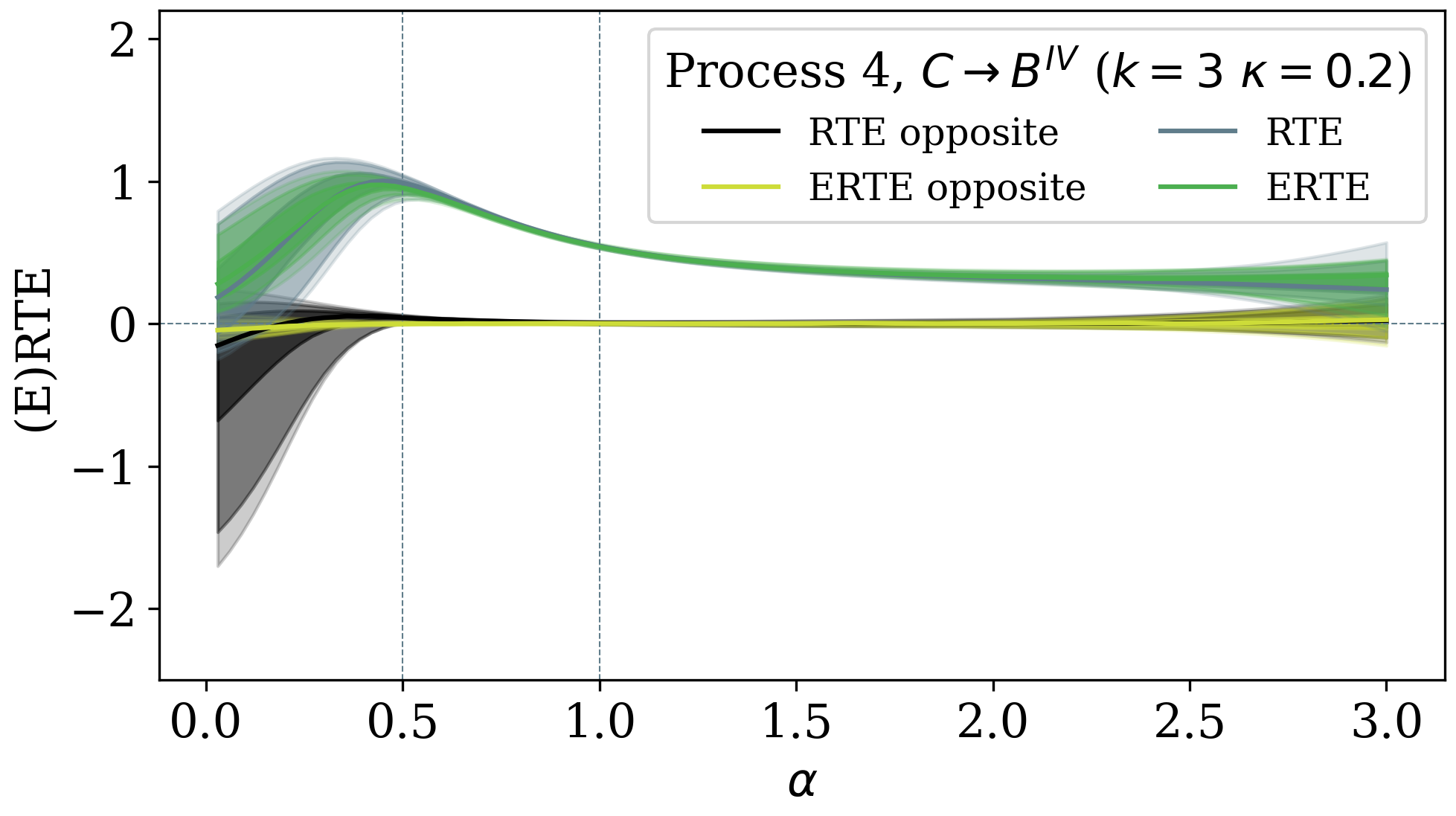}}
    \end{subcaptionbox}

    \begin{subcaptionbox}{\justifying \footnotesize{Conditional RTE and ERTE in both directions for $k=50$ and $\kappa=0.2$. }}
        {\includegraphics[width=0.48\textwidth]{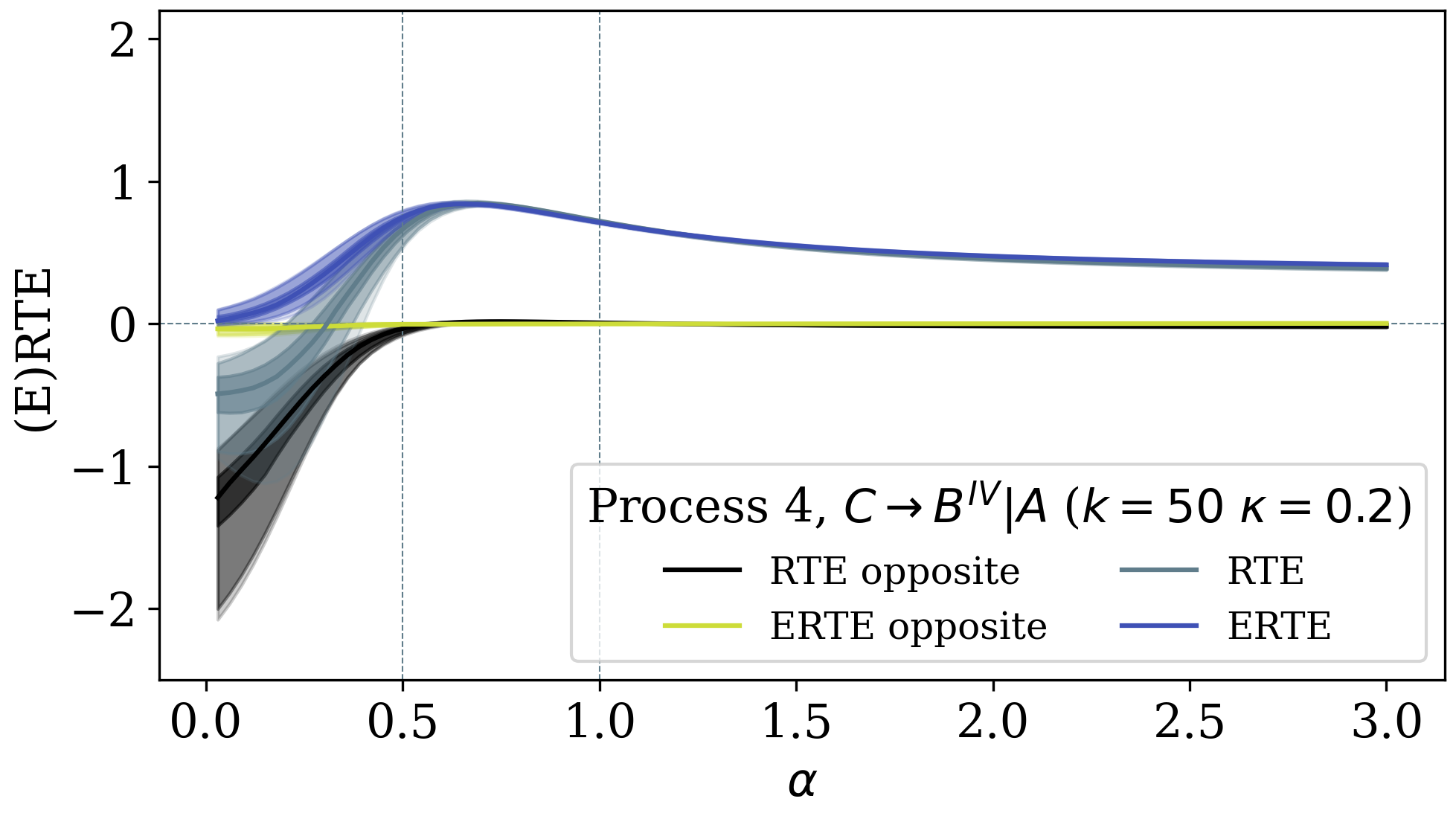}}
    \end{subcaptionbox}
    \hfill
    \begin{subcaptionbox}{\justifying \footnotesize{Conditional RTE and ERTE in both directions for $k=3$ and $\kappa=0.2$.} 
    }
        {\includegraphics[width=0.48\textwidth]{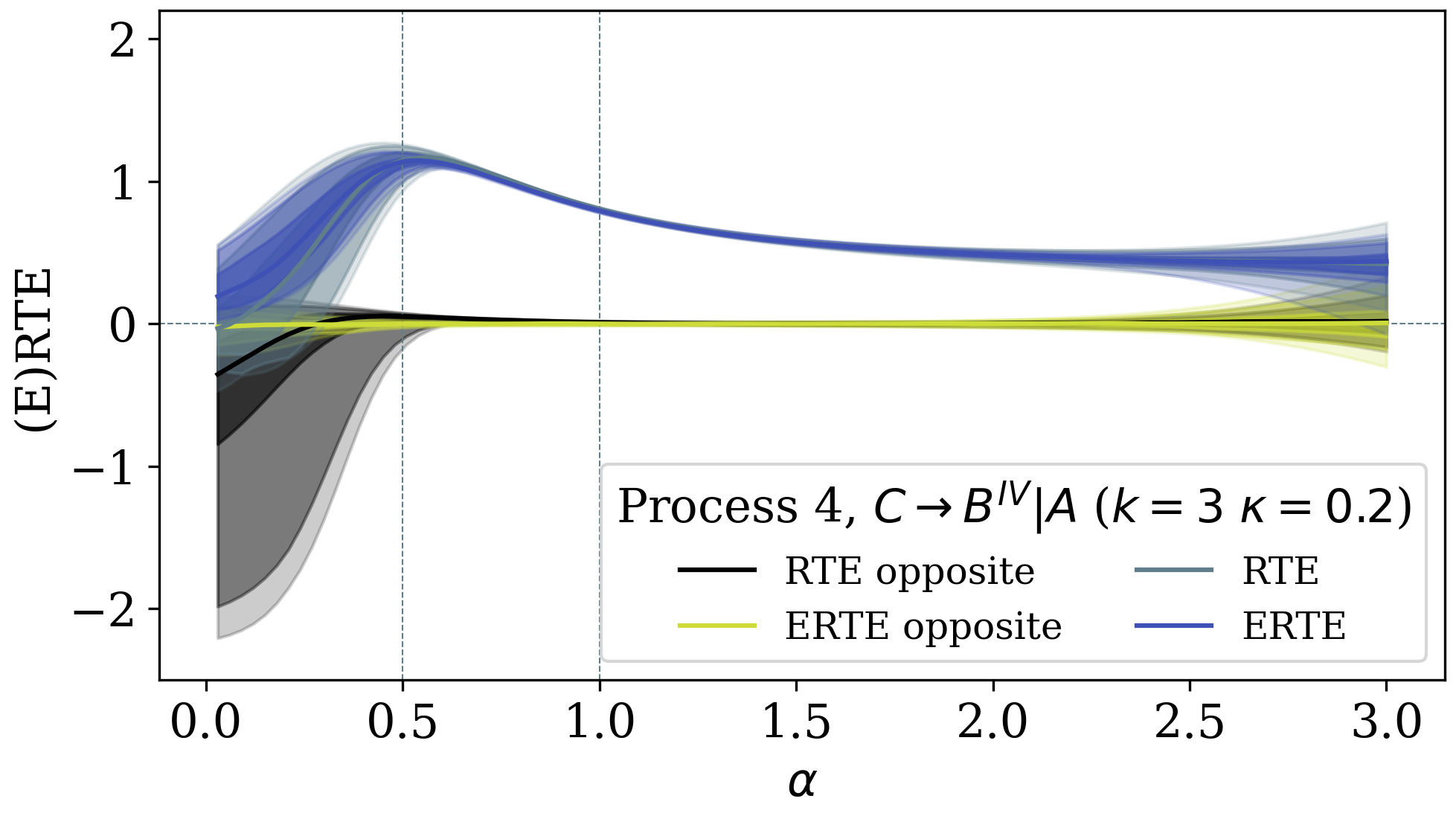}}
    \end{subcaptionbox}

    \caption{\justifying \footnotesize {\em RTE and ERTE of Process 4 in both directions for $k=50$ and $k=3$.} The $x$-axis in all panels shows the Rényi parameter $\alpha$, the $y$-axis shows estimated RTE and ERTE. The memory parameters are $r=l=1$ and sample size is $N=10^5$. The quantile bands -- 1st–99th, 5th–95th, and 25th–75th with a line indicating median  -- show results obtained over 30 runs. Across all panels, black bands represent RTE in the non-causal direction, dark gray bands indicate RTE in the causal direction, light yellow bands correspond to ERTE in the non-causal direction, and colored bands (orange, red, green and dark blue) show ERTE in the causal direction, as illustrated in Fig.~\ref{fig_rte_process4}. In most cases, RTE estimates (black and gray) exhibit substantial bias for small values of $\alpha$ in both directions. This bias is primarily due to the heavy-tailed distribution of the process $B^{IV}$ and is effectively corrected by subtracting RTE computed on the shuffled version of $B^{IV}$. }
    \label{figSI_effrte_process4}
\end{figure*}

\end{widetext}
\clearpage

\bibliography{main}

\providecommand{\noopsort}[1]{}\providecommand{\singleletter}[1]{#1}%
\begin{thebibliography}{45}%
\makeatletter
\providecommand \@ifxundefined [1]{%
 \@ifx{#1\undefined}
}%
\providecommand \@ifnum [1]{%
 \ifnum #1\expandafter \@firstoftwo
 \else \expandafter \@secondoftwo
 \fi
}%
\providecommand \@ifx [1]{%
 \ifx #1\expandafter \@firstoftwo
 \else \expandafter \@secondoftwo
 \fi
}%
\providecommand \natexlab [1]{#1}%
\providecommand \enquote  [1]{``#1''}%
\providecommand \bibnamefont  [1]{#1}%
\providecommand \bibfnamefont [1]{#1}%
\providecommand \citenamefont [1]{#1}%
\providecommand \href@noop [0]{\@secondoftwo}%
\providecommand \href [0]{\begingroup \@sanitize@url \@href}%
\providecommand \@href[1]{\@@startlink{#1}\@@href}%
\providecommand \@@href[1]{\endgroup#1\@@endlink}%
\providecommand \@sanitize@url [0]{\catcode `\\12\catcode `\$12\catcode `\&12\catcode `\#12\catcode `\^12\catcode `\_12\catcode `\%12\relax}%
\providecommand \@@startlink[1]{}%
\providecommand \@@endlink[0]{}%
\providecommand \url  [0]{\begingroup\@sanitize@url \@url }%
\providecommand \@url [1]{\endgroup\@href {#1}{\urlprefix }}%
\providecommand \urlprefix  [0]{URL }%
\providecommand \Eprint [0]{\href }%
\providecommand \doibase [0]{https://doi.org/}%
\providecommand \selectlanguage [0]{\@gobble}%
\providecommand \bibinfo  [0]{\@secondoftwo}%
\providecommand \bibfield  [0]{\@secondoftwo}%
\providecommand \translation [1]{[#1]}%
\providecommand \BibitemOpen [0]{}%
\providecommand \bibitemStop [0]{}%
\providecommand \bibitemNoStop [0]{.\EOS\space}%
\providecommand \EOS [0]{\spacefactor3000\relax}%
\providecommand \BibitemShut  [1]{\csname bibitem#1\endcsname}%
\let\auto@bib@innerbib\@empty
\bibitem [{\citenamefont {Kantz}\ and\ \citenamefont {Schreiber}(2003)}]{kantz2003nonlinear}%
  \BibitemOpen
  \bibfield  {author} {\bibinfo {author} {\bibfnamefont {H.}~\bibnamefont {Kantz}}\ and\ \bibinfo {author} {\bibfnamefont {T.}~\bibnamefont {Schreiber}},\ }\href@noop {} {\emph {\bibinfo {title} {Nonlinear time series analysis}}}\ (\bibinfo  {publisher} {Cambridge university press},\ \bibinfo {year} {2003})\BibitemShut {NoStop}%
\bibitem [{\citenamefont {Lizier}\ and\ \citenamefont {Prokopenko}(2010)}]{lizier2010differentiating}%
  \BibitemOpen
  \bibfield  {author} {\bibinfo {author} {\bibfnamefont {J.~T.}\ \bibnamefont {Lizier}}\ and\ \bibinfo {author} {\bibfnamefont {M.}~\bibnamefont {Prokopenko}},\ }\bibfield  {title} {\bibinfo {title} {Differentiating information transfer and causal effect},\ }\href@noop {} {\bibfield  {journal} {\bibinfo  {journal} {The European Physical Journal B}\ }\textbf {\bibinfo {volume} {73}},\ \bibinfo {pages} {605} (\bibinfo {year} {2010})}\BibitemShut {NoStop}%
\bibitem [{\citenamefont {Granger}(1969)}]{Granger}%
  \BibitemOpen
  \bibfield  {author} {\bibinfo {author} {\bibfnamefont {C.}~\bibnamefont {Granger}},\ }\bibfield  {title} {\bibinfo {title} {Investigating causal relations by econometric models and cross-spectral methods},\ }\href@noop {} {\bibfield  {journal} {\bibinfo  {journal} {Econometrica}\ }\textbf {\bibinfo {volume} {37}},\ \bibinfo {pages} {424} (\bibinfo {year} {1969})}\BibitemShut {NoStop}%
\bibitem [{\citenamefont {Wiener}(1956)}]{Wiener}%
  \BibitemOpen
  \bibfield  {author} {\bibinfo {author} {\bibfnamefont {N.}~\bibnamefont {Wiener}},\ }\href@noop {} {\emph {\bibinfo {title} {Modern Mathematics for Engineers; Beckenbach, E.F., Ed.}}}\ (\bibinfo  {publisher} {McGraw-Hill},\ \bibinfo {address} {New York, NY, USA},\ \bibinfo {year} {1956})\BibitemShut {NoStop}%
\bibitem [{\citenamefont {Quiroga}\ \emph {et~al.}(2000)\citenamefont {Quiroga}, \citenamefont {Arnhold},\ and\ \citenamefont {Grassberger}}]{Quiroga:2000}%
  \BibitemOpen
  \bibfield  {author} {\bibinfo {author} {\bibfnamefont {R.~Q.}\ \bibnamefont {Quiroga}}, \bibinfo {author} {\bibfnamefont {J.}~\bibnamefont {Arnhold}},\ and\ \bibinfo {author} {\bibfnamefont {P.}~\bibnamefont {Grassberger}},\ }\bibfield  {title} {\bibinfo {title} {Learning driver-response relationships from synchronization patterns},\ }\href@noop {} {\bibfield  {journal} {\bibinfo  {journal} {Physical Review E}\ }\textbf {\bibinfo {volume} {61}},\ \bibinfo {pages} {5142} (\bibinfo {year} {2000})}\BibitemShut {NoStop}%
\bibitem [{\citenamefont {Ye}\ \emph {et~al.}(2015)\citenamefont {Ye}, \citenamefont {Deyle}, \citenamefont {Gilarranz},\ and\ \citenamefont {Sugihara}}]{Ye:2015}%
  \BibitemOpen
  \bibfield  {author} {\bibinfo {author} {\bibfnamefont {H.}~\bibnamefont {Ye}}, \bibinfo {author} {\bibfnamefont {E.~R.}\ \bibnamefont {Deyle}}, \bibinfo {author} {\bibfnamefont {L.~J.}\ \bibnamefont {Gilarranz}},\ and\ \bibinfo {author} {\bibfnamefont {G.}~\bibnamefont {Sugihara}},\ }\bibfield  {title} {\bibinfo {title} {Distinguishing time-delayed causal interactions using convergent cross mapping},\ }\href@noop {} {\bibfield  {journal} {\bibinfo  {journal} {Scientific Reports}\ }\textbf {\bibinfo {volume} {5}},\ \bibinfo {pages} {14750} (\bibinfo {year} {2015})}\BibitemShut {NoStop}%
\bibitem [{\citenamefont {Paluš}\ \emph {et~al.}(2018)\citenamefont {Paluš}, \citenamefont {Krakovská}, \citenamefont {Jakubík},\ and\ \citenamefont {Chvosteková}}]{Palus:2018c}%
  \BibitemOpen
  \bibfield  {author} {\bibinfo {author} {\bibfnamefont {M.}~\bibnamefont {Paluš}}, \bibinfo {author} {\bibfnamefont {A.}~\bibnamefont {Krakovská}}, \bibinfo {author} {\bibfnamefont {J.}~\bibnamefont {Jakubík}},\ and\ \bibinfo {author} {\bibfnamefont {M.}~\bibnamefont {Chvosteková}},\ }\bibfield  {title} {\bibinfo {title} {Causality, dynamical systems and the arrow of time},\ }\href@noop {} {\bibfield  {journal} {\bibinfo  {journal} {Chaos}\ }\textbf {\bibinfo {volume} {28}},\ \bibinfo {pages} {075307} (\bibinfo {year} {2018})}\BibitemShut {NoStop}%
\bibitem [{\citenamefont {Kathpalia}\ \emph {et~al.}(2022)\citenamefont {Kathpalia}, \citenamefont {Manshour},\ and\ \citenamefont {Paluš}}]{Kathpalia:2022}%
  \BibitemOpen
  \bibfield  {author} {\bibinfo {author} {\bibfnamefont {A.}~\bibnamefont {Kathpalia}}, \bibinfo {author} {\bibfnamefont {P.}~\bibnamefont {Manshour}},\ and\ \bibinfo {author} {\bibfnamefont {M.}~\bibnamefont {Paluš}},\ }\bibfield  {title} {\bibinfo {title} {Compression complexity with ordinal patterns for robust causal inference in irregularly sampled time series},\ }\href@noop {} {\bibfield  {journal} {\bibinfo  {journal} {Scientific Reports}\ }\textbf {\bibinfo {volume} {12}},\ \bibinfo {pages} {14170} (\bibinfo {year} {2022})}\BibitemShut {NoStop}%
\bibitem [{\citenamefont {Huang}\ \emph {et~al.}(2020)\citenamefont {Huang}, \citenamefont {Fu},\ and\ \citenamefont {Franzke}}]{Huang:2020}%
  \BibitemOpen
  \bibfield  {author} {\bibinfo {author} {\bibfnamefont {Y.}~\bibnamefont {Huang}}, \bibinfo {author} {\bibfnamefont {Z.}~\bibnamefont {Fu}},\ and\ \bibinfo {author} {\bibfnamefont {C.~L.~E.}\ \bibnamefont {Franzke}},\ }\bibfield  {title} {\bibinfo {title} {Detecting causality from time series in a machine learning framework},\ }\href@noop {} {\bibfield  {journal} {\bibinfo  {journal} {Chaos}\ }\textbf {\bibinfo {volume} {30}},\ \bibinfo {pages} {063116} (\bibinfo {year} {2020})}\BibitemShut {NoStop}%
\bibitem [{\citenamefont {Ay}\ and\ \citenamefont {Polani}(2008)}]{ay2008information}%
  \BibitemOpen
  \bibfield  {author} {\bibinfo {author} {\bibfnamefont {N.}~\bibnamefont {Ay}}\ and\ \bibinfo {author} {\bibfnamefont {D.}~\bibnamefont {Polani}},\ }\bibfield  {title} {\bibinfo {title} {Information flows in causal networks},\ }\href@noop {} {\bibfield  {journal} {\bibinfo  {journal} {Advances in complex systems}\ }\textbf {\bibinfo {volume} {11}},\ \bibinfo {pages} {17} (\bibinfo {year} {2008})}\BibitemShut {NoStop}%
\bibitem [{\citenamefont {Jizba}\ \emph {et~al.}(2012)\citenamefont {Jizba}, \citenamefont {Kleinert},\ and\ \citenamefont {Shefaat}}]{jizba2012renyi}%
  \BibitemOpen
  \bibfield  {author} {\bibinfo {author} {\bibfnamefont {P.}~\bibnamefont {Jizba}}, \bibinfo {author} {\bibfnamefont {H.}~\bibnamefont {Kleinert}},\ and\ \bibinfo {author} {\bibfnamefont {M.}~\bibnamefont {Shefaat}},\ }\bibfield  {title} {\bibinfo {title} {$\mbox{R}$\'{e}nyi’s information transfer between financial time series},\ }\href@noop {} {\bibfield  {journal} {\bibinfo  {journal} {Physica A: Statistical Mechanics and its Applications}\ }\textbf {\bibinfo {volume} {391}},\ \bibinfo {pages} {2971} (\bibinfo {year} {2012})}\BibitemShut {NoStop}%
\bibitem [{\citenamefont {Jizba}\ \emph {et~al.}(2022)\citenamefont {Jizba}, \citenamefont {Lavička},\ and\ \citenamefont {Tabachová}}]{JLT:22}%
  \BibitemOpen
  \bibfield  {author} {\bibinfo {author} {\bibfnamefont {P.}~\bibnamefont {Jizba}}, \bibinfo {author} {\bibfnamefont {H.}~\bibnamefont {Lavička}},\ and\ \bibinfo {author} {\bibfnamefont {Z.}~\bibnamefont {Tabachová}},\ }\bibfield  {title} {\bibinfo {title} {Causal inference in time series in terms of $\mbox{R}$\'{e}nyi transfer entropy},\ }\href@noop {} {\bibfield  {journal} {\bibinfo  {journal} {Entropy}\ }\textbf {\bibinfo {volume} {24}} (\bibinfo {year} {2022})}\BibitemShut {NoStop}%
\bibitem [{\citenamefont {Hlav{\'a}{\v{c}}kov{\'a}-Schindler}\ \emph {et~al.}(2007)\citenamefont {Hlav{\'a}{\v{c}}kov{\'a}-Schindler}, \citenamefont {Palu{\v{s}}}, \citenamefont {Vejmelka},\ and\ \citenamefont {Bhattacharya}}]{hlavavckova2007causality}%
  \BibitemOpen
  \bibfield  {author} {\bibinfo {author} {\bibfnamefont {K.}~\bibnamefont {Hlav{\'a}{\v{c}}kov{\'a}-Schindler}}, \bibinfo {author} {\bibfnamefont {M.}~\bibnamefont {Palu{\v{s}}}}, \bibinfo {author} {\bibfnamefont {M.}~\bibnamefont {Vejmelka}},\ and\ \bibinfo {author} {\bibfnamefont {J.}~\bibnamefont {Bhattacharya}},\ }\bibfield  {title} {\bibinfo {title} {Causality detection based on information-theoretic approaches in time series analysis},\ }\href@noop {} {\bibfield  {journal} {\bibinfo  {journal} {Physics Reports}\ }\textbf {\bibinfo {volume} {441}},\ \bibinfo {pages} {1} (\bibinfo {year} {2007})}\BibitemShut {NoStop}%
\bibitem [{\citenamefont {Schreiber}(2000)}]{schreiber2000measuring}%
  \BibitemOpen
  \bibfield  {author} {\bibinfo {author} {\bibfnamefont {T.}~\bibnamefont {Schreiber}},\ }\bibfield  {title} {\bibinfo {title} {Measuring information transfer},\ }\href@noop {} {\bibfield  {journal} {\bibinfo  {journal} {Physical Review letters}\ }\textbf {\bibinfo {volume} {85}},\ \bibinfo {pages} {461} (\bibinfo {year} {2000})}\BibitemShut {NoStop}%
\bibitem [{\citenamefont {Marschinski}\ and\ \citenamefont {Kantz}(2002)}]{marschinski2002analysing}%
  \BibitemOpen
  \bibfield  {author} {\bibinfo {author} {\bibfnamefont {R.}~\bibnamefont {Marschinski}}\ and\ \bibinfo {author} {\bibfnamefont {H.}~\bibnamefont {Kantz}},\ }\bibfield  {title} {\bibinfo {title} {Analysing the information flow between financial time series: An improved estimator for transfer entropy},\ }\href@noop {} {\bibfield  {journal} {\bibinfo  {journal} {The European Physical Journal B-Condensed Matter and Complex Systems}\ }\textbf {\bibinfo {volume} {30}},\ \bibinfo {pages} {275} (\bibinfo {year} {2002})}\BibitemShut {NoStop}%
\bibitem [{\citenamefont {Palu{\v{s}}}\ \emph {et~al.}(2001)\citenamefont {Palu{\v{s}}}, \citenamefont {Kom{\'a}rek}, \citenamefont {Hrn{\v{c}}{\'\i}{\v{r}}},\ and\ \citenamefont {{\v{S}}t{\v{e}}rbov{\'a}}}]{paluvs2001synchronization}%
  \BibitemOpen
  \bibfield  {author} {\bibinfo {author} {\bibfnamefont {M.}~\bibnamefont {Palu{\v{s}}}}, \bibinfo {author} {\bibfnamefont {V.}~\bibnamefont {Kom{\'a}rek}}, \bibinfo {author} {\bibfnamefont {Z.}~\bibnamefont {Hrn{\v{c}}{\'\i}{\v{r}}}},\ and\ \bibinfo {author} {\bibfnamefont {K.}~\bibnamefont {{\v{S}}t{\v{e}}rbov{\'a}}},\ }\bibfield  {title} {\bibinfo {title} {Synchronization as adjustment of information rates: Detection from bivariate time series},\ }\href@noop {} {\bibfield  {journal} {\bibinfo  {journal} {Physical Review E}\ }\textbf {\bibinfo {volume} {63}},\ \bibinfo {pages} {046211} (\bibinfo {year} {2001})}\BibitemShut {NoStop}%
\bibitem [{\citenamefont {Gencaga}(2018)}]{Gencaga:18}%
  \BibitemOpen
  \bibfield  {author} {\bibinfo {author} {\bibfnamefont {D.}~\bibnamefont {Gencaga}},\ }\bibfield  {title} {\bibinfo {title} {Transfer entropy},\ }\href@noop {} {\bibfield  {journal} {\bibinfo  {journal} {Entropy}\ }\textbf {\bibinfo {volume} {20}},\ \bibinfo {pages} {288} (\bibinfo {year} {2018})}\BibitemShut {NoStop}%
\bibitem [{\citenamefont {Bossomaier}\ \emph {et~al.}(2016)\citenamefont {Bossomaier}, \citenamefont {Barnett}, \citenamefont {Harré},\ and\ \citenamefont {Lizier}}]{Bossomaier}%
  \BibitemOpen
  \bibfield  {author} {\bibinfo {author} {\bibfnamefont {T.}~\bibnamefont {Bossomaier}}, \bibinfo {author} {\bibfnamefont {L.}~\bibnamefont {Barnett}}, \bibinfo {author} {\bibfnamefont {M.}~\bibnamefont {Harré}},\ and\ \bibinfo {author} {\bibfnamefont {J.~T.}\ \bibnamefont {Lizier}},\ }\href@noop {} {\emph {\bibinfo {title} {An Introduction to Transfer Entropy: Information Flow in Complex Systems}}}\ (\bibinfo  {publisher} {Springer},\ \bibinfo {address} {Cham, Switzerland},\ \bibinfo {year} {2016})\BibitemShut {NoStop}%
\bibitem [{\citenamefont {Palu{\v{s}}}\ \emph {et~al.}(2024)\citenamefont {Palu{\v{s}}}, \citenamefont {Chvostekov{\'a}},\ and\ \citenamefont {Manshour}}]{paluvs2024causes}%
  \BibitemOpen
  \bibfield  {author} {\bibinfo {author} {\bibfnamefont {M.}~\bibnamefont {Palu{\v{s}}}}, \bibinfo {author} {\bibfnamefont {M.}~\bibnamefont {Chvostekov{\'a}}},\ and\ \bibinfo {author} {\bibfnamefont {P.}~\bibnamefont {Manshour}},\ }\bibfield  {title} {\bibinfo {title} {Causes of extreme events revealed by $\mbox{R}$\'{e}nyi information transfer},\ }\href@noop {} {\bibfield  {journal} {\bibinfo  {journal} {Science Advances}\ }\textbf {\bibinfo {volume} {10}},\ \bibinfo {pages} {eadn1721} (\bibinfo {year} {2024})}\BibitemShut {NoStop}%
\bibitem [{\citenamefont {Jajcay}\ \emph {et~al.}(2018)\citenamefont {Jajcay}, \citenamefont {Kravtsov}, \citenamefont {Sugihara},\ and\ \citenamefont {et~al.}}]{Jajcay:2018}%
  \BibitemOpen
  \bibfield  {author} {\bibinfo {author} {\bibfnamefont {N.}~\bibnamefont {Jajcay}}, \bibinfo {author} {\bibfnamefont {S.}~\bibnamefont {Kravtsov}}, \bibinfo {author} {\bibfnamefont {G.}~\bibnamefont {Sugihara}},\ and\ \bibinfo {author} {\bibnamefont {et~al.}},\ }\bibfield  {title} {\bibinfo {title} {Synchronization and causality across time scales in el niño southern oscillation},\ }\href@noop {} {\bibfield  {journal} {\bibinfo  {journal} {npj Climate and Atmospheric Science}\ }\textbf {\bibinfo {volume} {1}},\ \bibinfo {pages} {33} (\bibinfo {year} {2018})}\BibitemShut {NoStop}%
\bibitem [{\citenamefont {Pereda}\ \emph {et~al.}(2005)\citenamefont {Pereda}, \citenamefont {Quiroga},\ and\ \citenamefont {Bhattacharya}}]{PEREDA20051}%
  \BibitemOpen
  \bibfield  {author} {\bibinfo {author} {\bibfnamefont {E.}~\bibnamefont {Pereda}}, \bibinfo {author} {\bibfnamefont {R.~Q.}\ \bibnamefont {Quiroga}},\ and\ \bibinfo {author} {\bibfnamefont {J.}~\bibnamefont {Bhattacharya}},\ }\bibfield  {title} {\bibinfo {title} {Nonlinear multivariate analysis of neurophysiological signals},\ }\href@noop {} {\bibfield  {journal} {\bibinfo  {journal} {Progress in Neurobiology}\ }\textbf {\bibinfo {volume} {77}},\ \bibinfo {pages} {1} (\bibinfo {year} {2005})}\BibitemShut {NoStop}%
\bibitem [{\citenamefont {Vicente}\ \emph {et~al.}(2011)\citenamefont {Vicente}, \citenamefont {Wibral}, \citenamefont {Lindner},\ and\ \citenamefont {et~al.}}]{Vicente:2011}%
  \BibitemOpen
  \bibfield  {author} {\bibinfo {author} {\bibfnamefont {R.}~\bibnamefont {Vicente}}, \bibinfo {author} {\bibfnamefont {M.}~\bibnamefont {Wibral}}, \bibinfo {author} {\bibfnamefont {M.}~\bibnamefont {Lindner}},\ and\ \bibinfo {author} {\bibnamefont {et~al.}},\ }\bibfield  {title} {\bibinfo {title} {Transfer entropy --- a model-free measure of effective connectivity for the neurosciences},\ }\href@noop {} {\bibfield  {journal} {\bibinfo  {journal} {Journal of Computational Neuroscience}\ }\textbf {\bibinfo {volume} {30}},\ \bibinfo {pages} {45–67} (\bibinfo {year} {2011})}\BibitemShut {NoStop}%
\bibitem [{\citenamefont {Permuter}\ \emph {et~al.}(2011)\citenamefont {Permuter}, \citenamefont {Kim},\ and\ \citenamefont {Weissman}}]{Permuter:2011}%
  \BibitemOpen
  \bibfield  {author} {\bibinfo {author} {\bibfnamefont {H.~H.}\ \bibnamefont {Permuter}}, \bibinfo {author} {\bibfnamefont {Y.-H.}\ \bibnamefont {Kim}},\ and\ \bibinfo {author} {\bibfnamefont {T.}~\bibnamefont {Weissman}},\ }\bibfield  {title} {\bibinfo {title} {Interpretations of directed information in portfolio theory, data compression, and hypothesis testing},\ }\href@noop {} {\bibfield  {journal} {\bibinfo  {journal} {IEEE Transactions on Information Theory}\ }\textbf {\bibinfo {volume} {57}},\ \bibinfo {pages} {3248} (\bibinfo {year} {2011})}\BibitemShut {NoStop}%
\bibitem [{\citenamefont {R{\'{e}}nyi}(1976)}]{Renyi:1976a}%
  \BibitemOpen
  \bibfield  {author} {\bibinfo {author} {\bibfnamefont {A.}~\bibnamefont {R{\'{e}}nyi}},\ }\href@noop {} {\emph {\bibinfo {title} {Selected Papers of $\mbox{A}$lfr\'{e}d $\mbox{R}$\'{e}nyi, 2nd Vol.}}}\ (\bibinfo  {publisher} {Akademia Kiado},\ \bibinfo {address} {Budapest},\ \bibinfo {year} {1976})\BibitemShut {NoStop}%
\bibitem [{\citenamefont {R{\'e}nyi}(1961)}]{renyi1961measures}%
  \BibitemOpen
  \bibfield  {author} {\bibinfo {author} {\bibfnamefont {A.}~\bibnamefont {R{\'e}nyi}},\ }\bibfield  {title} {\bibinfo {title} {On measures of entropy and information},\ }in\ \href@noop {} {\emph {\bibinfo {booktitle} {Proceedings of the fourth Berkeley symposium on mathematical statistics and probability, volume 1: contributions to the theory of statistics}}},\ Vol.~\bibinfo {volume} {4}\ (\bibinfo {organization} {University of California Press},\ \bibinfo {year} {1961})\ pp.\ \bibinfo {pages} {547--562}\BibitemShut {NoStop}%
\bibitem [{\citenamefont {Jizba}\ and\ \citenamefont {Arimitsu}(2004{\natexlab{a}})}]{JA}%
  \BibitemOpen
  \bibfield  {author} {\bibinfo {author} {\bibfnamefont {P.}~\bibnamefont {Jizba}}\ and\ \bibinfo {author} {\bibfnamefont {T.}~\bibnamefont {Arimitsu}},\ }\bibfield  {title} {\bibinfo {title} {The world according to $\mbox{R}$\'{e}nyi: thermodynamics of multifractal systems},\ }\href@noop {} {\bibfield  {journal} {\bibinfo  {journal} {Annals of Physics}\ }\textbf {\bibinfo {volume} {312}},\ \bibinfo {pages} {17} (\bibinfo {year} {2004}{\natexlab{a}})}\BibitemShut {NoStop}%
\bibitem [{\citenamefont {Zhang}\ \emph {et~al.}(2023)\citenamefont {Zhang}, \citenamefont {Cao}, \citenamefont {Wu}, \citenamefont {Huang}, \citenamefont {Ma},\ and\ \citenamefont {Zhou}}]{ZHANG2023}%
  \BibitemOpen
  \bibfield  {author} {\bibinfo {author} {\bibfnamefont {J.}~\bibnamefont {Zhang}}, \bibinfo {author} {\bibfnamefont {J.}~\bibnamefont {Cao}}, \bibinfo {author} {\bibfnamefont {T.}~\bibnamefont {Wu}}, \bibinfo {author} {\bibfnamefont {W.}~\bibnamefont {Huang}}, \bibinfo {author} {\bibfnamefont {T.}~\bibnamefont {Ma}},\ and\ \bibinfo {author} {\bibfnamefont {X.}~\bibnamefont {Zhou}},\ }\bibfield  {title} {\bibinfo {title} {A novel adaptive multi-scale rényi transfer entropy based on kernel density estimation},\ }\href@noop {} {\bibfield  {journal} {\bibinfo  {journal} {Chaos, Solitons $\&$ Fractals}\ }\textbf {\bibinfo {volume} {175}},\ \bibinfo {pages} {113972} (\bibinfo {year} {2023})}\BibitemShut {NoStop}%
\bibitem [{\citenamefont {Behrendt}\ \emph {et~al.}(2019)\citenamefont {Behrendt}, \citenamefont {Dimpfl}, \citenamefont {Peter},\ and\ \citenamefont {Zimmermann}}]{BEHRENDT2019}%
  \BibitemOpen
  \bibfield  {author} {\bibinfo {author} {\bibfnamefont {S.}~\bibnamefont {Behrendt}}, \bibinfo {author} {\bibfnamefont {T.}~\bibnamefont {Dimpfl}}, \bibinfo {author} {\bibfnamefont {F.~J.}\ \bibnamefont {Peter}},\ and\ \bibinfo {author} {\bibfnamefont {D.~J.}\ \bibnamefont {Zimmermann}},\ }\bibfield  {title} {\bibinfo {title} {Rtransferentropy — quantifying information flow between different time series using effective transfer entropy},\ }\href@noop {} {\bibfield  {journal} {\bibinfo  {journal} {SoftwareX}\ }\textbf {\bibinfo {volume} {10}},\ \bibinfo {pages} {100265} (\bibinfo {year} {2019})}\BibitemShut {NoStop}%
\bibitem [{\citenamefont {Korbel}\ \emph {et~al.}(2019)\citenamefont {Korbel}, \citenamefont {Jiang},\ and\ \citenamefont {Zheng}}]{Korbel:2019}%
  \BibitemOpen
  \bibfield  {author} {\bibinfo {author} {\bibfnamefont {J.}~\bibnamefont {Korbel}}, \bibinfo {author} {\bibfnamefont {X.}~\bibnamefont {Jiang}},\ and\ \bibinfo {author} {\bibfnamefont {B.}~\bibnamefont {Zheng}},\ }\bibfield  {title} {\bibinfo {title} {Transfer entropy between communities in complex financial networks},\ }\href@noop {} {\bibfield  {journal} {\bibinfo  {journal} {Entropy}\ }\textbf {\bibinfo {volume} {21}} (\bibinfo {year} {2019})}\BibitemShut {NoStop}%
\bibitem [{\citenamefont {Leonenko}\ \emph {et~al.}(2008)\citenamefont {Leonenko}, \citenamefont {Pronzato},\ and\ \citenamefont {Savani}}]{leonenko2008class}%
  \BibitemOpen
  \bibfield  {author} {\bibinfo {author} {\bibfnamefont {N.}~\bibnamefont {Leonenko}}, \bibinfo {author} {\bibfnamefont {L.}~\bibnamefont {Pronzato}},\ and\ \bibinfo {author} {\bibfnamefont {V.}~\bibnamefont {Savani}},\ }\bibfield  {title} {\bibinfo {title} {A class of r\'enyi information estimators for multidimensional densities},\ }\href@noop {} {\bibfield  {journal} {\bibinfo  {journal} {The Annals of Statistics}\ }\textbf {\bibinfo {volume} {36}},\ \bibinfo {pages} {2153} (\bibinfo {year} {2008})}\BibitemShut {NoStop}%
\bibitem [{\citenamefont {Palu{\v{s}}}\ and\ \citenamefont {Stefanovska}(2003)}]{palus2003}%
  \BibitemOpen
  \bibfield  {author} {\bibinfo {author} {\bibfnamefont {M.}~\bibnamefont {Palu{\v{s}}}}\ and\ \bibinfo {author} {\bibfnamefont {A.}~\bibnamefont {Stefanovska}},\ }\bibfield  {title} {\bibinfo {title} {Direction of coupling from phases of interacting oscillators: An information-theoretic approach},\ }\href@noop {} {\bibfield  {journal} {\bibinfo  {journal} {Physical Review E}\ }\textbf {\bibinfo {volume} {67}},\ \bibinfo {pages} {055201(R)} (\bibinfo {year} {2003})}\BibitemShut {NoStop}%
\bibitem [{\citenamefont {Sun}\ and\ \citenamefont {Bollt}(2014)}]{sun2014causation}%
  \BibitemOpen
  \bibfield  {author} {\bibinfo {author} {\bibfnamefont {J.}~\bibnamefont {Sun}}\ and\ \bibinfo {author} {\bibfnamefont {E.~M.}\ \bibnamefont {Bollt}},\ }\bibfield  {title} {\bibinfo {title} {Causation entropy identifies indirect influences, dominance of neighbors and anticipatory couplings},\ }\href@noop {} {\bibfield  {journal} {\bibinfo  {journal} {Physica D: Nonlinear Phenomena}\ }\textbf {\bibinfo {volume} {267}},\ \bibinfo {pages} {49} (\bibinfo {year} {2014})}\BibitemShut {NoStop}%
\bibitem [{Note1()}]{Note1}%
  \BibitemOpen
  \bibinfo {note} {While the estimator uses Euclidean distance by default, alternative distance measures such as the Mahalanobis distance can be applied when appropriate, particularly for spherically distributed data.}\BibitemShut {Stop}%
\bibitem [{\citenamefont {Cover}(1999)}]{cover1999elements}%
  \BibitemOpen
  \bibfield  {author} {\bibinfo {author} {\bibfnamefont {T.~M.}\ \bibnamefont {Cover}},\ }\href@noop {} {\emph {\bibinfo {title} {Elements of information theory}}}\ (\bibinfo  {publisher} {John Wiley \& Sons},\ \bibinfo {year} {1999})\BibitemShut {NoStop}%
\bibitem [{\citenamefont {Van~Trees}\ and\ \citenamefont {Bell}(2013)}]{van2013detection}%
  \BibitemOpen
  \bibfield  {author} {\bibinfo {author} {\bibfnamefont {H.~L.}\ \bibnamefont {Van~Trees}}\ and\ \bibinfo {author} {\bibfnamefont {K.~L.}\ \bibnamefont {Bell}},\ }\href@noop {} {\emph {\bibinfo {title} {Detection estimation and modulation theory, part I: detection, estimation, and filtering theory}}}\ (\bibinfo  {publisher} {John Wiley \& Sons},\ \bibinfo {year} {2013})\BibitemShut {NoStop}%
\bibitem [{\citenamefont {Jizba}\ and\ \citenamefont {Arimitsu}(2004{\natexlab{b}})}]{jizba2004world}%
  \BibitemOpen
  \bibfield  {author} {\bibinfo {author} {\bibfnamefont {P.}~\bibnamefont {Jizba}}\ and\ \bibinfo {author} {\bibfnamefont {T.}~\bibnamefont {Arimitsu}},\ }\bibfield  {title} {\bibinfo {title} {The world according to r{\'e}nyi: thermodynamics of multifractal systems},\ }\href@noop {} {\bibfield  {journal} {\bibinfo  {journal} {Annals of Physics}\ }\textbf {\bibinfo {volume} {312}},\ \bibinfo {pages} {17} (\bibinfo {year} {2004}{\natexlab{b}})}\BibitemShut {NoStop}%
\bibitem [{\citenamefont {Fisher}\ and\ \citenamefont {Yates}(1939)}]{fischer-yates:1938}%
  \BibitemOpen
  \bibfield  {author} {\bibinfo {author} {\bibfnamefont {R.~A.}\ \bibnamefont {Fisher}}\ and\ \bibinfo {author} {\bibfnamefont {F.}~\bibnamefont {Yates}},\ }\href@noop {} {\emph {\bibinfo {title} {Statistical Tables for Biological, Agricultural, and Medical Research}}}\ (\bibinfo  {publisher} {Oliver and Boyd},\ \bibinfo {year} {1939})\BibitemShut {NoStop}%
\bibitem [{\citenamefont {Bellman}(1961)}]{bellman1961adaptive}%
  \BibitemOpen
  \bibfield  {author} {\bibinfo {author} {\bibfnamefont {R.}~\bibnamefont {Bellman}},\ }\href@noop {} {\emph {\bibinfo {title} {Adaptive Control Processes: A Guided Tour}}}\ (\bibinfo  {publisher} {Princeton University Press},\ \bibinfo {address} {Princeton, NJ},\ \bibinfo {year} {1961})\BibitemShut {NoStop}%
\bibitem [{\citenamefont {Bishop}(2006)}]{bishop2006pattern}%
  \BibitemOpen
  \bibfield  {author} {\bibinfo {author} {\bibfnamefont {C.~M.}\ \bibnamefont {Bishop}},\ }\href@noop {} {\emph {\bibinfo {title} {Pattern Recognition and Machine Learning}}}\ (\bibinfo  {publisher} {Springer},\ \bibinfo {address} {New York},\ \bibinfo {year} {2006})\BibitemShut {NoStop}%
\bibitem [{\citenamefont {Hastie}\ \emph {et~al.}(2009)\citenamefont {Hastie}, \citenamefont {Tibshirani},\ and\ \citenamefont {Friedman}}]{hastie2009elements}%
  \BibitemOpen
  \bibfield  {author} {\bibinfo {author} {\bibfnamefont {T.}~\bibnamefont {Hastie}}, \bibinfo {author} {\bibfnamefont {R.}~\bibnamefont {Tibshirani}},\ and\ \bibinfo {author} {\bibfnamefont {J.}~\bibnamefont {Friedman}},\ }\href@noop {} {\emph {\bibinfo {title} {The Elements of Statistical Learning: Data Mining, Inference, and Prediction}}},\ \bibinfo {edition} {2nd}\ ed.\ (\bibinfo  {publisher} {Springer},\ \bibinfo {address} {New York, NY, USA},\ \bibinfo {year} {2009})\BibitemShut {NoStop}%
\bibitem [{\citenamefont {Palu{\v{s}}}(2007)}]{paluvs2007nonlinearity}%
  \BibitemOpen
  \bibfield  {author} {\bibinfo {author} {\bibfnamefont {M.}~\bibnamefont {Palu{\v{s}}}},\ }\bibfield  {title} {\bibinfo {title} {From nonlinearity to causality: statistical testing and inference of physical mechanisms underlying complex dynamics},\ }\href@noop {} {\bibfield  {journal} {\bibinfo  {journal} {Contemporary physics}\ }\textbf {\bibinfo {volume} {48}},\ \bibinfo {pages} {307} (\bibinfo {year} {2007})}\BibitemShut {NoStop}%
\bibitem [{\citenamefont {Lancaster}\ \emph {et~al.}(2018)\citenamefont {Lancaster}, \citenamefont {Iatsenko}, \citenamefont {Pidde}, \citenamefont {Ticcinelli},\ and\ \citenamefont {Stefanovska}}]{lancaster2018surrogate}%
  \BibitemOpen
  \bibfield  {author} {\bibinfo {author} {\bibfnamefont {G.}~\bibnamefont {Lancaster}}, \bibinfo {author} {\bibfnamefont {D.}~\bibnamefont {Iatsenko}}, \bibinfo {author} {\bibfnamefont {A.}~\bibnamefont {Pidde}}, \bibinfo {author} {\bibfnamefont {V.}~\bibnamefont {Ticcinelli}},\ and\ \bibinfo {author} {\bibfnamefont {A.}~\bibnamefont {Stefanovska}},\ }\bibfield  {title} {\bibinfo {title} {Surrogate data for hypothesis testing of physical systems},\ }\href@noop {} {\bibfield  {journal} {\bibinfo  {journal} {Physics Reports}\ }\textbf {\bibinfo {volume} {748}},\ \bibinfo {pages} {1} (\bibinfo {year} {2018})}\BibitemShut {NoStop}%
\bibitem [{\citenamefont {Vejmelka}\ \emph {et~al.}(2015)\citenamefont {Vejmelka}, \citenamefont {Pokorn{\'a}}, \citenamefont {Hlinka}, \citenamefont {Hartman}, \citenamefont {Jajcay},\ and\ \citenamefont {Palu{\v{s}}}}]{vejmelka2015non}%
  \BibitemOpen
  \bibfield  {author} {\bibinfo {author} {\bibfnamefont {M.}~\bibnamefont {Vejmelka}}, \bibinfo {author} {\bibfnamefont {L.}~\bibnamefont {Pokorn{\'a}}}, \bibinfo {author} {\bibfnamefont {J.}~\bibnamefont {Hlinka}}, \bibinfo {author} {\bibfnamefont {D.}~\bibnamefont {Hartman}}, \bibinfo {author} {\bibfnamefont {N.}~\bibnamefont {Jajcay}},\ and\ \bibinfo {author} {\bibfnamefont {M.}~\bibnamefont {Palu{\v{s}}}},\ }\bibfield  {title} {\bibinfo {title} {Non-random correlation structures and dimensionality reduction in multivariate climate data},\ }\href@noop {} {\bibfield  {journal} {\bibinfo  {journal} {Climate Dynamics}\ }\textbf {\bibinfo {volume} {44}},\ \bibinfo {pages} {2663} (\bibinfo {year} {2015})}\BibitemShut {NoStop}%
\bibitem [{\citenamefont {Runge}\ \emph {et~al.}(2015)\citenamefont {Runge}, \citenamefont {Petoukhov}, \citenamefont {Donges}, \citenamefont {Hlinka}, \citenamefont {Jajcay}, \citenamefont {Vejmelka}, \citenamefont {Hartman}, \citenamefont {Marwan}, \citenamefont {Palu{\v{s}}},\ and\ \citenamefont {Kurths}}]{runge2015identifying}%
  \BibitemOpen
  \bibfield  {author} {\bibinfo {author} {\bibfnamefont {J.}~\bibnamefont {Runge}}, \bibinfo {author} {\bibfnamefont {V.}~\bibnamefont {Petoukhov}}, \bibinfo {author} {\bibfnamefont {J.~F.}\ \bibnamefont {Donges}}, \bibinfo {author} {\bibfnamefont {J.}~\bibnamefont {Hlinka}}, \bibinfo {author} {\bibfnamefont {N.}~\bibnamefont {Jajcay}}, \bibinfo {author} {\bibfnamefont {M.}~\bibnamefont {Vejmelka}}, \bibinfo {author} {\bibfnamefont {D.}~\bibnamefont {Hartman}}, \bibinfo {author} {\bibfnamefont {N.}~\bibnamefont {Marwan}}, \bibinfo {author} {\bibfnamefont {M.}~\bibnamefont {Palu{\v{s}}}},\ and\ \bibinfo {author} {\bibfnamefont {J.}~\bibnamefont {Kurths}},\ }\bibfield  {title} {\bibinfo {title} {Identifying causal gateways and mediators in complex spatio-temporal systems},\ }\href@noop {} {\bibfield  {journal} {\bibinfo  {journal} {Nature communications}\ }\textbf {\bibinfo {volume} {6}},\ \bibinfo {pages} {8502} (\bibinfo {year} {2015})}\BibitemShut {NoStop}%
\bibitem [{\citenamefont {Shannon}(1948)}]{Shannon:1948}%
  \BibitemOpen
  \bibfield  {author} {\bibinfo {author} {\bibfnamefont {C.~E.}\ \bibnamefont {Shannon}},\ }\bibfield  {title} {\bibinfo {title} {A mathematical theory of communication},\ }\href@noop {} {\bibfield  {journal} {\bibinfo  {journal} {Bell Sys. Tech. J.}\ }\textbf {\bibinfo {volume} {27}},\ \bibinfo {pages} {379} (\bibinfo {year} {1948})}\BibitemShut {NoStop}%
\end{thebibliography}%

\end{document}